%% file: paper.tex
\documentclass{article}
\usepackage{graphicx}
\input axel
\def\tx{\tilde x}
\def\ty{\tilde y}
\def\tr{\tilde r}
\def\xtx{{\partial\ln\tx\over\partial\ln x}}
\def\ytx{{\partial\ln\ty\over\partial\ln x}}
\def\rtx{{\partial\ln\tr\over\partial\ln x}}
\def\xty{{\partial\ln\tx\over\partial\ln y}}
\def\yty{{\partial\ln\ty\over\partial\ln y}}
\def\rty{{\partial\ln\tr\over\partial\ln y}}
\def\xt{\left({\tx\over\tr}\right)^2}
\def\yt{\left({\ty\over\tr}\right)^2}

\def\mxtx{\displaystyle{\partial\ln\tx\over\partial\ln x}}
\def\mytx{\displaystyle{\partial\ln\ty\over\partial\ln x}}
\def\mxty{\displaystyle{\partial\ln\tx\over\partial\ln y}}
\def\myty{\displaystyle{\partial\ln\ty\over\partial\ln y}}
\def\mxt{\left(\displaystyle{\tx\over\tr}\right)^2}
\def\myt{\left(\displaystyle{\ty\over\tr}\right)^2}

\usepackage{url}
\usepackage{fancybox}
\usepackage{psboxit}

\title{Computational aspects of astrophysical MHD and turbulence}
\author{Axel Brandenburg\\
Nordita, Blegdamsvej 17, DK 2100 Copenhagen \O, Denmark;\\
Department of Mathematics, University of Newcastle upon Tyne NE1 7RU, UK}
\date{}
\begin{document}
\maketitle

\begin{abstract}
The advantages of high-order finite difference scheme for astrophysical
MHD and turbulence simulations are highlighted. A number of
one-dimensional test cases are presented ranging from various shock
tests to Parker-type wind solutions. Applications to magnetized
accretion discs and their associated outflows are discussed. Particular
emphasis is placed on the possibility of dynamo action in
three-dimensional turbulent convection and shear flows, which is relevant to
stars and astrophysical discs. The generation of large scale fields is
discussed in terms of an inverse magnetic cascade and the consequences
imposed by magnetic helicity conservation are reviewed with particular
emphasis on the issue of $\alpha$-quenching.
\end{abstract}

\section{Introduction}

Over the past 20 years multidimensional astrophysical gas simulations
have become a primary tool to understand the formation, evolution, and
the final fate of stars, galaxies, and their surrounding medium. The
assumption that those processes happen smoothly and in a non-turbulent
manner can at best be regarded as a first approximation. This is evidenced
by the ever improving quality of direct imaging techniques using space
telescopes for example. At the same time not only have computers become
large enough to run three-dimensional simulations with relatively little
effort, there have also been substantial improvements in the algorithms
that are used. In fact, there is now a vast literature on numerical
astrophysics. An excellent book was published recently by LeVeque \ea
(1998) where both numerical methods and astrophysical applications were
discussed in great detail. Most of the applications focused however
on rather more `violent' processes such as supersonic jets, supernova
explosions, core collapse, and on radiative transfer problems, while
hydromagnetic phenomena and turbulence problems where only touched
upon briefly.  Meanwhile, hydromagnetic turbulence simulations have
become crucial for understanding viscous dissipation in accretion discs
(Hawley, Gammie, \& Balbus 1995), and for understanding magnetic field
generation by dynamo action in discs (Brandenburg \ea 1995, 1996a, Hawley,
Gammie, \& Balbus 1996, Stone \ea 1996), stars (Nordlund \ea 1992,
Brandenburg \ea 1996b), and planets (Glatzmaier \& Roberts 1995, 1996).

Much of the present day astrophysical hydrodynamic work is based on
the ZEUS code, which has been documented in great detail and described
with a number of test cases in a series of papers by Stone \& Norman
(1992a,b). The main advantage is its flexibility in dealing with arbitrary
orthogonal coordinates which makes the code applicable to a wide variety
of astrophysical systems. The code, which is freely available on the net,
uses artificial viscosity for stability and shock capturing, and is based
on an operator split method with second-order finite differences on a
staggered mesh. Another approach used predominantly in turbulence
research are spectral methods (e.g.\ Canuto \ea 1988), which have the
advantage of possessing high accuracy.  Although these methods are most
suitable for incompressible flows (imposing the solenoidality condition
is then straightforward), they have also been applied to compressible flows
(e.g.\ Passot \& Pouquet 1987). As a compromise one may resort to high
order finite difference methods, which have the advantage of being easy
to implement and yet have high accuracy. Compact methods (e.g.\
Lele 1992) are a special variety of high order finite difference methods,
but the truncation error is smaller than for an explicit scheme of the
same order. Compact schemes have been used by Nordlund \& Stein (1990)
in simulations of solar convection (Stein \& Nordlund 1989, 1998) and
convective dynamos (Nordlund \ea 1992, Brandenburg \ea 1996b), for example.

The use of compact methods involves solving tridiagonal matrix equations,
making this method essentially nonlocal in that all points are now coupled
at once. This is problematic for massively parallel computations, which is
why Nordlund \& Galsgaard (1995, see also Nordlund, Galsgaard, \&
Stein 1994) began to use explicit high order schemes for their work on
coronal heating by reconnection (Galsgaard \& Nordlund 1996, 1997a,b). In
their code the equations are solved in a semi-conservative fashion using
a staggered mesh. This code was also used by Padoan, Nordlund, \& Jones
(1997) and Padoan, \& Nordlund (1999) in models of isothermal interstellar
turbulence in molecular clouds, and by R\"ognvaldsson, Nordlund, \&
Sommer-Larsen (2001) in simulations of cooling flows and galaxy formation.

A somewhat different code was used by Brandenburg (1999) and Bigazzi
(1999) in simulations of the inverse magnetic cascade, by Kerr \&
Brandenburg (1999) in work on the possibility of a singularity of
the nonresistive and inviscid MHD equations, and by Sanchez-Salcedo
\& Brandenburg (1999, 2001) in simulations of dynamical friction. A
two-dimensional version of the code modelling outflows from magnetized
accretion discs has been described by Brandenburg \ea (2000). This code
uses sixth order explicit finite differences in space and third order
Runge-Kutta timestepping. It employs central finite differences,
so the extra cost of recentering a large number of variables between
staggered meshes each timestep is avoided.

Apart from high numerical accuracy, another important requirement for
astrophysical gas simulations is the capability to deal with a large
dynamical range in density and temperature. This requirement favors
the use of non-conservative schemes, because then logarithmic variables
can be used which vary much less than linear density and energy density
per unit volume. Solving the nonconservative form of the equations can
be more accurate than solving the conservative form. The conservation
properties can then be used as an indicator for the overall accuracy.

In this chapter we concentrate on numerical astrophysical turbulence
aspects starting with a discussion of different numerical methods and
a description of the results of various numerical test problems. This is a
good way of assessing the quality of a numerical scheme and of comparing
with other methods; see Stone \& Norman (1992a,b) for a series of tests
using the ZEUS code. After that we discuss particular astrophysical
applications including stellar convection, accretion disc turbulence and
associated outflows, as well as the generation of magnetic fields (small
scale and large scale) from turbulence in various astrophysical settings.

\section{The Navier-Stokes equations}
\label{NavierStokes}

The discussion of magnetic fields will be postponed until later, because
the inclusion of the Lorentz force in the momentum equation is
straightforward. We begin by writing down the Navier-Stokes equations in
nonconservative form and rewrite them such that the main thermodynamical
variables are entropy and either logarithmic density or potential
enthalpy. These variables have the advantage of varying spatially much
less than for example linear pressure and density.

The primitive form of the continuity equation is
\EQ
{\partial\rho\over\partial t}=-\nab\cdot(\rho\uu),
\EN
which means that the local change of density is given by the divergence of
the mass flux at that point. The Navier-Stokes equation can be written as
\EQ
\rho{\DD\uu\over\DD t}=-\nab p
-\rho\nab\Phi+\FF+\nab\cdot\ttau,
\EN
where $\DD/\DD t=\partial/\partial t+\uu\cdot\nab$ is the advective
derivative, $p$ is the pressure, $\Phi$ is the gravitational potential,
$\FF$ is a body force (e.g., the Lorentz force), and $\ttau$ is the
stress tensor.

The Navier-Stokes equation is here written in terms of forces per unit volume.
As argued above, if the density contrast is large it is advantageous to write
it in terms of forces per unit mass and to divide by $\rho$. Before we can
replace $p$ and $\rho$ by entropy and logarithmic density or potential entropy
we first have to define some thermodynamic quantities.

Internal energy, $e$, and specific enthalpy, $h$ are related to each other by
\EQ
h=e+pv,
\EN
where $v=1/\rho$ is the specific volume and $\rho$ the density.
The specific entropy is defined by
\EQ
T\dd s=\dd e+p\dd v,
\EN
where $T$ is temperature.
The specific heats at constant pressure and constant volume are
defined as $c_p=\dd h/\dd T|_p$ and $c_v=\dd e/\dd T|_v$, their ratio
is $\gamma=c_p/c_v$, and their difference is ${\cal R}/\mu=c_p-c_v$,
where ${\cal R}$ is the universal gas constant and $\mu$ the specific
molecular weight.

In the following we assume $c_p$ and $c_v$ to be constant for all processes
considered. Ionization and recombination processes are therefore ignored
here, although this is not a major obstacle; see, e.g., simulations of
Nordlund (1982, 1985), Steffen, Ludwig, \& Kr\"u{\ss} (1989), Stein \&
Nordlund (1989, 1998), Rast \ea (1993), and Rast \& Toomre (1993a,b)
where realistic equations of state have been used.

We now assume that $c_p$ and $c_v$ are constant, so internal energy and
specific enthalpy are given by
\EQ
h=c_p T\quad\mbox{and}\quad e=c_v T,
\EN
This allows us then to write
the specific entropy (up to an additive constant) as
\EQ
s=c_v\ln p-c_p\ln\rho.
\EN
The pressure gradient term in the momentum equation can then be
written as
\EQ
{1\over\rho}\nab p=
{p\over\rho}\nab\ln p=
{\gamma p\over\rho}(\nab\ln\rho+\nab s/c_p)=
c_{\rm s}^2(\nab\ln\rho+\nab s/c_p),
\EN
where we have used
\EQ
c_{\rm s}^2=\gamma p/\rho
={\gamma p_0\over\rho_0}\exp[(\gamma-1)\ln(\rho/\rho_0)+\gamma s/c_p]
=c_{\rm s0}^2\left({\rho\over\rho_0}\right)^{\gamma-1}
\exp(\gamma s/c_p),
\label{adSoundSpeed}
\EN
where $c_{\rm s}$ is the adiabatic sound speed, and
$c_{\rm s0}^2=\gamma p_0/\rho_0$.

With these preparations the evolution of velocity $\uu$, logarithmic
density $\ln\rho$, and specific entropy $s$ can be expressed as follows:
\\

\SHADOWBOX{
\EQ
{\DD\uu\over\DD t}=-c_{\rm s}^2(\nab\ln\rho+\nab s/c_p)
-\nab\Phi+\ff+{1\over\rho}\nab\cdot(2\nu\rho\SSSS),
\label{dudt}
\EN
\EQ
{\DD\ln\rho\over\DD t}=-\nab\cdot\uu,
\label{drhodt}
\EN
\EQ
T{\DD s\over\DD t}=2\nu\SSSS^2+\Gamma-\rho\Lambda,
\label{Tdsdt}
\EN
}\\

\noindent
where $\ff=\FF/\rho$ is the body force per unit mass,
$\Gamma$ and $\Lambda$ are heating
and cooling functions, $\nu$ kinematic viscosity and $\SSSS$ is the
(traceless) strain tensor with the components
\EQ
{\sf S}_{ij}=\half(u_{i,j}+u_{j,i}-\twothird\delta_{ij}u_{k,k}).
\label{rate-of-strain}
\EN
In the presence of an additional kinematic bulk viscosity, $\zeta$,
the term $2\nu{\sf S}_{ij}$ under the divergence in \Eq{dudt} would
need to be replaced by $2\nu{\sf S}_{ij}+\zeta\delta_{ij}\nab\cdot\uu$,
and the viscous heating term, $2\nu\SSSS^2$, in \Eq{Tdsdt} would need
to be replaced by $2\nu\SSSS^2+\zeta(\nab\cdot\uu)^2$.

Instead of using $\rho$ as a dependent variable an can also use the
specific enthalpy $h$, which allows us to write the pressure gradient as
\EQ
-{1\over\rho}\nab p=-\nab h+T\nab s.
\EN
This formulation is particularly useful if the entropy is nearly constant
(or if the gas is barotropic, i.e.\ $p=p(\rho)$) and if there is a gravitational potential
$\Phi$, so that the potential enthalpy $H\equiv h+\Phi$ can be used
as dependent variable. In order to express \Eq{drhodt} in terms of $h$
we write down the total differential of the specific entropy,
\EQ
\dd s/c_p={1\over\gamma}\dd\ln p-\dd\ln\rho=
{1\over\gamma}\dd\ln h-\left(1-{1\over\gamma}\right)\dd\ln\rho,
\EN
so
\EQ
{\DD\ln h\over\DD t}=\gamma{\DD s/c_p\over\DD t}
+(\gamma-1){\DD\ln\rho\over\DD t}.
\EN
Furthermore, $T\nab s=h\nab s/c_p$, and so the
final set of equations is
\\

\SHADOWBOX{
\EQ
{\DD\uu\over\DD t}=-\nab H+h\nab s/c_p+\ff+\nab\cdot(2\nu\rho\SSSS),
\label{basic_u}
\EN
\EQ
{\DD s/c_p\over\DD t}=
{1\over h}\left(2\nu\SSSS^2+\Gamma-\rho\Lambda\right),
\label{basic_s}
\EN
\EQ
{\DD H\over\DD t}=\uu\cdot\nab\Phi+\gamma h{\DD s/c_p\over\DD t}
-c_{\rm s}^2\nab\cdot\uu,
\label{basic_h}
\EN
}\\

\noindent
where we have absorbed $\Phi$ in the potential enthalpy $H=h+\Phi$. In
this formulation the density can be recovered as
\EQ
\rho=\rho_0\left[\left(1-{1\over\gamma}\right)
\left({h\over c_{\rm s0}^2}\right)
\,e^{-\gamma s/c_p}\right]^{1/(\gamma-1)}
\EN
(in dimensional form) or, for $\gamma=5/3$ and in nondimensional form
(where $\rho_0=p_0=c_p=1$),
\EQ
\rho=(0.4h)^{1.5}\,e^{-2.5s}.
\EN
We shall use either of the two sets of the equations, \eqss{dudt}{Tdsdt}
or \eqss{basic_u}{basic_h}, in some of the
following sections, especially in connection with shock tests and stellar
wind problems. In these cases the gravity potential $\Phi$ is important
and it turns out that the potential enthalpy $H\equiv h+\Phi$ varies
only very little near the central object even though $\Phi$ itself tends
to become singular. 

The heating and cooling terms ($\Gamma$ and $\Lambda$) are important
for example in the case of interstellar turbulence which is driven
primarily by supernova explosions which inject a certain amount of
thermal energy ($\int\rho\Gamma\dd V$) with each supernova explosion. MHD
turbulence simulations of this type were performed recently by Korpi
\ea (1999). At the same time there is cooling through various processes
(e.g.\ bremsstrahlung at high temperatures) which transports energy either
nonlocally via a cooling term $\Lambda(T)$, or locally via
thermal conduction or radiative diffusion. In the radiative diffusion approximation
we express $\Lambda$ as
$-\rho^2\Lambda=\nab\cdot K\nab T$, where $K$ is the radiative
conductivity which is in general a function of temperature and density.
The radiative diffusivity (which has the same dimensions
as the kinematic viscosity $\nu$) is given by $\chi=K/(\rho c_p)$, so
\EQ
-\rho\Lambda/(c_pT)={1\over\rho c_pT}\nab\cdot\rho\chi c_p\nab T.
\EN
Since we shall use a nonconservative scheme with centered finite
differences is it important to isolate second derivative terms, so
\EQ
-\rho\Lambda/h=\chi(\nabla^2\ln T+\nab\ln p\cdot\nab\ln T),
\EN
where we have assumed for simplicity that $\chi$ is constant. In terms
of $s/c_p$ and $\ln\rho$ we have
\EQ
-\rho\Lambda/h=\chi\gamma\left[\nabla^2s/c_p+\nabla_{\rm ad}\nabla^2\ln\rho
+\gamma(\nab s/c_p+\nab\ln\rho)\cdot
(\nab s/c_p+\nabla_{\rm ad}\nab\ln\rho)\right],
\label{Lambdacond}
\EN
where $\nabla_{\rm ad}=1-1/\gamma$ is a commonly used abbreviation
in stellar astrophysics. For $\gamma=5/3$ we have $\nabla_{\rm
ad}=2/5=0.4$. We shall use \Eq{Lambdacond} later in connection with shock
and wind calculations. However, we begin by discussing first a suitable
numerical scheme which will be used in most of the cases presented below.

\section{The advantage of higher-order derivative schemes}
\label{Scentered}

Spectral methods are commonly used in almost all studies of ordinary
(usually incompressible) turbulence. The use of this method is justified
mainly by the high numerical accuracy of spectral schemes. Alternatively,
one may use high order finite differences that are faster to compute
and that can possess almost spectral accuracy.  Nordlund \& Stein
(1990) and Brandenburg \ea (1995) use high order finite difference
methods, for example fourth and sixth order compact schemes (Lele
1992).\footnote{The fourth order compact scheme is really identical to
calculating derivatives from a cubic spline, as was done in Nordlund \&
Stein (1990). In the book by Collatz (1966) the compact methods are also
referred to as {\it Hermitian methods} or as {\it Mehrstellen-Verfahren},
because the derivative in one point is calculated using the derivatives
in neighboring points.}

In this section we demonstrate, using simple test problems, some of
the advantages of high order schemes. We begin by defining various
schemes including their truncation errors and their high wavenumber
characteristics. We consider centered finite differences of 2nd, 4th,
6th, 8th, and 10th order, which are given respectively by the formulae
\EQ
f'_i=(-f_{i-1}+f_{i+1})/(2\delta x),
\EN
\EQ
f'_i=(f_{i-2}-8f_{i-1}+8f_{i+1}-f_{i+2})/(12\delta x),
\label{first_4th}
\EN
\EQ
f'_i=(-f_{i-3}+9f_{i-2}-45f_{i-1}
+45f_{i+1}-9f_{i+2}+f_{i+3})/(60\delta x),
\EN
\EQ
f'_i=(3f_{i-4}-32f_{i-3}+168f_{i-2}-672f_{i-1}
+672f_{i+1}-168f_{i+2}+32f_{i+3}-3f_{i+4})/(840\delta x),
\EN
\EQ
f'_i=(-2f_{i-5}+25f_{i-4}-150f_{i-3}+600f_{i-2}-2100f_{i-1}
+2100f_{i+1}-...)/(2520\delta x),
\label{first_10th}
\EN
for the first derivative, and
\EQ
f''_i=(f_{i-1}-2f_i+f_{i+1})/(\delta x^2),
\EN
\EQ
f''_i=(-f_{i-2}+16f_{i-1}-30f_i
+16f_{i+1}-f_{i+2})/(12\delta x^2),
\EN
\EQ
f''_i=(2f_{i-3}-27f_{i-2}+270f_{i-1}-490f_i
+270f_{i+1}-27f_{i+2}+2f_{i+3})/(180\delta x^2),
\EN
\EQ
f''_i=(-9f_{i-4}+128f_{i-3}-1008f_{i-2}+8064f_{i-1}-14350f_i
+8064f_{i+1}-1008f_{i+2}+...)/(5040\delta x^2),
\EN
\EQ
f''_i=(8f_{i-5}-125f_{i-4}+1000f_{i-3}-6000f_{i-2}+42000f_{i-1}
-73766f_i+42000f_{i+1}-...)/(25200\delta x^2),
\EN
for the second derivative. The expressions for one-sided and semi-onesided
finite difference formulae are given in \App{Sonesided}.

\subsection{High wavenumber characteristics}

The chief advantage of high order schemes is their high fidelity at high
wavenumber. Suppose we differentiate the function $\sin kx$, we are
supposed to get $k\cos kx$, but when $k$ is close to the Nyquist
frequency, $k_{\rm Ny}\equiv\pi/\delta x$, where $\delta x$ is the mesh
spacing, numerical schemes yield {\it effective} wavenumbers, $k_{\rm
eff}$, that can be significantly less than the actual wavenumber $k$.
Here we calculate $k_{\rm eff}$ from
\EQ
(\cos kx)'_{\rm num}=-k_{\rm eff}\sin kx.
\EN
When $k=k_{\rm Ny}$, every centered difference scheme will give $k_{\rm
eff}=0$, because then the function values of $\cos kx$ are just $-1, +1,
-1, ...$, so the function values on the left and the right are the same,
and the difference that enters the scheme gives therefore zero.

If is useful to mention at this point that for a staggered mesh, where
the first derivative is evaluated {\it between} mesh points, the value
of the first derivative remains finite at the Nyquist frequency, provided
one does not need to remesh back to the original mesh. Especially in the
context with magnetic fields, however, remeshing needs to be done quite
frequently, which therefore diminishes the advantage of a staggered mesh.

\begin{figure}[h!]\begin{center}\includegraphics[width=.99\textwidth]{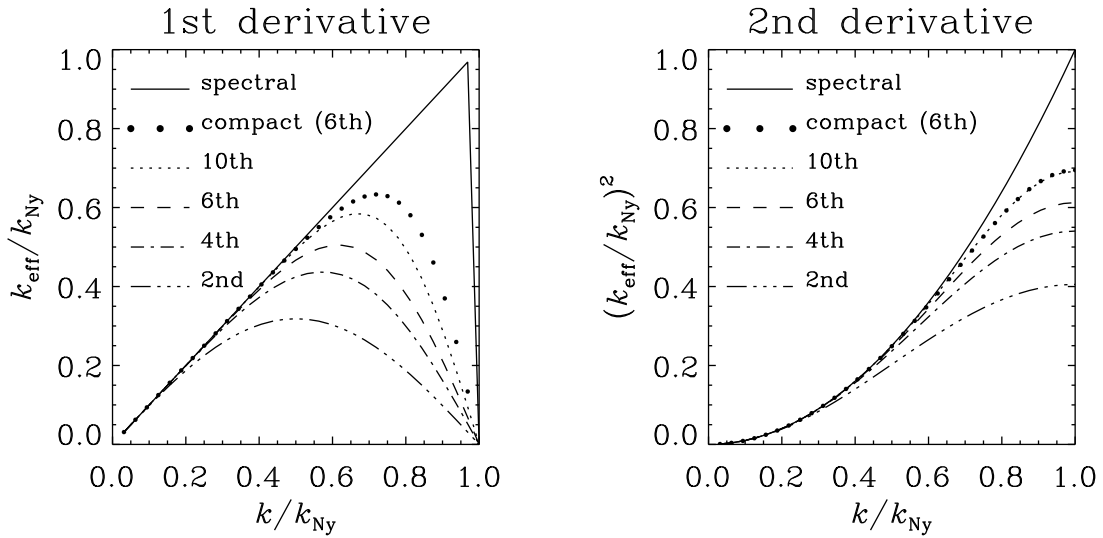}\end{center}\caption[]{
Effective wave numbers for first and second derivatives using different
schemes. Note that for the second derivatives the sixth order compact
scheme is almost equivalent to the tenth order explicit scheme. For the
first derivative the sixth order compact scheme is still superior to the
tenth order explicit scheme.
}\label{Fpkeff}\end{figure}

In \Fig{Fpkeff} we plot effective wavenumbers for different schemes.
Apart from the different {\it explicit} finite difference schemes
given above, we also consider a {\it compact} scheme of 6th order,
which can be written in the form
\EQ
{\textstyle{1\over3}}f'_{i-1}+f'_i+{\textstyle{1\over3}}f'_{i+1}
=(f_{i-2}-28f_{i-1}+28f_{i+1}-f_{i+2})/(36\delta x),
\EN
for the first derivative, and
\EQ
{\textstyle{2\over11}}f''_{i-1}+f''_i+{\textstyle{2\over11}}f''_{i+1}
=(3f_{i-2}+48f_{i-1}-102f_i+48f_{i+1}+3f_{i+2})/(44\delta x^2).
\EN
for the second derivative. As we have already mentioned in the introduction, this
scheme involves obviously solving tridiagonal matrix equations and is
therefore effectively nonlocal.

In the second panel of \Fig{Fpkeff} we have plotted effective
wavenumbers for second derivatives, which were calculated as
\EQ
(\cos kx)''_{\rm num}=-k_{\rm eff}^2\cos kx.
\EN
Of particular interest is the behavior of the second derivative at the
Nyquist frequency, because that is relevant for damping zig-zag modes.
For a second-order finite difference scheme $k_{\rm eff}^2$ is only 4,
which is less than half the theoretical value of $\pi^2=9.87$. For fourth,
sixth, and tenth order schemes this value is respectively 5.33, 6.04,
6.83. The last value is almost the same as for the 6th order compact
scheme, which is 6.86. Significantly stronger damping at the Nyquist
frequency can be obtained by using hyperviscosity, which Nordlund \&
Galsgaard (1995) treat as a quenching factor that diminishes the value
of the second derivative for wavenumbers that are small compared with
the Nyquist frequency. Accurate high order second derivatives (with no
quenching factors) are important when calculating the current $\JJ$ in
the Lorentz force $\JJ\times\BB$ from a vector potential $\AAA$ using
$-\mu_0\JJ=\nabla^2\AAA-\nab\nab\cdot\AAA$. This will be important in
the MHD calculations presented below.

\subsection{The truncation error}
\label{Struncation}

One can express $f_{i-1}$, $f_{i+1}$, etc, in terms of the derivatives
of $f$ at point $i$, so
\EQ
f_{i-1}=f_i-\delta x f_i'+{1\over2}\delta x^2f_i''-{1\over6}\delta x^3f_i'''+...
\EN
\EQ
f_{i+1}=f_i+\delta x f_i'+{1\over2}\delta x^2f_i''+{1\over6}\delta x^3f_i'''+...
\EN
Inserting this into the finite difference expressions yields for the
second order formula
\EQ
(f_i')_{\rm 2nd}\equiv(f_{i+1}-f_{i-1})/(2\delta x)
=f_i'+{1\over6}\delta x^2f_i'''.
\EN
The error scales quadratically with the mesh size, which is why the
method is called second order. The
truncation error is proportional to the third derivative of the function.
Because this is an odd derivative it corresponds to a dispersive (as
opposed to diffusive) error. Schemes that are only first order (or of
any odd order) have diffusive errors, and it is this what is sometimes
referred to as {\it numerical diffusivity}, which is not to be confused
with artificial diffusivity that is sometimes used for stability and shock
capturing. For the other schemes given in \Eqss{first_4th}{first_10th}
the truncation errors are
\EQ
(f_i')_{\rm 4th}=f_i'+3\times10^{-2}\,\delta x^4f_i^{\sf(v)},
\EN
\EQ
(f_i')_{\rm 6th}=f_i'+7\times10^{-3}\,\delta x^6f_i^{\sf(vii)},
\EN
\EQ
(f_i')_{\rm 10th}=f_i'+3\times10^{-4}\,\delta x^{10}f_i^{\sf(xi)}.
\EN
For the sixth order compact scheme the error scales like for the
sixth order explicit scheme, but the coefficient in front of the
truncation error is about ten times smaller, so
\EQ
(f_i')_{\rm 6th}^{\rm compact}=f_i'+5.1\times10^{-4}\,\delta x^6f_i^{\sf(vii)}.
\EN
For the second derivatives we have
\EQ
(f_i'')_{\rm 2nd}=f_i''+8\times10^{-2}\,\delta x^2f_i^{\sf(iv)},
\EN
\EQ
(f_i'')_{\rm 4th}=f_i''-1\times10^{-2}\,\delta x^4f_i^{\sf(vi)},
\EN
\EQ
(f_i'')_{\rm 6th}=f_i''+2\times10^{-3}\,\delta x^6f_i^{\sf(viii)},
\EN
\EQ
(f_i'')_{\rm 10th}=f_i''-5\times10^{-5}\,\delta x^{10}f_i^{\sf(xii)}.
\EN
Again, for the sixth order compact scheme the scaling is the same as for
the sixth order explicit scheme, but the coefficient in front of the
truncation error is about 5 times less, so
\EQ
(f_i'')_{\rm 6th}^{\rm compact}=f_i''+3.2\times10^{-4}\,\delta x^6f_i^{\sf(viii)}.
\EN
This information about the accuracy of schemes would obviously be of
little use if the various schemes did not perform well when
applied to real problems. For this reason we now begin by carrying out
various tests, including advection and shock tests.

\subsection{Advection tests}

As a first test we compare the various schemes by performing inviscid advection tests
and solve the equation $\DD f/\DD t=0$, i.e.\
\EQ
\dot{f}=-uf',
\label{advtest}
\EN
on a periodic mesh. It is advantageous to use a relatively small number
of meshpoints (here we use $N_x=8$ meshpoints), because that way we see
deficiencies most clearly. This case is actually also relevant to real
applications, because in practice one will always have small scale
structures that are just barely resolved.

After some time an initially sinusoidal signal will suffer a change in amplitude
and phase. We have calculated the amplitude and phase errors for schemes
of different spatial order. For the time integration we use high order
Runge-Kutta methods of 3rd or 4th order, RK3 and RK4, respectively. In
most cases considered below we use the RK3 scheme that allows reasonable
use of storage. It can be written in three steps (Rogallo 1981)
\EQ
\begin{array}{lll}
\mbox{1st step:}&
\quad f=f+\gamma_1\delta t\dot{f},&
\quad g=f+\zeta_1\delta t\dot{f},\\
\mbox{2nd step:}&
\quad f=g+\gamma_2\delta t\dot{f},&
\quad g=f+\zeta_2\delta t\dot{f},\\
\mbox{3rd step:}&
\quad f=g+\gamma_3\delta t\dot{f},&
\end{array}
\EN
where
\EQ
\gamma_1={8\over15},\quad
\gamma_2={5\over12},\quad
\gamma_3={3\over4},\quad
\zeta_1=-{17\over60},\quad
\zeta_2=-{5\over12}.
\EN
where, $f$ and $g$ always refer to the current value (so the same space
in memory can be used), but $\dot{f}$ is evaluated only once at the
beginning of each of the three steps at $t=t_0$,
$t_{1/3}=t_0+\gamma_1\delta t\approx t_0+0.5333\delta t$, and at
$t_{2/3}=t_0+(\gamma_1+\zeta_1+\gamma_2)\delta t=t_0+\twothird\delta t$.
Even more memory-effective are the so-called $2N$-schemes that require
one set of variables less to be hold in memory. Such schemes work for
arbitrarily high order, although not all Runge-Kutta schemes can be
written as $2N$-schemes (Williamson 1980, Stanescu \& Habashi 1998).
These schemes work iteratively according to the formula
\EQ
w_i=\alpha_i w_{i-1}+\delta t\,F(t_{i-1},u_{i-1}),\quad
u_i=u_{i-1}+\beta_i w_i.
\label{iterform0}
\EN
For a three-step scheme we have $i=1,...,3$.
In order to advance the variable $u$ from $u^{(n)}$ at time $t^{(n)}$
to $u^{(n+1)}$ at time $t^{(n+1)}=t^{(n)}+\delta h$ we set in \Eq{iterform0}
\EQ
u_0=u^{(n)}\quad\mbox{and}\quad u^{(n+1)}=u_3,
\EN
with $u_1$ and $u_2$ being intermediate steps. In order to be able to
calculate the first step, $i=1$, for which no $w_{i-1}\equiv w_0$ exists,
we have to require $\alpha_1=0$. Thus, we are left with 5 unknowns,
$\alpha_2$, $\alpha_3$, $\beta_1$, $\beta_2$, and $\beta_3$. Three
conditions follow from the fact that the scheme be third order, so we
have to have two more conditions. One possibility is the choose the
fractional times at which the right hand side is evaluated, for
example (0,~1/3,~2/3) or even (0,~1/2,~1). In the latter case the right hand
side is evaluated twice at the same time. It is therefore some sort of
`predictor-corrector' scheme. In the following these two schemes are
therefore referred to as `symmetric' and `predictor/corrector' schemes.
Yet another possibility is to require that
inhomogeneous equations of the form $\dot{u}=t^n$ with $n=1$ and 2 are
solved exactly. Such schemes are abbreviated as `inhomogeneous' schemes.
The detailed method of calculating the coefficients
for such third order Runge-Kutta schemes with $2N$-storage is discussed
in detail in \App{S2NRK3}. Several possible sets of coefficients are
listed in \Tab{Ttab_2N-RK3} and compared with the favorite scheme of Williamson (1980).
Note that the first order Euler scheme corresponds to $\beta_1=1$ and the
classic second order to $\alpha_2=-1/2$, $\beta_1=1/2$, and $\beta_2=1$.

\begin{table}[htb]\caption{
Possible coefficients for different $2N$-RK3 schemes.
}\vspace{12pt}\centerline{\begin{tabular}{lccccccc}
\hline
label & $\alpha_2$ & $\alpha_3$ & $\beta_1$ & $\beta_2$ & $\beta_3$ &
$(t_1-t_0)/\delta t$ & $(t_2-t_0)/\delta t$ \\
\hline
symmetric (i)       &  $-2/3$  &   $-1$   & 1/3 &  1  & 1/2 & 1/3 & 2/3 \\
symmetric (ii)      &  $-1/3$  &   $-1$   & 1/3 & 1/2 &  1  & 1/3 & 2/3 \\
predictor/corrector &  $-1/4$  &  $-4/3$  & 1/2 & 2/3 & 1/2 & 1/2 &  1  \\
inhomogeneous       & $-17/32$ & $-32/27$ & 1/4 & 8/9 & 3/4 & 1/4 & 2/3 \\
quadratic           & $-0.367$ & $-1.028$ &0.308&0.540&  1  &0.308&0.650\\
Williamson (1980)   &  $-5/9$  &$-153/128$& 1/3 &15/16& 8/15& 1/3 & 3/4 \\
\label{Ttab_2N-RK3}\end{tabular}}\end{table}

We estimate the accuracy of these schemes by solving the homogeneous
differential equation
\EQ
\dot{u}=nu^{1-1/n},\quad u(1)=1.
\label{ode_error_check}
\EN
The exact solution is $u=t^n$. In \Tab{Ttab_2N-RK3-err}
we list the rms error with respect to the exact solution,
for the range $1<t\leq 4$ and fixed time step $\delta t=0.1$
using $n=-1$, 2 or 3.

\begin{table}[htb]\caption{
Errors (in units of $10^{-6}$) for different $2N$-RK3 schemes, obtained by solving
\Eq{ode_error_check} in the range $1<t\leq 4$ with $\delta t=0.1$
and different values of $n$.
}\vspace{12pt}\centerline{\begin{tabular}{lrrr}
\hline
label & $n=-1$ & $n=2$ & $n=3$ \\
\hline
symmetric (i)       &  69 &  103  &   193 \\
symmetric (ii)      & 226 &  119  &   411 \\
predictor/corrector & 469 &  346  &  1068 \\
inhomogeneous       &  84 &{\bf6} &{\bf97}\\
quadratic           & 197 &   94  &   339 \\
Williamson (1980)   &{\bf68}& 10  &   123 \\
for comparison: RK3 &  66 &   13  &   134 \\
\label{Ttab_2N-RK3-err}\end{tabular}}\end{table}

The length of the time step must
always be a certain fraction of the Courant-Friedrich-Levy condition,
i.e.\ $\delta t=k_{\rm CFL}\delta x/U_{\max}$, where $k_{\rm CFL}={\cal O}(1)$
and $U_{\max}$ is the maximum transport speed in the system (taking into
account advection, sound waves, viscous transport, etc). Too long a time
step can not only lead to instability, but it also increases the error.

In \Tab{Ttab1} we give amplitude and phase errors for the various
schemes. The most important conclusion to be drawn from this is the fact
that low order spatial schemes result in large {\it phase errors}. In
the case of a second order scheme the phase error is $36^\circ$ after a
single passage of a barely resolved wave through a periodic mesh. Higher
order schemes have easily a hundred times smaller phase errors. The
amplitude error, on the other hand, is virtually not affected by the
spatial order of the scheme. The amplitude error is mainly affected both
by the temporal order of the scheme and by the length of the timestep;
see also \Tab{Ttab2}. Therefore, high order schemes with low dissipation
and dispersion are particularly important in computational acoustics
(Stanescu \& Habashi 1998). However, in applications to turbulence a
certain amount of viscosity is always necessary. This would decrease the
amplitude of the wave further and would eventually be even more important.
(This additional viscosity could be the real one, an explicit
artificial, or an implicit numerical viscosity that would result from
the discretisation error or the numerical scheme; see \Sec{Struncation}.)

\begin{table}[htb]\caption{
Amplitude and phase errors for inviscid advection of the function
$f=\cos kx$ with $k=2\pi$ and $N=8$ meshpoints after $t=20$, corresponding
to 20 revolutions in a periodic mesh. The amplitude error is counted
positive when the amplitude decreases. A positive phase error means that
the solution lags behind the theoretical one.
}\vspace{12pt}\centerline{\begin{tabular}{cccccc}
\hline
$\delta t$ scheme & 2nd order & 4th order & 6th order & 10th order & spectral \\
\hline
\hline
RK4               &     1\%   &     2\%   &     2\%   &     2\%   &     2\%   \\
$c_{\delta t}=0.6$&$720^\circ$& $87^\circ$& $13^\circ$&$2.9^\circ$&$2.8^\circ$\\
\hline
RK3               &    10\%   &    14\%   &    14\%   &    15\%   &    15\%   \\
$c_{\delta t}=0.4$&$716^\circ$& $83^\circ$&$8^\circ$&$-2.1^\circ$&$-2.3^\circ$\\
\hline
RK3               &     4\%   &   6.3\%   &   6.6\%   &   6.6\%   &   6.6\%   \\
$c_{\delta t}=0.3$&$717^\circ$& $84^\circ$&$10^\circ$&$-0.6^\circ$&$-0.8^\circ$\\
\hline
\label{Ttab1}\end{tabular}}\end{table}

\begin{table}[htb]\caption{
Dependence of the amplitude and phase errors
on the length of the timestep and the scheme used for the timestep.
In all cases spectral $x$-derivatives are used.
}\vspace{12pt}\centerline{\begin{tabular}{cccccc}
\hline
$\delta t$ scheme &    RK3     &    RK3     &   RK4     &   RK4    &    RK4  \\
$c_{\delta t}$    &    0.3     &    0.4     &   0.4     &   0.6    &    1.0  \\ 
\hline
\hline
amplitude error   &    6.6\%   &     15\%   &   0.3\%   &   2\%    &    21\% \\  
phase error       &$-0.8^\circ$&$-2.3^\circ$&$0.6^\circ$&$2.8^\circ$&$18^\circ$\\  
\hline
\label{Ttab2}\end{tabular}}\end{table}

\begin{figure}[h!]\begin{center}\includegraphics[width=.99\textwidth]{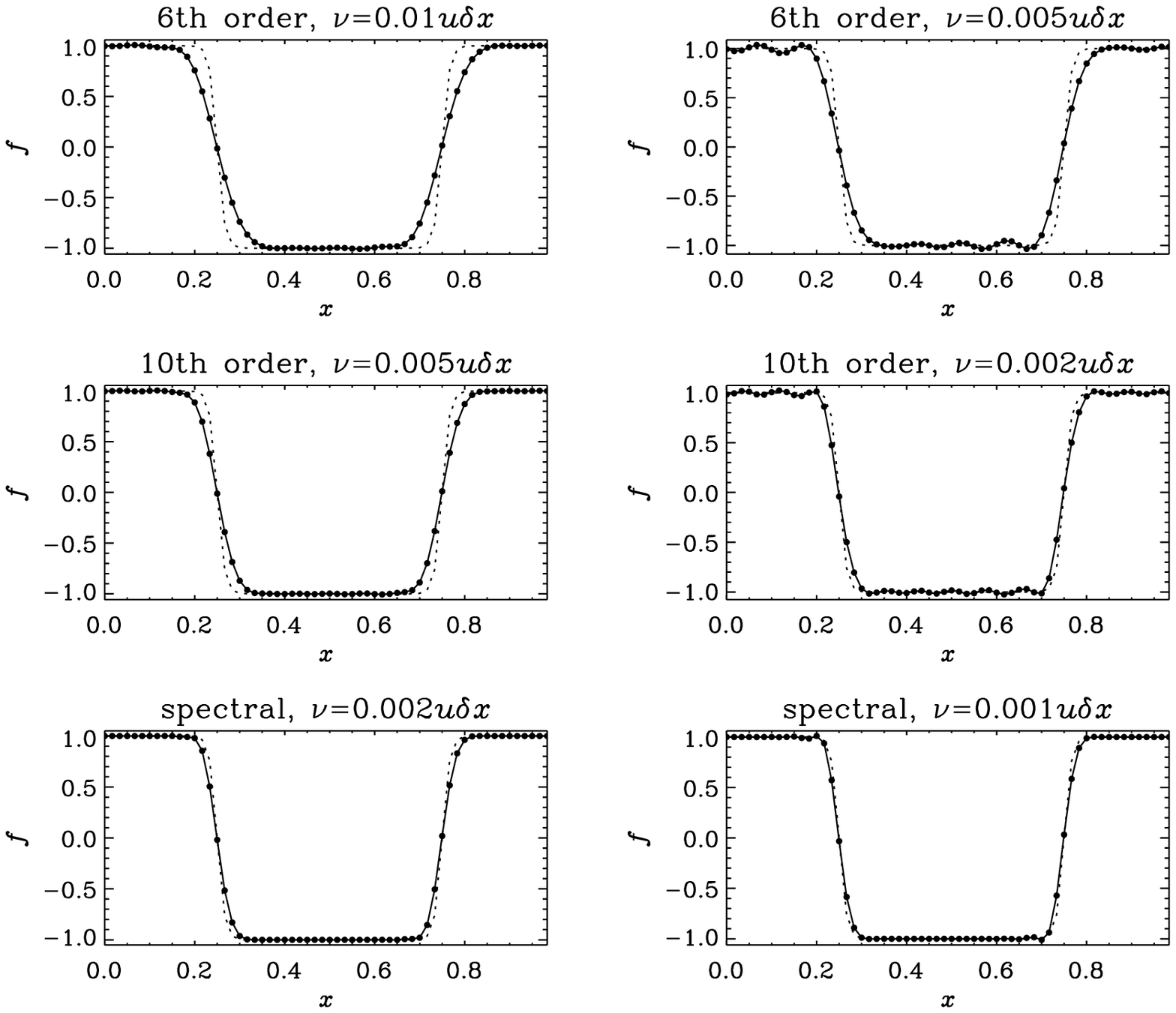}\end{center}\caption[]{
Resulting profile after advecting a step-like function 5 times
through the periodic mesh. The dots on the solid line give the location
of the function values at the computed meshpoints and the dotted line
gives the original profile. For the panels on the right hand side the
diffusion coefficient is too small and the profile shows noticeable
wiggles. $\delta x=1/60$.
}\label{Fpadvect_all}\end{figure}

A common criticism of high order schemes is their tendency to produce Gibbs
phenomena (ripples) near discontinuities. Consequently one needs a small
amount of diffusion to damp out the modes near the Nyquist frequency. Thus,
one needs to replace \Eq{advtest} by the equation
\EQ
\dot{f}=-uf'+\nu f''.
\label{advdiff}
\EN
The question is now how much diffusion is necessary, and how this
depends on the spatial order of the scheme.

A perfect step function would produce large start-up errors; it is better
to use a smoothed profile, for example one of the form
\EQ
f(x)=\tanh\left({\cos x\over\delta x}\right),
\EN
where $\delta x$ is the mesh width. For a periodic mesh of length $L$
one would obviously use $f=f(kx)$, where $k=2\pi/L$. In that case the step
width would be $k\delta x$. In the following we consider a periodic domain
of size $L_x=1$ with $N_x=60$ meshpoints, so we use $k=2\pi/L_x=2\pi$.

\begin{figure}[h!]\begin{center}\includegraphics[width=.99\textwidth]{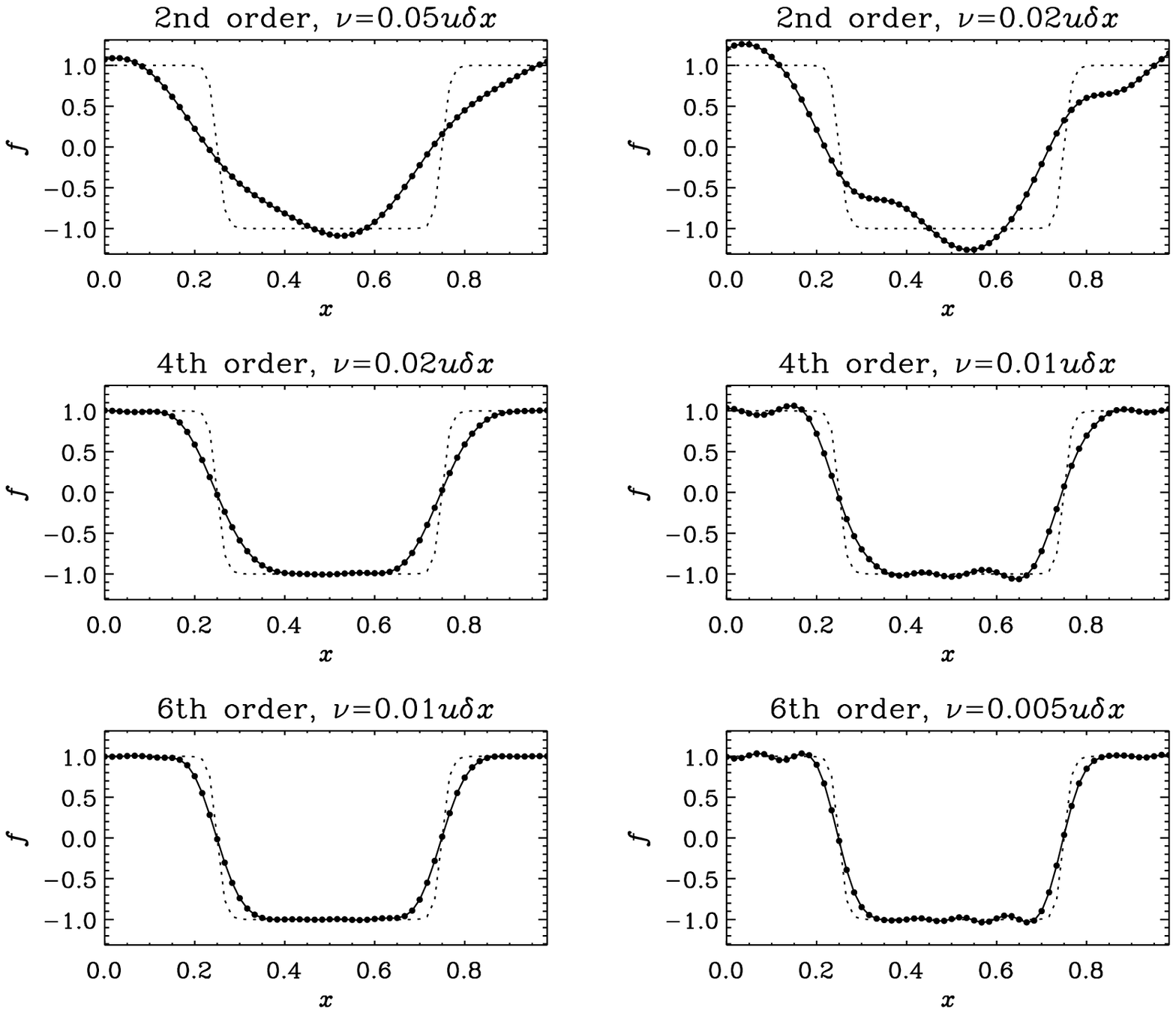}\end{center}\caption[]{
Like \Fig{Fpadvect_all}, but for low-order schemes. Again, for the
panels on the right hand side the diffusion coefficient is too small
and the profile shows noticeable wiggles. For the 2nd-order scheme
one needs a viscosity of $\nu=0.05\,u\delta x$ to prevent wiggles, but
then the resulting distortion of the original profile becomes rather
unacceptable. $\delta x=1/60$.
}\label{Fpadvect_all_2nd}\end{figure}

In \Fig{Fpadvect_all} we plot the result of advecting the periodic
step-like function, $f(kx)$, over 5 wavelengths, corresponding to a
time $T=L/u$.  The goal is to find the minimum diffusion coefficient
$\nu$ necessary to avoid wiggles in the solution. In the first two
panels one sees that for a 6th order scheme the diffusion coefficient
has to be approximately $\nu=0.01\,u\delta x$. For $\nu=0.005\,u\delta
x$ there are still wiggles.  For a 10th order scheme one can still use
$\nu=0.005\,u\delta x$ without producing wiggles, while for a spectral
scheme of nearly infinite order one can go down to $\nu=0.002\,u\delta
x$ without any problems.

We may thus conclude that all these schemes need some diffusion, but that
the diffusion coefficient can be much reduced when the spatial order of the
scheme is high. In that sense it is therefore not true that high order
schemes are particularly vulnerable to Gibbs phenomena, but rather the
contrary!

In \Fig{Fpadvect_all_2nd} we compare the corresponding results of
advection tests for second and fourth order schemes with the sixth order
scheme. It is evident that a second order scheme requires a relatively
high diffusion coefficient, typically around $\nu=0.05\,u\delta x$, but
this leads to rather unacceptable distortions of the original profile. (It
may be noted that, if one uses at the same time a 1st order temporal
scheme, which has antidiffusive properties, and a time step which is
not too short, then the antidiffusive error of the timestep scheme would
partially compensate the actual diffusion and one could reduce the value
of $\nu$, but this would be a matter of tuning and hence not generally
useful for arbitrary profiles.)

\subsection{Burgers equation}

In the special case where the velocity itself is being advected, i.e.\
$f=u$, \Eq{advdiff} turns into the Burgers equation,
\EQ
\dot{u}=-uu'+\nu u''.
\label{burgers}
\EN
In one dimension there is an analytic solution for a kink,
\EQ
u=U\left(1-\tanh{x-Ut\over2\delta}\right),
\EN
where $\delta=\nu/U$ is the shock thickness (e.g., Dodd \ea 1982).
Note that the amplitude of the kink is twice its propagation speed.
Expressed in terms of the Reynolds number, $\mbox{Re}=UL/\nu$, we have
$\delta=L\mbox{Re}^{-1}$. (We note in passing that the dissipative
cutoff scale in ordinary turbulence is somewhat larger;
$\delta=L\mbox{Re}^{-3/4}$.)

\begin{figure}[h!]\begin{center}\includegraphics[width=.99\textwidth]{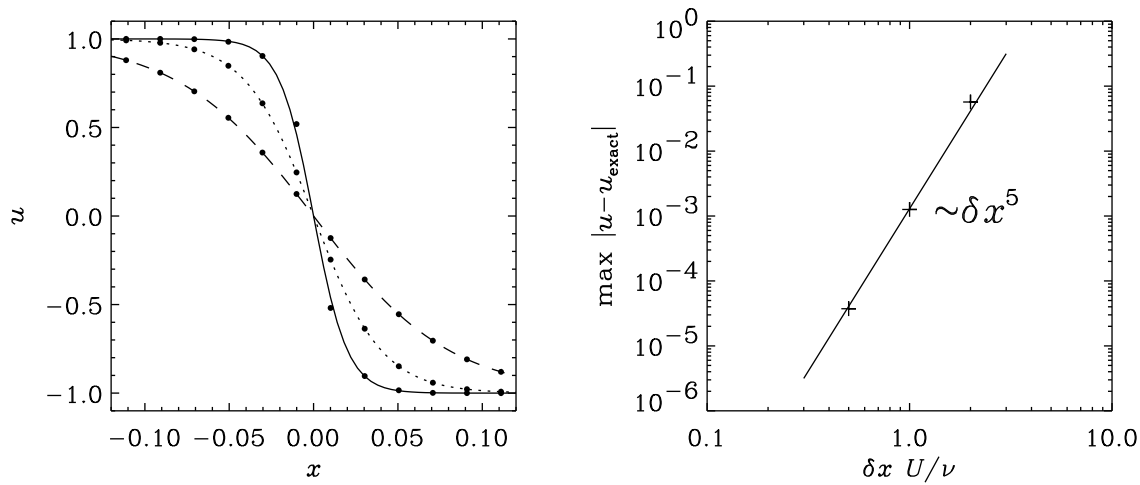}\end{center}\caption[]{
Numerical solution of the Burgers equation using the sixth order
explicit scheme with \Eq{burgers_sol} as initial condition. In the
left hand panel the lines give the exact solution \eq{burgers_sol}
and the dots give the numerical solution for the corresponding value
of the mesh Reynolds number: $\delta x\,U/\nu=0.5$ (solid line)
1.0 (dotted line), and 2.0 (dashed line). In the right hand panel
the scaling of the error with $\delta x$ is shown.
}\label{Fpburg}\end{figure}

In order to have a stationary shock we use the initial condition
\EQ
u=-U\tanh(x/2\delta).
\label{burgers_sol}
\EN
In \Fig{Fpburg} we present numerical solutions using the sixth order
explicit scheme with different values of the mesh Reynolds number,
$\delta x\,U/\nu$, which was varied by changing the value of $\nu$. Here
we used $N_x=100$ meshpoints in the range $-1\leq x\leq1$. Note that the
overall error, defined here as $\max(|f-f_{\rm exact}|)$, decreases with
decreasing mesh width like $\delta x^5$.

The test cases considered so far were not directly related to the
Navier-Stokes equation, which permits sound waves that can pile up
to form shocks, for example. This will be considered in the next
section.

\subsection{Shock tube tests}

A popular test problem for compressible codes is the shock tube problem
of Sod (1951). On the one hand, one can assess the sharpness of the
various fronts. On the other hand, and perhaps most importantly, it
allows one to test the conversion of kinetic energy to thermal energy
via viscous heating.

\begin{figure}[h!]\begin{center}\includegraphics[width=.99\textwidth]{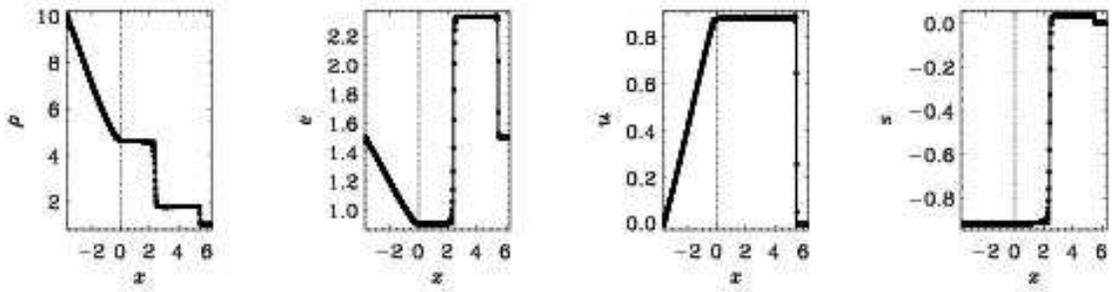}\end{center}\caption[]{
Standard shock tube test with an initial pressure jump of 1:10 and
$\nu=0.6\delta x\,c_{\rm s}$ and $\chi/\nu=0.05$. The solid line indicates
the analytic solution (in the limit $\tilde\nu\rightarrow0$) and the dots the numerical solution.  Note the
small entropy excess on the right of the initial entropy discontinuity.
}\label{Fshock1}\end{figure}

In the following we use the formulation of the compressible Navier-Stokes
equations in terms of entropy and enthalpy, \eq{basic_u}--\eq{basic_h}. We
use units where $p_0=\rho_0=c_p=1$ and adopt the abbreviations
$\Lambda=\ln\rho$ (not to be confused with the cooling function used in
\Sec{NavierStokes}). In one dimension (with $\nu=\mbox{const}$)
these equations reduce to
\EQ
\dot{u}=-uu'-c_{\rm s}^2(\Lambda'+s')+\tilde\nu(u''+\Lambda'u'),
\label{shock1}
\EN
\EQ
\dot{s}=-us'+(\gamma-1)\tilde\nu u'^2/c_{\rm s}^2+Q_s,
\label{shock2}
\EN
\EQ
\dot{\Lambda}=-u\Lambda'-u',
\label{shock3}
\EN
where dots and primes refer respectively to time and space derivatives,
$Q_s$ describes the change of entropy due to radiative diffusion, and
$\Lambda=\ln\rho$ is the logarithmic density. In \Eqss{shock1}{shock3}
we have used the abbreviation
\EQ
c_{\rm s}^2=\gamma p/\rho=\gamma\exp[(\gamma-1)\Lambda+\gamma s]
\EN
for the adiabatic sound speed squared, and $\tilde\nu={4\over3}\nu$ is the
effective viscosity for compressive motions. This 4/3 factor comes
from the fact that in 1-D
\EQ
\SSSS=\diag\left({\textstyle{2\over3}},-{\textstyle{1\over3}},
-{\textstyle{1\over3}}\right)u_{x,x},
\EN
and therefore
\EQ
{1\over\rho}\nab\cdot(2\nu\rho\SSSS)
={\textstyle{4\over3}}\nu[u_{x,xx}+(\ln\rho)_{,x}u_{x,x}],
\EN
so $\SSSS^2={2\over3}u_{x,x}^2$, or $2\nu\SSSS^2=\tilde\nu u_{x,x}^2$. In
the radiative diffusion approximation we have $Q_s=-\Lambda_{\rm
cond}/(c_pT)$, and so \Eq{Lambdacond} gives in one dimension
\EQ
Q_s=\chi\gamma[s''+\nabla_{\rm ad}\Lambda''
+\gamma(\Lambda'+s')(s'+\nabla_{\rm ad}\Lambda')]
\EN
In \Fig{Fshock1} we show the solution for an initial density and pressure
jump of 1:10 and the the viscosity is now $\nu=0.6\delta x\,c_{\rm s}$. In
this case a small amount of thermal diffusion (with Prandtl number
$\chi/\nu=0.05$) has been adopted to remove wiggles in the entropy.

\begin{figure}[h!]\begin{center}\includegraphics[width=.99\textwidth]{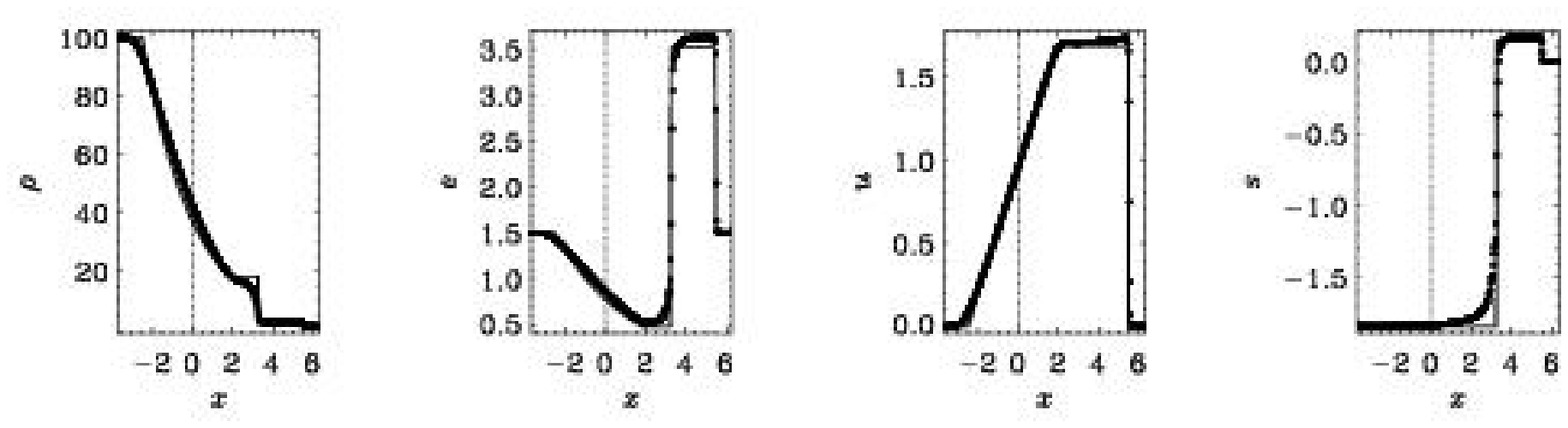}\end{center}\caption[]{
Same as \Fig{Fshock1}, but for an initial pressure jump of 1:100 and
$\nu=1.6\delta x\,c_{\rm s}$.
}\label{Fshock2}\end{figure}

For stronger shocks velocity and entropy excess increase; see
\Figs{Fshock2}{Fshock3}, where the initial pressure jumps are 1:100 and
1:1000, respectively, and the viscosities are chosen to be $\nu=1.6\delta
x\,c_{\rm s}$ and $\nu=2.4\delta x\,c_{\rm s}$. (For the stationary shock
problem considered below we also find that the viscosity must increase
with the Mach number and, moreover, that the two should be proportional
to each other.) In the cases shown in \Figs{Fshock2}{Fshock3} we were
able to put $\chi=0$ without getting any wiggles in $s$. However, in the
case of strong shocks (pressure ratio 1:1000) the discrepancy between
numerical and analytical solutions becomes quite noticeable.

\begin{figure}[h!]\begin{center}\includegraphics[width=.99\textwidth]{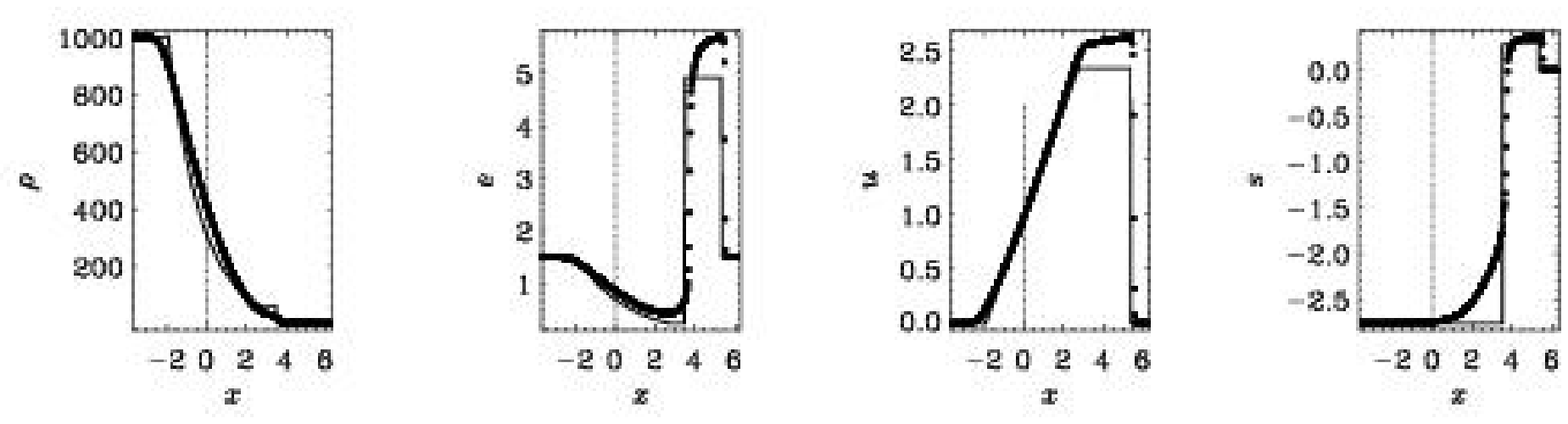}\end{center}\caption[]{
Same as \Fig{Fshock2}, but for an initial pressure jump of 1:1000 and
$\nu=2.4\delta x\,c_{\rm s}$.
}\label{Fshock3}\end{figure}

In many practical applications shocks occur only in a small portion
of space. One can therefore reduce the viscosity outside shocks or,
conversely, use a small viscosity everywhere except in the locations
of shock, i.e.\ where the flow is convergent (negative divergent). This
leads to the concept of an artificial (Neumann-Richtmyer) shock viscosity,
\EQ
\nu_{\rm shock}=c_{\rm shock}\delta x^2\,\bra{(-\nab\cdot\uu)_+}_{\rm n.n.},
\EN
where $\bra{...}_{\rm n.n.}$ indicates averaging over nearest neighbors
and the subscript $+$ means that only the positive part is taken.

The last panel in \Figs{Fshock2}{Fshock3} shows quite clearly how the
entropy increases behind the shock. This entropy increase is just a
consequence of the viscous heating term, $\tilde\nu u'^2/h$. Without this
term the solution would obviously be wrong everywhere behind the shock,
especially when the shock is strong.

\begin{figure}[h!]\begin{center}\includegraphics[width=.99\textwidth]{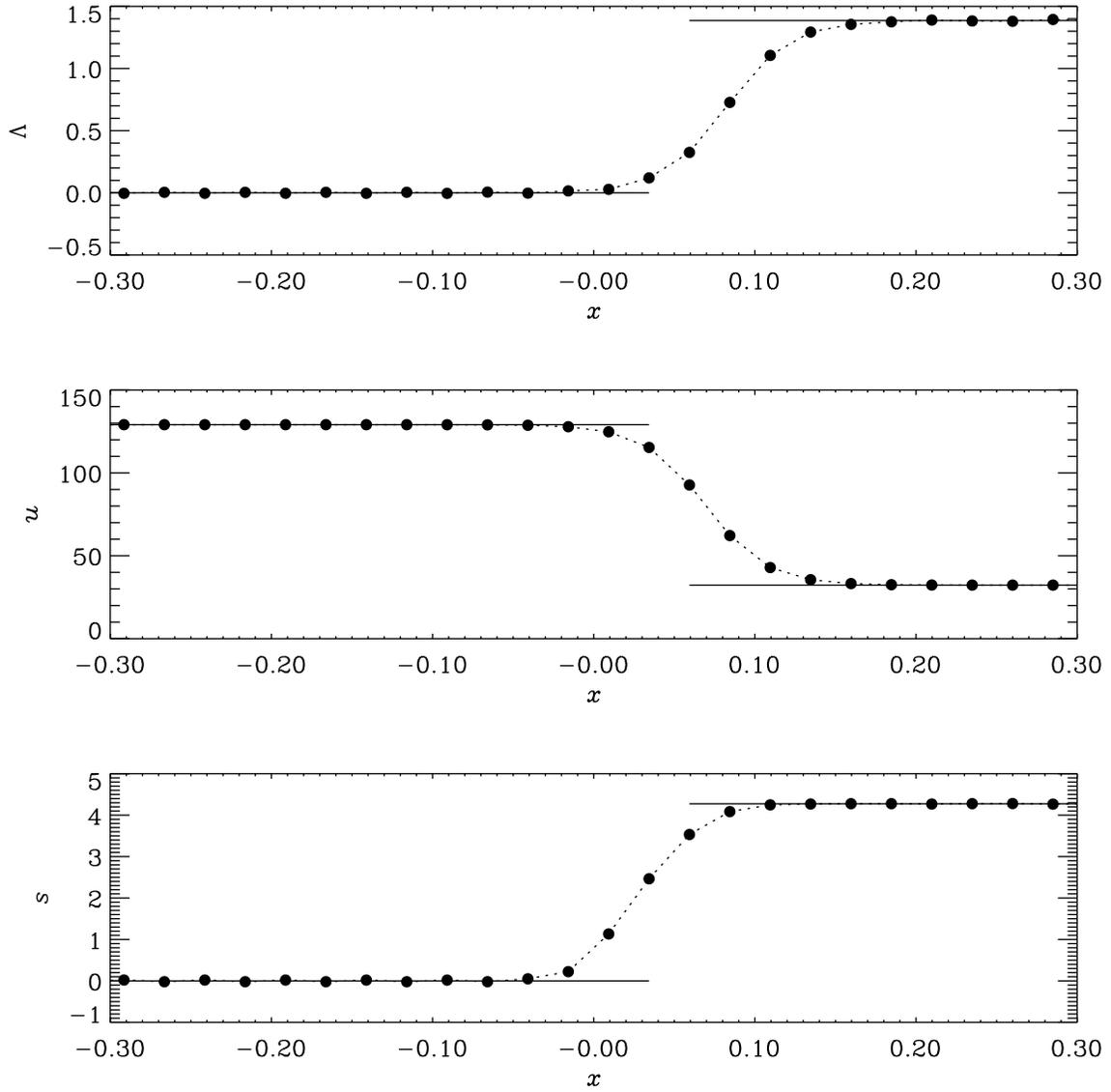}\end{center}\caption[]{
Example of a very strong standing shock with Ma=100. Note the relative
shift of the position where $s$ increases relative to where $\Lambda=\ln\rho$
increases. The viscosity is chosen to be $\nu=\mbox{Ma}\times\delta x$.
}\label{Fshock1_standing}\end{figure}

A somewhat simpler situation is encountered with standing shocks. In
\Fig{Fshock1_standing} we give an example of a numerically determined
solution at Ma=100. The agreement in the jump for the numerically
determined solution (dotted line and dots) and the theoretical solution
(solid lines) is very good, although the position of the jump has moved
away somewhat from the initial location ($x=0$), but this is merely a
consequence of having used more-or-less arbitrarily a tanh profile to
smooth the initial jumps. After some initial adjustment phase the profiles
do indeed remain stationary. Note also, however, that the entropy profile
is slightly shifted relative to the profiles of $u$ and $\Lambda$.

It is interesting to note that when solving the Rankine-Hugionot jump
conditions for shocks one is allowed to use the inviscid equations
provided they are written in conservative form. Sometimes one finds
in the literature the inviscid Navier-Stokes equations written in
nonconservative form. This is not strictly correct, because without
viscosity there would be no viscous heating and hence no entropy increase
behind the shock. Moreover, it is quite common to consider a polytropic
equation of state, $p=K\rho^\Gamma$.  Again, in this case the entropy
is constant, and so energy conservation is violated. Nevertheless, given
that polytropic equations of state are often considered in astrophysics
we consider this case in more detail in the next subsection.

\subsection{Polytropic and isothermal shocks}
\label{Spoly}

For polytropes with $p=K\rho^\Gamma$, but $\Gamma\neq\gamma$ in general,
we can write
\EQ
-\nab h+h\nab s=-{1\over\rho}\nab p
=-\nab\left({\Gamma K\over\Gamma-1}\rho^{\Gamma-1}\right)
\equiv\nab\tilde{h},
\EN
so we can introduce a pseudo enthalpy $\tilde{h}$ as
\EQ
\tilde{h}={\Gamma K\over\Gamma-1}\rho^{\Gamma-1}
=\left[\left(1-{1\over\gamma}\right)
\left/\left(1-{1\over\Gamma}\right)\right]\,h\right..
\EN
This is consistent with a fixed entropy dependence,
where $s$ only depends on $\rho$ like
\EQ
s=s(\rho)={1\over\gamma}\ln\left(p/\rho^\gamma\right)
={1\over\gamma}\ln\left(K\,\rho^{\Gamma-\gamma}\right)
={1\over\gamma}\ln K+\left({\Gamma\over\gamma}-1\right)\ln\rho,
\EN
which implies that in the polytropic case \Eq{shock2} is discarded. In
the adiabatic case, $\Gamma=\gamma$, entropy is constant. In the
isothermal case, $p=c_{\rm s}^2\rho$, we have $\Gamma\rightarrow1$,
so entropy is not constant, but it varies only in direct relation
to $-\ln\rho$ and not as a consequence of viscous heating behind
the shock.

\begin{figure}[h!]\begin{center}\includegraphics[width=.99\textwidth]{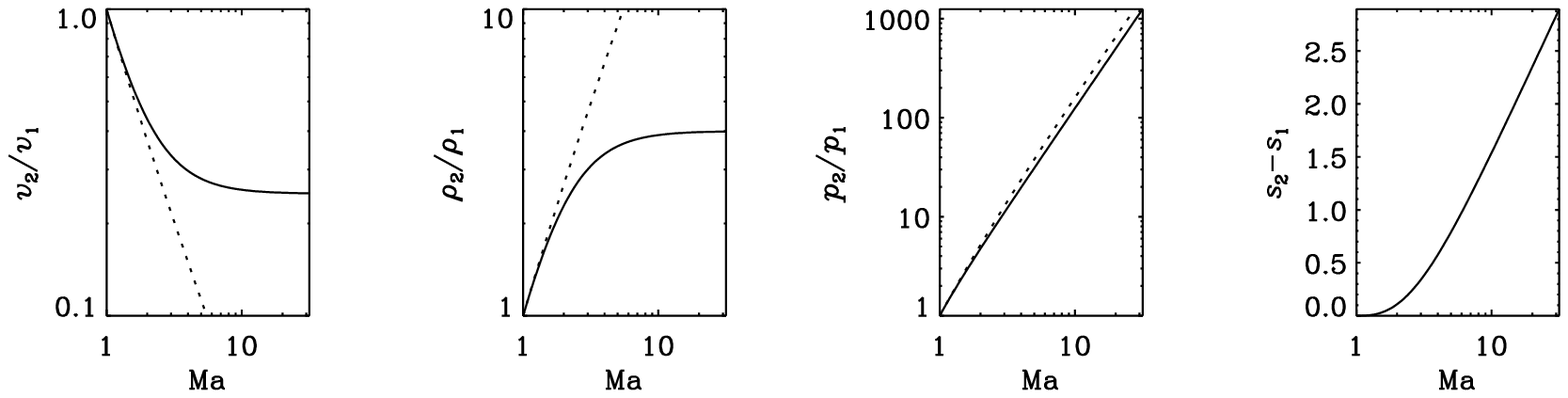}\end{center}\caption[]{
The dependence of the velocity, density, pressure, and entropy jumps
on the upstream Mach number for the case $\gamma=5/3$ (solid line)
and comparison with the polytropic case using $\Gamma=\gamma$ (dotted
line). Note that the velocity and density jumps saturate at 1/4 and 4,
respectively, while there is no such saturation for the polytropic shock.
}\label{Fpshock_gamma=1.6667}\end{figure}

In deriving the Rankine-Hugionot jump conditions one uses the conservation
of mass, momentum, and energy in a comoving frame, where the following
three quantities are constants of motion:
\EQ
J=\rho u,\quad
I=\rho u^2+p,\quad
E=\half u^2+{\gamma\over\gamma-1}{p\over\rho}.
\EN
The values of these three constants can be calculated when all three variables,
$u$, $p$, and $\rho$, are known on one side of the shock. For polytropic
equations of state, with $p=K\rho^\gamma$, the energy equation is no longer
used, so there are only the following two conserved quantities,
\EQ
J=\rho u,\quad
I=\rho u^2+K\rho^\gamma.
\EN
The dependence of the velocity, density, pressure, and entropy jumps
on the upstream Mach number is plotted in \Fig{Fpshock_gamma=1.6667}
for the case $\gamma=5/3$ and compared with the polytropic case using
$\Gamma=\gamma$.

\begin{figure}[h!]\begin{center}\includegraphics[width=.99\textwidth]{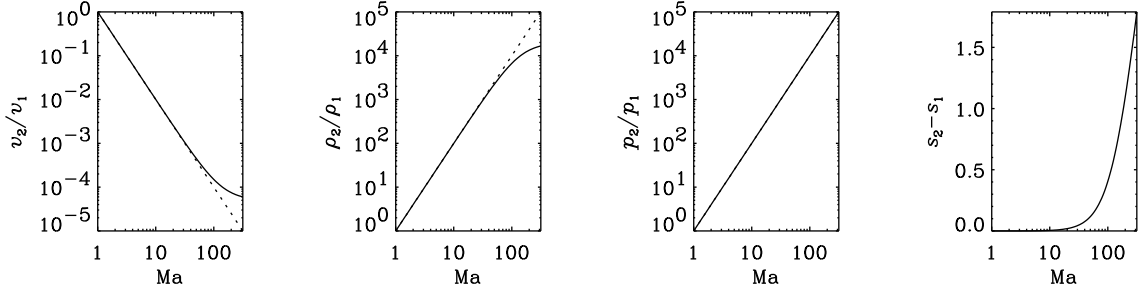}\end{center}\caption[]{
Like \Fig{Fpshock_gamma=1.6667}, but for $\gamma=\Gamma=1.0001$. Note that
the velocity and density jumps saturate at much more extreme values than for
$\gamma=5/3$. Thus, the results using polytropic and perfect gas equations
agree up to much larger Mach numbers than for $\gamma=5/3$.
}\label{Fpshock_gamma=1.0001}\end{figure}

Note that the pressure jump, $p_2/p_1$, is almost independent of
the value of $\gamma$ and does also not significantly depend on the
polytropic assumption.

\section{Nonuniform and lagrangian meshes}

In many cases it is useful to consider nonuniform meshes, either by
adding more points in places where large gradients are expected, or by
letting the points move with the flow (lagrangian mesh). The lagrangian
mesh is particularly useful in one-dimensional cases, because then the
mesh topology (i.e.\ the ordering of mesh points) remains unchanged.
Another method that gains constantly in popularity is adaptive mesh
refinement (e.g., Grauer, Marliani, \& Germaschewski 1998), which will
not be discussed here.

\subsection{Nonuniform topologically cartesian meshes}
\label{nonunicart}

Nonuniform meshes can be implemented relative easily when each of
the new coordinates depend on only one variable, for example when
$x=x(\tilde{x})$, $y=y(\tilde{y})$, and $z=z(\tilde{z})$. Here,
$\tilde{x}$, $\tilde{y}$ and $\tilde{z}$ are cartesian coordinates on
a uniform mesh. In the more general case, however, we have
\EQ
x=x(\tilde{x},\tilde{y},\tilde{z}),\quad
y=y(\tilde{x},\tilde{y},\tilde{z}),\quad
z=z(\tilde{x},\tilde{y},\tilde{z}),
\EN
so that $x$, $y$ and $z$ derivatives of a function $f$ can be calculated
using the chain rule,
\EQ
{\partial f\over\partial x}
={\partial f\over\partial\tilde{x}}{\partial\tilde{x}\over\partial x}
+{\partial f\over\partial\tilde{y}}{\partial\tilde{y}\over\partial x}
+{\partial f\over\partial\tilde{z}}{\partial\tilde{z}\over\partial x}
\equiv{\partial f\over\partial\tilde{x}_i}{\partial\tilde{x}_i\over\partial x}.
\label{dertrans}
\EN
Corresponding formulae apply obviously for the other two directions, so
in general we can write
\EQ
{\partial f\over\partial x_j}=
{\partial f\over\partial\tilde{x}_i} J_{ij},\quad\mbox{where}\quad
J_{ij}={\partial\tilde{x}_i\over\partial x_j}
\EN
is the jacobian of this coordinate transformation. This method allows
one to have high resolution for example near a central object, without
however having high resolution anywhere else far away from the central
object. This is useful in connection with outflows from jets.

\begin{figure}[h!]\begin{center}\includegraphics[width=.99\textwidth]{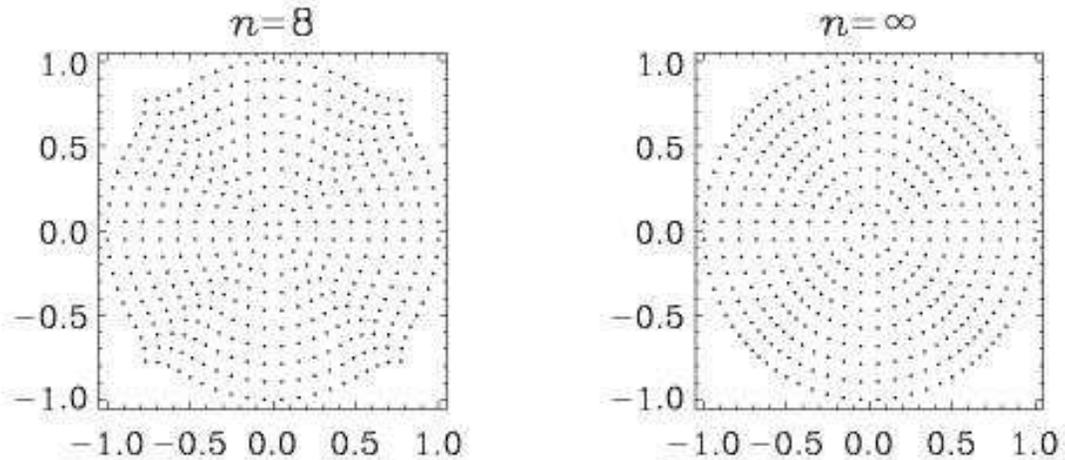}\end{center}\caption[]{
Examples of the resulting meshes for $n=8$ and $n\rightarrow\infty$
in which case \Eq{infty_formula} is used.
}\label{Fpmesh}\end{figure}

We discuss here one particular application that is relevant for simulating
flows in a sphere. It is possible to transform a cartesian mesh to cover a
sphere without a coordinate singularity. It will turn out, however, that
there is a discontinuity in the jacobian. We discuss this here in 2-D.
We denote the coordinate mesh by a tilde, so $(\tilde{x},\tilde{y})$ are
the coordinates in a uniform cartesian mesh.  We want to stretch the mesh
such that points on the $\tilde x$ and $\tilde y$ axes are not affected,
and that the distance of points on the diagonal is reduced by a factor
$1/\sqrt{2}$ (or by $1/\sqrt{3}$ in 3-D). This can be accomplished by
introducing new coordinates $(x,y)$ as
\EQ
\pmatrix{x\cr y}=\pmatrix{\tilde x\cr \tilde y}\,
{(\tilde x^n+\tilde y^n)^{1/n}\over(\tilde x^2+\tilde y^2)^{1/2}},
\EN
where $n$ is a large even number. In the limit $n\rightarrow\infty$
we may substitute
\EQ
(\tilde x^n+\tilde y^n)^{1/n}\rightarrow\max(|\tilde x|,|\tilde y|).
\label{infty_formula}
\EN
Examples of the resulting meshes for two different values of $n$ are
given in \Fig{Fpmesh}.

\begin{figure}[h!]\begin{center}\includegraphics[width=.5\textwidth]{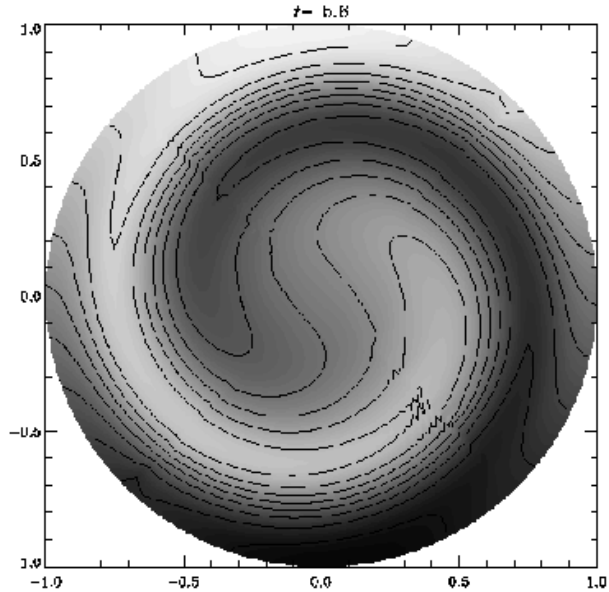}\end{center}\caption[]{
Example of an advection experiment on a $n=8$ mesh.
}\label{Fnonuni_s}\end{figure}

In order to obtain the jacobian of this transformation,
$\partial\tilde x_i/\partial x_j$, we have to consider separately
the cases $\tx\ge\ty$ and $\tx\le\ty$. The derivation is given in
\App{Sjacobian}, and is most concisely expressed
in terms of the logarithmic derivative, so
\EQ
\pmatrix{\mxtx&\mxty\cr\mytx&\myty}=
\pmatrix{+1-\myt&+1-\mxt\cr-1+\mxt&+1+\myt}
\quad\mbox{if $|\tx|\ge|\ty|$},
\label{transform_sphere1}
\EN
\EQ
\pmatrix{\mxtx&\mxty\cr\mytx&\myty}=
\pmatrix{+1+\mxt&-1+\myt\cr+1-\myt&+1-\mxt}
\quad\mbox{if $|\tx|\le|\ty|$},
\label{transform_sphere2}
\EN
where $\tr^2=\tx^2+\ty^2$.
Note that the jacobian is discontinuous on the diagonals.
This is a somewhat unfortunate feature of this transformation.
It is not too surprising however that something like this happens,
because the diagonals are the locations where a rotating flow
must turn direction by $90^\circ$ in the coordinate mesh.
Nevertheless, it is possible to obtain reasonably well behaved
solutions; see \Fig{Fnonuni_s} for an advection experiment using
a prescribed differentially rotating flow.

The fluid equations are still solved in rectangular cartesian coordinates,
so for example the equation $\DD s/\DD t=0$ is solved in the form
\EQ
{\partial s\over\partial t}=
-u_x{\partial s\over\partial x}
-u_y{\partial s\over\partial y},
\EN
where the spatial derivatives are evaluated according to \Eq{dertrans}.
For the velocity field, stress-free boundary conditions, for example,
would be written in the form
\EQ
\hat{r}_j u_j=0,\quad\hat{\phi}_i S_{ij}\hat{r}_j=0,
\EN
where $S_{ij}$ is the rate of strain tensor,
$\hat{r}_j$ and $\hat{\phi}_i$ are the cartesian components
($i=x,y,z$) of the radial and azimuthal unit vectors, i.e.\
\EQ
\rr={1\over r}\pmatrix{x\cr y},\quad\mbox{and}\quad
\pp={1\over r}\pmatrix{-y\cr x}
\EN
are unit vectors in the $r$ and $\phi$ directions and $r=\sqrt{x^2+y^2}$
is the distance from the rotation axis. The stress-free boundary
conditions are then
\EQ
xu_x+yu_y=0
\EN
and
\EQ
(x^2-y^2)(u_{x,y}+u_{y,x})-2xy(u_{x,x}-u_{y,y})=0.
\EN

\subsection{Lagrangian meshes}

We now consider a simple one-dimensional lagrangian mesh problem.
Assume that $\ell$ labels the particle, then the lagrangian derivative is
\EQ
{\DD s\over\DD t}\equiv\left({\partial s\over\partial t}\right)_{\ell
={\rm const}}=\left({\partial s\over\partial t}\right)_{x={\rm const}}
+\left({\partial x\over\partial t}\right)_{\ell={\rm const}}
{\partial s\over\partial x}
\EN
Now, because
\EQ
\left({\partial x\over\partial t}\right)_{\ell={\rm const}}={\DD x\over\DD t}=u
\EN
we have the well-known equation
\EQ
{\DD s\over\DD t}={\partial s\over\partial t}+u{\partial s\over\partial x}.
\EN
As an example we now consider the Burgers equation,
\EQ
{\DD u\over\DD t}=\tilde\nu{\partial^2 u\over\partial x^2}.
\EN
We now take $u(x,t)=u(\ell(x),t)$ to be a function of the coordinate
variable $\ell$ which, in turn, is a function of $x$. The $x$-derivatives
are obtained using the chain rule, i.e.\
\EQ
{\partial u\over\partial x}=
{\partial\ell\over\partial x}
{\partial u\over\partial\ell}={u'\over x'},
\EN
and likewise for the second derivative
\EQ
{\partial^2 u\over\partial x^2}={u''x'-u'x''\over x'^3}.
\EN
\\

\SHADOWBOX{
Thus, the Burgers equation can then be written as
\EQ
{\partial u\over\partial t}=\tilde\nu\,{u''x'-u'x''\over x'^3},
\EN
where the $x$ variable is given by
\EQ
{\partial x\over\partial t}=u.
\label{dxdt}
\EN
}\\

\begin{figure}[h!]\begin{center}\includegraphics[width=.99\textwidth]{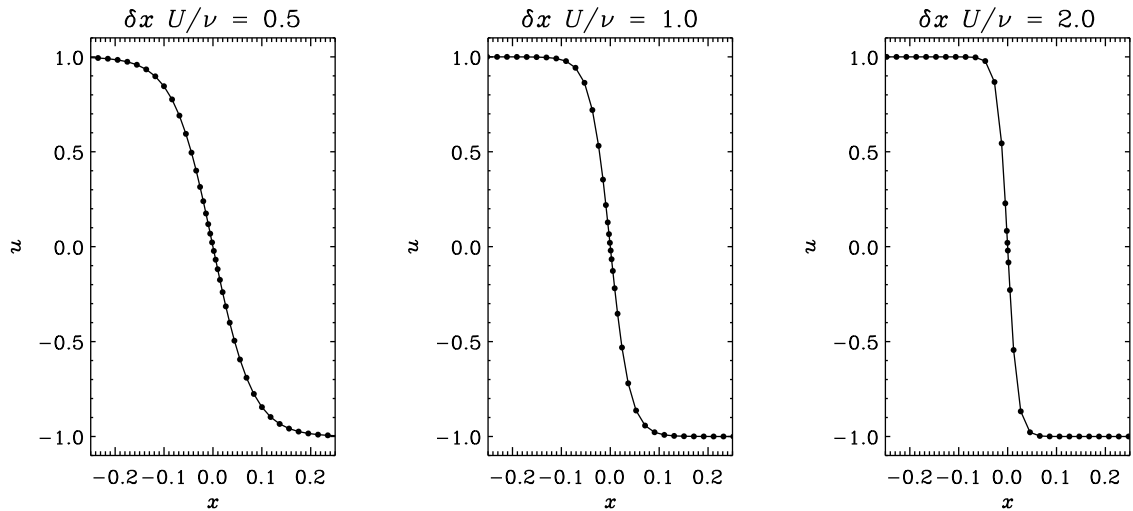}\end{center}\caption[]{
Solution of the Burgers equation using a lagrangian mesh combined with
a sixth order explicit scheme. The values of the mesh Reynolds number
vary between $\delta x\,U/\nu=0.5$ and 2.0, where $\delta x$ refers to
the initially uniform mesh spacing.
}\label{Fpburg_lagrange}\end{figure}

\noindent
A solution of these two equations is given in \Fig{Fpburg_lagrange}.

In the test problem above the initial meshpoint distribution was
uniform. Although this is not quite suitable for this problem, it shows
that subsequently the mesh spacing became narrower still, which means
that the timestep in now governed by viscosity,
$\delta t\leq0.06\delta x^2_{\min}/\tilde\nu$, where the numerical
factor is empirical. However, the mesh spacing does not need to be
governed by \Eq{dxdt}, so it is quite possible to come up with other
prescriptions for the mesh spacing.

Consider as another example the isothermal eulerian equations
\EQ
{\DD u\over\DD t}=-c_{\rm s}^2{\partial\ln\rho\over\partial x}
+\tilde\nu\left({\partial^2 u\over\partial x^2}
+{\partial\ln\rho\over\partial x}{\partial u\over\partial x}\right),
\EN
\EQ
{\DD\ln\rho\over\DD t}=-{\partial u\over\partial x}.
\EN
In lagrangian form they take the form\\

\SHADOWBOX{
\EQ
{\partial u\over\partial t}=-c_{\rm s}^2{(\ln\rho)'\over x'}+\tilde\nu\,
\left({u''x'-u'x''\over x'^3}+{(\ln\rho)'\over x'}{u'\over x'}\right),
\EN
\EQ
{\partial\ln\rho\over\partial t}=-{u'\over x'},
\EN
\EQ
{\partial x\over\partial t}=u.
\EN
}\\

The example above demonstrates clearly the problem that lagrangian
mesh points can continue to pile up near convergence points of the
flow. This is a general problem with fully lagrangian schemes. One
possible alternative is to use {\it lagrangian-eulerian} schemes (e.g.,
Benson 1992, Peterkin, Frese, \& Sovinec 1998, Arber \ea 2001), which
combine the advantages of lagrangian and eulerian codes, but involve
obviously some kind of interpolation. Another alternative is to use
a semi-lagrangian code which advects the mesh points not with the
actual gas velocity $\uu$, but with a more independent mesh velocity
$\UU$. Clearly, we want to avoid too small distances between neighboring
points, so one could artificially lower the effective mesh velocity
by involving for example the modulus of the jacobian, $|\JJJJ|$, which
becomes large when the concentration of mesh points is high. Thus, one
could choose for example $\UU=\uu/(1+|\JJJJ|)$. In the present case,
$|\JJJJ|=|x'|^{-1}$. In the following we discuss the formalism that
needs to be invoked in order to calculate first and second derivatives
on an advected mesh.

\subsection{Non-lagrangian mesh advection}

The main advantage of a lagrangian mesh is that it allows higher
resolution locally. Another advantage, which is however less crucial,
is that the nonlinear advection term drops out. The main disadvantage
is however that a lagrangian mesh may become too distorted and
overconcentrated, as seen in the previous section. In this subsection
we address the possibility of advecting the mesh with a velocity $\UU$
that can be different from the fluid velocity. This way one can remove
the swirl of the mesh by taking a velocity that is the gradient of some
other quantity, i.e.\
\EQ
\UU=-\nab\Phi_{\rm mesh},
\EN
where $\Phi_{\rm mesh}$ should be large in those regions where many
points are needed. One possible criterion would be to require that the
number of scale heights per meshpoint, $|\delta\xx\cdot\nab\ln\rho|$,
does not exceed an empirical value of $1/3$, say. Thus,
$3|\delta\xx\cdot\nab\ln\rho|<1$ would be a necessary condition.
Another possibility would be to let $\Phi_{\rm mesh}$ evolve itself
according to some suitable advection-diffusion equation. However,
no generally satisfactory method seems to be available as yet. In
order to calculate the jacobian for the coordinate transformation one
can make use of the fact that the mesh evolves only gradually from one
timestep to the next. For a more extended discussion of mesh advection
schemes we refer to the article by Dorfi in the book by LeVeque \ea
(1998).

\subsubsection{Calculating the jacobian}

Initially, at $t=0$, we have $\xx=\tilde{\xx}$. After the $n$th timestep,
at $t=n\delta t$, we calculate the new $\xx$-mesh, $\xx^{(n+1)}$, from
the previous one, $\xx^{(n)}$, i.e.\
\EQ
\xx^{(n+1)}=\xx^{(n)}+\mbox{$\UU\left(\xx^{(n)},t\right)\,\delta t$}.
\EN
Here, $\xx^{(0)}=\tilde{\xx}$ is just the original coordinate mesh.
Differentiating the $i$th component with respect to the $j$th component,
as we have done in \Sec{nonunicart}, we obtain
\EQ
\delta_{ij}
={\partial x^{(n+1)}_i\over\partial x^{(n+1)}_j}
={\partial x^{(n)}_i\over\partial x^{(n+1)}_j}
+{\partial U^{(n)}_i\over\partial x^{(n+1)}_j}\,\delta t,
\EN
where $U^{(n)}_i=U_i\left(\xx^{(n)},t\right)$. In the expression above
we have $U_i$ on the mesh $\xx^{(n)}$, but we need to differentiate with
respect to the new mesh $\xx^{(n+1)}$. This can be fixed by another
factor $\partial x^{(n)}_i/\partial x^{(n+1)}_j$. Thus, we have
\EQ
\delta_{ij}
={\partial x^{(n)}_i\over\partial x^{(n+1)}_j}
+{\partial x^{(n)}_k\over\partial x^{(n+1)}_j}
{\partial U_i\left(\xx^{(n)},t\right)\over\partial x^{(n)}_k}\,\delta t
=\left[\delta_{ik}+
{\partial U^{(n)}_i\over\partial x^{(n)}_k}\,\delta t\right]
{\partial x^{(n)}_k\over\partial x^{(n+1)}_j}.
\label{1st_jacobi2}
\EN
This can be written in matrix form,
\EQ
\delta_{ij}={\sf M}_{ik} {\sf J}^{(n)}_{kj},
\EN
where
\EQ
{\sf M}_{ik}=\delta_{ik}+
{\partial U_i\left(\xx^{(n)},t\right)\over\partial x^{(n)}_k}\,\delta t
\label{Mmatrix}
\EN
is a transformation matrix and
\EQ
{\sf J}^{(n)}_{ij}={\partial x^{(n)}_i\over\partial x^{(n+1)}_j}
\EN
is the incremental jacobian, so $\JJJJ^{(n)}=\MMMM^{-1}$.
To obtain the jacobian at $t=2\delta t$, for example, we calculate
\EQ
{\partial x^{(0)}_i\over\partial x^{(2)}_j}=
{\partial x^{(0)}_i\over\partial x^{(1)}_k}
{\partial x^{(1)}_k\over\partial x^{(2)}_j}=
{\sf J}^{(0)}_{ik} {\sf J}^{(1)}_{kj}
\equiv\left(\JJJJ^{(0)}\JJJJ^{(1)}\right)_{ij}.
\EN
The jacobian at $t=n\delta t$ is then obtained by successive matrix
multiplication from the right, so
\EQ
{\sf J}_{ij}^{\rm(0\rightarrow n+1)}=
{\sf J}_{ik}^{\rm(0\rightarrow n)}{\sf J}_{kj}^{(n)},
\EN
where ${\sf J}_{ij}^{\rm(0\rightarrow n+1)}$ and ${\sf
J}_{ij}^{\rm(0\rightarrow n)}$ are the full (as opposed to incremental)
jacobians at the new and previous timesteps, respectively.

\subsubsection{Calculating the second order jacobian}

A corresponding calculation (see \App{Sjacobi2nd}) for the
second derivatives of a function $f$ shows that
\EQ
{\partial^2 f\over\partial x_i\partial x_j}=
{\partial^2 f\over\partial\tilde x_p\partial\tilde x_q}{\sf J}_{pi}{\sf J}_{qj}
+{\partial f\over\partial\tilde x_k}{\sf K}_{kij},
\label{2nd_jacobi1}
\EN
where 
\EQ
{\sf K}_{kij}={\partial^2\tilde x_k\over\partial x_i\partial x_j}
\label{2nd_jacobi2}
\EN
is the second order jacobian. Like for the first derivative the
second order jacobian can be obtained by successive tensor multiplication,
\EQ
{\sf K}_{kij}^{\rm(0\rightarrow n+1)}=
{\sf K}_{kpq}^{\rm(0\rightarrow n)}{\sf J}_{pi}^{(n)}{\sf J}_{qj}^{(n)}+
{\sf J}_{kl}^{\rm(0\rightarrow n)}{\sf K}_{lij}^{(n)},
\label{2nd_jacobi3}
\EN
where ${\sf K}_{kij}^{\rm(0\rightarrow n+1)}$ and
${\sf K}_{kij}^{\rm(0\rightarrow n)}$ are the second order jacobians
at the new and previous timesteps, respectively, and
\EQ
{\sf K}^{(n)}_{kij}=
{\partial^2 x^{(n)}_k\over\partial x^{(n+1)}_i\partial x^{(n+1)}_j}
\label{2nd_jacobi4}
\EN
is the incremental second order jacobian, which is calculated at each
timestep as
\EQ
{\sf K}^{(n)}_{kij}=-\delta t\left(\MMMM^{-1}\right)_{kl}\,U_{l,pq}
{\sf J}_{pi}^{(n)}{\sf J}_{qj}^{(n)},
\label{2nd_jacobi5}
\EN
where $\MMMM$ was defined in \Eq{Mmatrix} and
\EQ
U_{l,pq}=
{\partial^2U_l\left(\xx^{(n)},t\right)\over\partial x^{(n)}_p\partial x^{(n)}_q}
\label{2nd_jacobi6}
\EN
is the second order velocity gradient matrix on the physical mesh. Since
$\MMMM^{-1}=\JJJJ^{(n)}$ is just the incremental jacobian, we can write
\eq{2nd_jacobi5} as
\EQ
{\sf K}^{(n)}_{kij}=-{\sf J}_{kl}^{(n)}\delta t\,U_{l,pq}
{\sf J}_{pi}^{(n)}{\sf J}_{qj}^{(n)}.
\label{2nd_jacobi5b}
\EN
Since the expressions \eq{2nd_jacobi3} and \eq{2nd_jacobi5b} involve
both a multiplication with ${\sf J}_{pi}^{(n)}{\sf J}_{qj}^{(n)}$, we
can simplify \Eq{2nd_jacobi3} to give
\EQ
{\sf K}_{kij}^{\rm(0\rightarrow n+1)}=
\left[{\sf K}_{kpq}^{\rm(0\rightarrow n)}-
{\sf J}_{kl}^{\rm(0\rightarrow n)}{\sf J}_{lm}^{(n)}\delta t\,U_{m,pq}
\right]{\sf J}_{pi}^{(n)}{\sf J}_{qj}^{(n)}.
\label{2nd_jacobi3b}
\EN
Here the expression ${\sf J}_{kl}^{\rm(0\rightarrow n)}{\sf J}_{lm}^{(n)}$
is of course the new jacobian, ${\sf J}_{kl}^{\rm(0\rightarrow n+1)}$.
\\

\SHADOWBOX{
So in summary, the new first and second order jacobians are obtained
from the previous ones via the formulae
\EQ
{\sf J}_{ij}^{\rm(0\rightarrow n+1)}=
{\sf J}_{ik}^{\rm(0\rightarrow n)}{\sf J}_{kj}^{(n)},
\EN
\EQ
{\sf K}_{kij}^{\rm(0\rightarrow n+1)}=
\left[{\sf K}_{kpq}^{\rm(0\rightarrow n)}-
{\sf J}_{kl}^{\rm(0\rightarrow n+1)}\delta t\,U_{l,pq}
\right]{\sf J}_{pi}^{(n)}{\sf J}_{qj}^{(n)},
\label{2nd_jacobi3b2}
\EN
\EQ
{\partial f\over\partial x_i}=
{\partial f\over\partial\tilde x_p}{\sf J}_{pi},
\EN
\EQ
{\partial^2 f\over\partial x_i\partial x_j}=
{\partial^2 f\over\partial\tilde x_p\partial\tilde x_q}{\sf J}_{pi}{\sf J}_{qj}
+{\partial f\over\partial\tilde x_k}{\sf K}_{kij},
\EN
where $\JJJJ\equiv\JJJJ^{\rm(0\rightarrow n+1)}$ and
$\KKKK\equiv\KKKK^{\rm(0\rightarrow n+1)}$ has been assumed.
}\\

\begin{figure}[h!]\begin{center}\includegraphics[width=.99\textwidth]{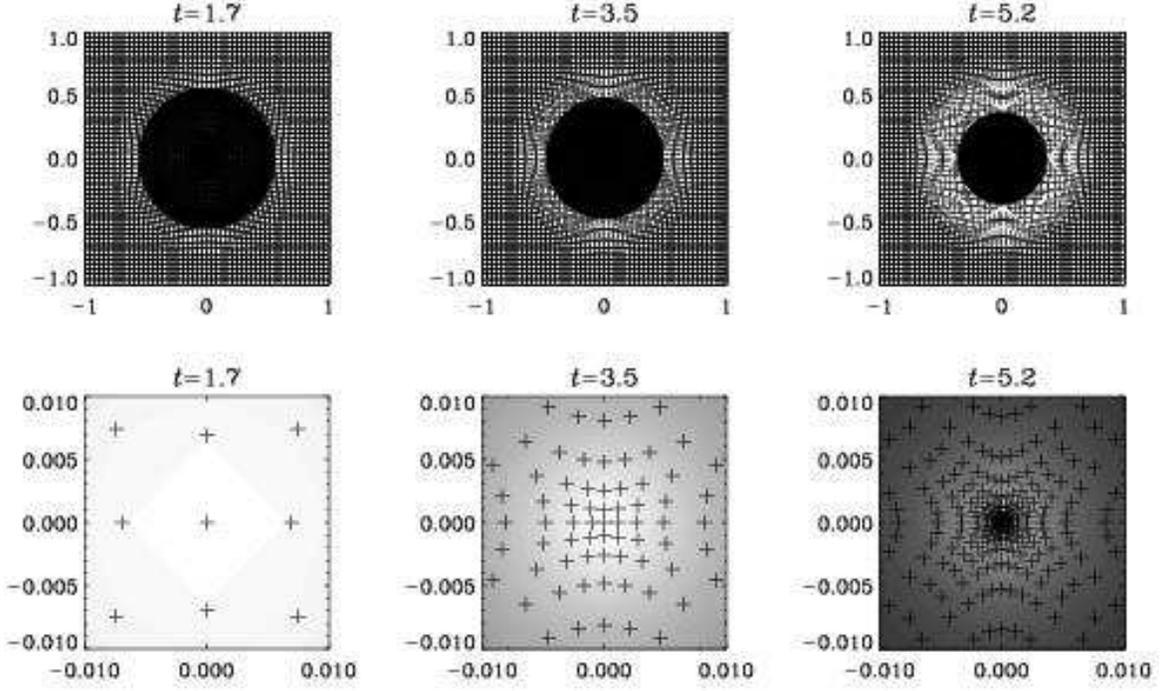}\end{center}\caption[]{
Example of a collapse calculation. The second row shows only the inner
parts with $|x|,\,|y|\leq0.01$ at the same times as in the upper row.
}\label{Fpcollapse}\end{figure}

\begin{figure}[h!]\begin{center}\includegraphics[width=.99\textwidth]{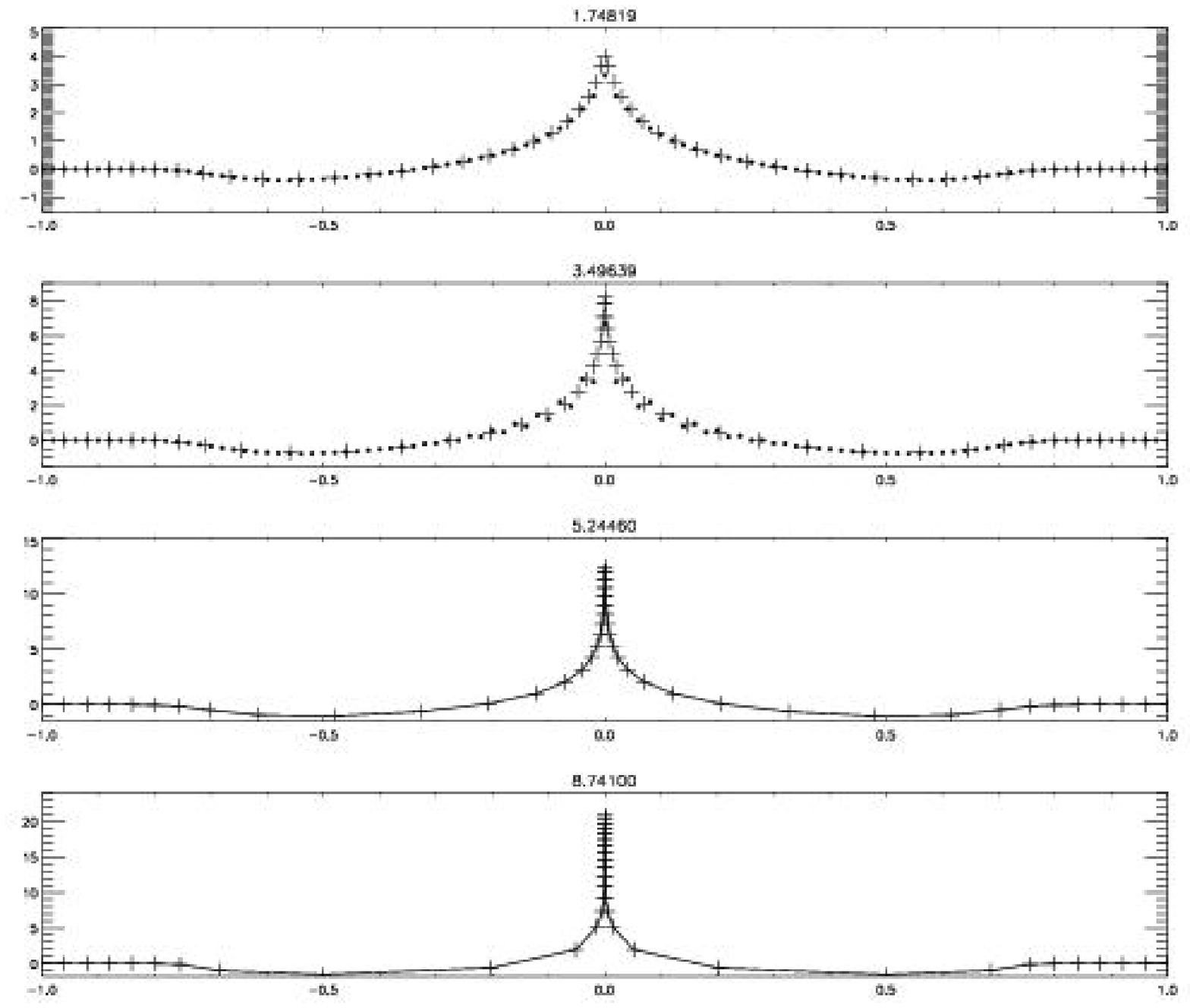}\end{center}\caption[]{
Comparison of lagrangian (+ signs) and eulerian (dots) calculations
in the first two plots, and later development (last two plots) where
the eulerian no longer works. Note that already in the second plot the
eulerian calculation has developed noticeable wiggles which the lagrangian
proceeds without problems.
}\label{Fpcomp}\end{figure}

Since now the mesh is moving in time with the local speed $\UU$ which is
different from the gas velocity $\uu$, the lagrangian derivative is
\EQ
{\DD\over\DD t}={\partial\over\partial t}+(\uu-\UU)\cdot\nab.
\EN
In all other respects the basic equations, written in cartesian form,
are still unchanged, provided all $x$, $y$, and $z$ derivatives (first and
second) are evaluated, as in \eq{dertrans} and \eq{2nd_jacobi1}, with the
components of the jacobian. As an example we show in \Fig{Fpcollapse} the
result of a kinematic collapse calculation where $\DD\uu/\DD t=-\nab\phi$
and $\DD s/\DD t=0$ with a smoothed but localized gravitational potential
$\phi$. In \Fig{Fpcomp} we compare the results of an eulerian and a
lagrangian calculation using the same number of meshpoints. Already after
some short time the eulerian calculation begins to become underresolved
and develops wiggles while the lagrangian calculation proceeds without
problems.

\subsection{Unstructured meshes}

\begin{figure}[h!]\begin{center}\includegraphics[width=.99\textwidth]{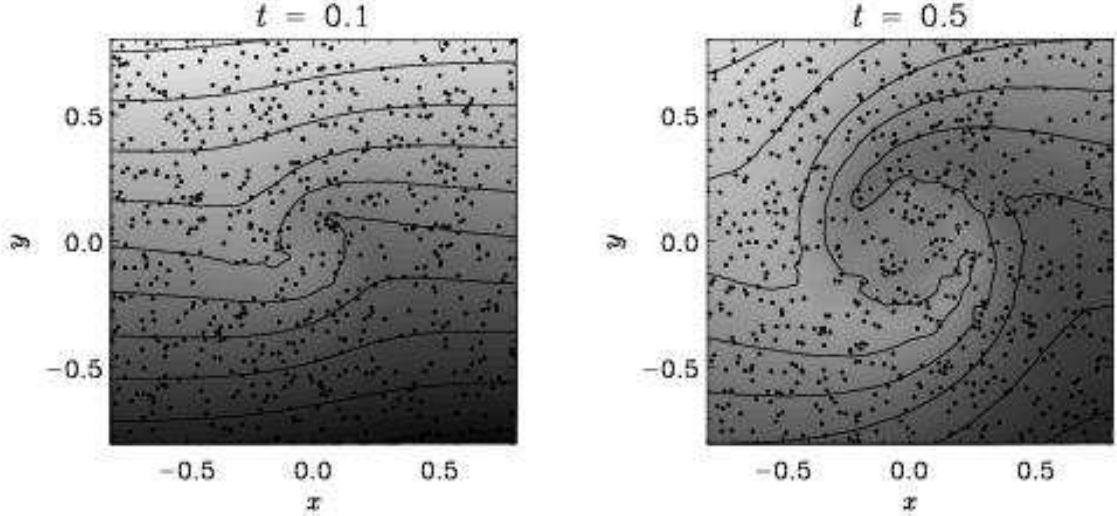}\end{center}\caption[]{
Two-dimensional advection problem on an unstructured lagrangian mesh.
The dots indicate the 1000 lagrangian particles which
constitute the unstructured mesh.
}\label{Fremesh}\end{figure}

We now discuss how we can calculate spatial derivatives of our variables
from a nonuniformly spaced ensemble of points.
Consider the function $f(x,y,z)$, which stand for one of the components
of a vector (velocity or magnetic vector potential) or a scalar,
such as $\ln\rho$. We approximate the function $f(x,y,z)$ in the
neighborhood of the point $\xx_i=(x_i,y_i,z_i)$ by a multidimensional
polynomial of degree $N$,
\EQ
f(x,y,z)=f(x_i,y_i,z_i)
+\sum_{l+m+n\leq N}c_{lmn}(x-x_i)^l(y-y_i)^m(z-z_i)^n,
\label{expand}
\EN
where $l$, $m$, and $n$ are non-negative integers and $c_{lmn}$ are
coefficients that are to be determined separately for each point by
applying \Eq{expand} to all neighboring points $\xx_j$. Note that
$c_{000}=0$ and does not need to be considered. Thus, for each point $j$
we have a system of equations
\EQ
f_{ij}=\sum_{l+m+n\leq N}{1\over l!\,m!\,n!}
c_{lmn}x_{ij}^l\,y_{ij}^m\,z_{ij}^n,
\EN
where $f_{ij}=f(x_i,y_i,z_i)-f(x_j,y_j,z_j)$ and $\xx_{ij}=\xx_i-\xx_j$.
This system of equations can be written in matrix form
\EQ
f_\alpha={\sf M}_{\alpha\beta} C_\beta,
\EN
where $1\leq(\alpha,\beta)\leq M$ and $M$ is the spatial dimension of
the matrix, which is related to $N$ and the dimension as follows:
\EQ
M=\left\{
\begin{array}{ll}
N & \mbox{in 1-D}, \\
(N+1)(N+2)/2-1 & \mbox{in 2-D}, \\
(N+1)(3N/2) & \mbox{in 3-D}.
\end{array}
\right.
\EN
When $N=2$ the matrix $\MMMM$ is given by
\EQA
\MMMM=\pmatrix{
x_{ij_1}&y_{ij_1}&z_{ij_1}&\half x_{ij_1}^2&x_{ij_1}y_{ij_1}
&\half y_{ij_1}^2&y_{ij_1}z_{ij_1}&\half z_{ij_1}^2&z_{ij_1}x_{ij_1}\cr
x_{ij_2}&y_{ij_2}&z_{ij_2}&\half x_{ij_2}^2&x_{ij_2}y_{ij_2}
&\half y_{ij_2}^2&y_{ij_2}z_{ij_2}&\half z_{ij_2}^2&z_{ij_2}x_{ij_2}\cr
 ... & ... & ... & ... & ... & ... & ... & ... & ...\cr
x_{ij_M}&y_{ij_M}&z_{ij_M}&\half x_{ij_M}^2&x_{ij_M}y_{ij_M}
&\half y_{ij_M}^2&y_{ij_M}z_{ij_M}&\half z_{ij_M}^2&z_{ij_M}x_{ij_M}}
\ENA
and
\EQA
\CC=\left(c_{100},c_{010},c_{001},c_{200},c_{110},c_{020},
c_{011},c_{002},c_{101}\right)^T.
\ENA
Here, $j_n$ ($n=1,2,...,M$) are the $M$ nearest neighbors of
the point $i$. In general the matrix can be written in the form
\EQ
M_{\alpha\beta}^{(J)}=
x_{\alpha J}^{l(\beta)}
y_{\alpha J}^{m(\beta)}
z_{\alpha J}^{n(\beta)},
\EN
where $J$ is the index of the point at which the derivative
is to be calculated. The set of exponents $l(\beta)$, $m(\beta)$, and
$n(\beta)$ is given here for the case $N=4$ in 3-D:
\EQ
l(\beta)=(1,0,0|2,1,0,0,0,1|3,2,1,0,0,0,0,1,2|4,3,2,1,0,0,0,0,0,1,2,3),
\EN
\EQ
m(\beta)=(0,1,0|0,1,2,1,0,0|0,1,2,3,2,1,0,0,0|0,1,2,3,4,3,2,1,0,0,0,0),
\EN
\EQ
n(\beta)=(0,0,1|0,0,0,1,2,1|0,0,0,0,1,2,3,2,1|0,0,0,0,0,1,2,3,4,3,2,1),
\EN
where the vertical bars separate the sets of exponents that correspond
to increasing orders. Once the $C$ vector has been obtained, the first
derivatives of $f$ are simply given by
\EQ
{\partial f\over\partial x}=c_{100},\quad
{\partial f\over\partial y}=c_{010},\quad
{\partial f\over\partial z}=c_{001}.
\EN
Likewise, the second derivatives are given by
\EQ
{\partial^2 f\over\partial x^2}=c_{200},\quad
{\partial^2 f\over\partial y^2}=c_{020},\quad
{\partial^2 f\over\partial z^2}=c_{002},
\EN
and the mixed second derivatives are given by
\EQ
{\partial^2 f\over\partial x\partial y}=c_{110},\quad
{\partial^2 f\over\partial y\partial z}=c_{011},\quad
{\partial^2 f\over\partial z\partial x}=c_{101}.
\EN
Although this method can be used for meshes that are static in time,
it can also be used in connection with multi-dimensional lagrangian
schemes. In that case there may arise the problem that neighboring
points get very close together, and so small errors strongly
affect the coefficients. A good way out of this is to use a few more
points and to solve the linear matrix equation using singular value
decomposition. An example of such a calculation is shown in \Fig{Fremesh},
where a passive scalar, with the initial distribution $A(\xx,0)=x$, is
advected by the velocity, $\uu=\dot{\rrr}$, which in turn is obtained by
solving Kepler's equation, $\ddot{\rrr}=-GM\rrr/r^3$, using the normalization
$GM=1$. This windup problem corresponds to the windup of initially horizontal
magnetic field lines.

In diffusivity used in \Fig{Fremesh} was $\eta=0.02$, but due to the coarse
resolution and the implicit smoothing resulting from the singular value
decomposition technique the effective diffusivity is somewhat larger.

\section{Implementing magnetic fields}

As mentioned in \Sec{NavierStokes}, implementing magnetic fields is
relatively straightforward. On the one hand, the magnetic field causes
a Lorentz force, $\JJ\times\BB$, where $\BB$ is the flux density,
$\JJ=\nab\times\BB/\mu_0$ is the current density, and $\mu_0$ is the
vacuum permeability. Note, however, that $\JJ\times\BB$ is the force per
unit volume, so in \Eq{dudt} we need to add the term $\JJ\times\BB/\rho$
on the right hand side. On the other hand, $\BB$ itself evolves according
to the Faraday equation,
\EQ
{\partial\BB\over\partial t}=-\nab\times\EE
\label{induction}
\EN
where the electric field $\EE$ can be expressed in terms of $\JJ$ using
Ohm's law in the laboratory frame, $\EE=-\uu\times\BB+\JJ/\sigma$, where
$\sigma=(\eta\mu_0)^{-1}$ is the electric conductivity and $\eta$ is the
magnetic diffusivity.

In addition we have to satisfy the condition $\nab\cdot\BB=0$. This is
most easily done by solving not for $\BB$, but instead for the magnetic
vector potential $\AAA$, where $\BB=\nab\times\AAA$. The evolution
of $\AAA$ is governed by the uncurled form of \Eq{induction},
\EQ
{\partial\AAA\over\partial t}=-\EE-\nab\phi
\EN
where $\phi$ is the electrostatic potential, which takes the role of an
integration constant which does not affect the evolution of $\BB$.
The choice $\phi=0$ is most advantageous on numerical grounds. (By
contrast, the Coulomb gauge $\nabla\cdot\AAA=0$, which is very popular
in analytic considerations, would obviously be of no advantages, since
one still has the problem of solving a the solenoidality condition.

Solving for $\AAA$ instead of $\BB$ has significant advantages,
even though this involves taking another
derivative. However, the total number of derivatives taken in the code is
essentially the same. In fact, when centered finite differences are
employed, Alfv\'en waves are better resolved when $\AAA$ is used,
because then the system of equations for one-dimensional Alfv\'en waves
in the presence of a uniform $B_{x0}$ field in a medium of constant
density $\rho_0$ reduces to
\EQ
\dot{u}_z={1\over\mu_0\rho_0}B_{x0}A_y'',\quad
\dot{A}_y=B_{x0}u_z,
\EN
where a second derivative is taken only once (primes denote
$x$-derivatives). If, instead, one solves for the $B_z$ field, one has
\EQ
\dot{u}_z={1\over\mu_0\rho_0}B_{x0}B_z',\quad
\dot{B}_z=B_{x0}u_z',
\EN
where a first derivative is applied twice, which is far less accurate
at small scales if a centered finite difference scheme is used. At
the Nyquist frequency, for example, the first derivative is zero and
applying an additional first derivative gives still zero. By contrast,
taking a second derivative once gives of course not zero. The use of
a staggered mesh circumvents this difficulty. However, such an approach
introduces additional complications which hamper the ease with which
the code can be adapted to other problems.

Another advantage of using $\AAA$ is that it is straightforward to
evaluate the magnetic helicity, $\bra{\AAA\cdot\BB}$, which is a
particularly important quantity to monitor in connection with dynamo
and reconnection problems.

The main advantage of solving for $\AAA$ is of course that one does
not need to worry about the solenoidality of the $\BB$-field, even
though one may want to employ irregular meshes or complicated
boundary conditions.

As we have emphasized before, when centered meshes are used, it is
usually a good idea to avoid taking first derivatives of the same
variable twice, because it is more accurate to take instead a second
derivative only once. For this reason we calculate the current not
as $\JJ=\mu_0^{-1}\nab\times(\nab\times\AAA)$, but as
\EQ
\JJ=\mu_0^{-1}\left[-\nabla^2\AAA+\nab(\nab\cdot\AAA)\right].
\EN
Taking the gradient of $\nab\cdot\AAA$ involves of course also
taking first derivatives of the same variable twice, but these
contributions are canceled by corresponding components of the
$\nabla^2\AAA$ term. There are some advantages relying here on
the numerical cancellation, which is of course not exact. The
reason is that the full $\nabla^2\AAA$ term is important when
used in the magnetic diffusion term. If the diagonal terms,
$\partial^2A_x/\partial x^2$, $\partial^2A_y/\partial y^2$,
and $\partial^2A_z/\partial z^2$, which would all drop out
analytically, were taken out there would be no diffusion of $\AAA$
in the direction of $\AAA$.

There is one more aspect that is often useful keeping in mind. There is
a particular gauge that allows one to rewrite the uncurled induction
equation in such a form that the evolution of $\AAA$ is controlled
by the advective derivative of $\AAA$. The calculation is easy. Write
the induction term $\uu\times\BB$ in component form and express $\BB$
in terms of $\AAA$, so
\EQ
(\uu\times\nab\times\AAA)_i=u_j(\partial_i A_j-\partial_j A_i)
=\partial_i(u_j A_j)-A_j\partial_i u_j-u_j\partial_j A_i.
\EN
Here the last term contributes to the advective derivative, the
first term can be removed by a gauge transformation
and the middle term is a modified stretching term,
so the induction equation takes the form
\EQ
{\DD A_i\over\DD t}=-A_j\partial_i u_j-\eta\mu_0 J_i.
\label{special_gauge}
\EN
This gauge was used by Brandenburg \ea (1995) in order to treat a
linear velocity shear using pseudo-periodic (shearing box) boundary
conditions. The formulation \eq{special_gauge} can also be useful
when solving the induction equation using lagrangian methods. Note,
however, that the nonresistive evolution of $\AAA$ differs from
that of $\BB$ in that the indices of the matrix
${\sf U}_{ij}\equiv\partial u_i/\partial x_j$ are interchanged
and that the sign is different; positive for the $\BB$-equation,
\EQ
{\DD B_i\over\DD t}=+{\sf U}_{ij}B_j+\mbox{other terms},
\EN
and negative for the $\AAA$-equation,
\EQ
{\DD A_i\over\DD t}=-A_j{\sf U}_{ji}+\mbox{other terms}.
\EN
These two formulations are particularly advantageous when the velocity
has a constant gradient, as in the case of linear shear. In local
simulations of accretion discs, for example, the shear component is
$u_y(x)=-{3\over2}\Omega x$, so ${\sf U}_{yx}=-{3\over2}\Omega$, and
all other ${\sf U}_{ij}$ vanish. Hence
\EQ
{\DD A_x\over\DD t}=+{3\over2}\Omega A_y+\mbox{other terms}
\EN
for the $\AAA$-formulation, or
\EQ
{\DD B_y\over\DD t}=-{3\over2}\Omega B_x+\mbox{other terms}
\EN
for the $\BB$-formulation. In these two formulations all dependent
variables are clearly periodic (or rather pseudo-periodic), so there is
no term that is explicitly non-periodic such as $u_y(x)=-{3\over2}\Omega
x$. In the following, whenever magnetic fields are present, we use the
$\AAA$-formulation, mainly because it guarantees the solenoidality of
$\BB$ everywhere (including the boundaries), and also because it is easy
to use.

\subsection*{Cache-efficient coding}

Unlike the CRAY computers that dominated supercomputing in the 80ties and
early 90ties, all modern computers have a cache that constitutes a significant
bottleneck for many codes. This is the case if large three-dimensional
arrays are constantly used within each time step. The advantage of this
way of coding is clearly the conceptual simplicity of the code. A more
cache-efficient way of coding is to calculate an entire timestep (or
a corresponding substep in a three-stage $2N$ Runge-Kutta scheme) only
along a one-dimensional pencil of data within the box. On Linux and Irix
architectures, for example, this leads to a speed-up by 60\%. An additional
advantage is a drastic reduction in temporary storage that is needed for
auxiliary variables within each time step.

\section{Application to astrophysical outflows}

\subsection{The isothermal Parker wind}

\begin{figure}[h!]\begin{center}\includegraphics[width=.99\textwidth]{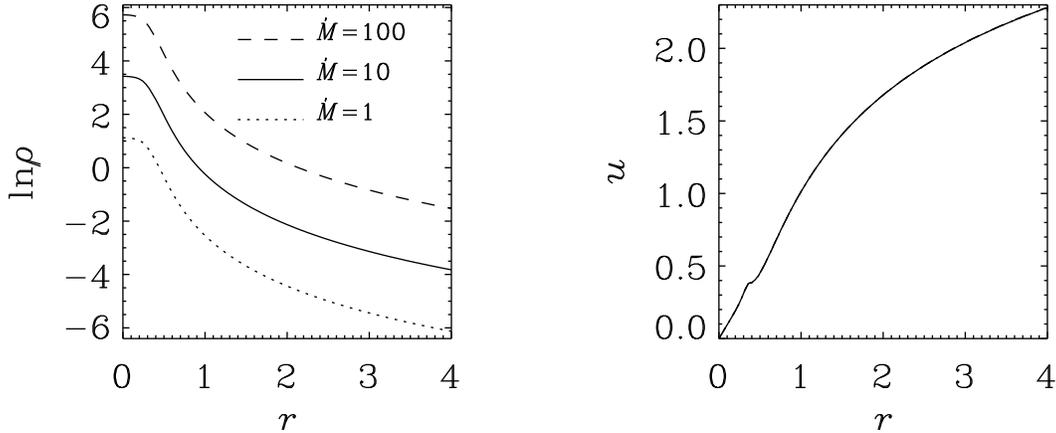}\end{center}\caption[]{
Isothermal Parker wind solutions for different values of $\dot{M}$. Note
that the $u$ profile is independent of the value of $\dot{M}$. $GM=2$,
$c_{\rm s}=1$, $r_0=0.4$.
}\label{Fwindq-visc-pall2_gam=1}\end{figure}

Before discussing outflows from accretion discs it is illuminating
to consider first the one-dimensional example of pressure-driven outflows
in spherical geometry. A particularly simple case is the
{\it isothermal} wind problem, which is governed by the equations
\EQ
{\partial u\over\partial t}+u{\partial u\over\partial r}=
-c_{\rm s}^2{\partial\ln\rho\over\partial r}
-{\partial\Phi\over\partial r},
\EN
\EQ
{\partial\ln\rho\over\partial t}+u{\partial\ln\rho\over\partial r}=
-{1\over r^2}{\partial\over\partial r}(r^2u)+{\dot{M}\xi(r)\over\rho},
\EN
where $c_{\rm s}$ is the isothermal sound speed (assumed constant),
$\dot{M}$ is the mass loss rate, and $\xi(r)$ is a prescribed function
of position, normalized such that $\int4\pi r^2\xi(r)\dd r=1$, and
non-vanishing only near $r=0$. For a point mass the gravity potential
$\Phi$ would be $-GM/r$, but this becomes singular at the origin.
Therefore we use the expression $\Phi=-GM/(r^n+r_0^n)^{1/n}$ instead,
where we choose $n=5$ in all cases, and $1/r_0$ gives the depth of the
potential well. In \Fig{Fwindq-visc-pall2_gam=1} we show radial velocity
and density profiles for different values of $\dot{M}$. Note that the
velocity profile is independent of the value of $\dot{M}$, but the density
profile changes by a constant factor. In the steady case the equations
can be combined to
\EQ
\left(u^2-c_{\rm s}^2\right){\dd\ln|u|\over\dd r}
={2c_{\rm s}^2\over r}-{GM\over r^2},
\EN
so the sonic point, $|u|=c_{\rm s}$, is at $r=r_*=GM/2c_{\rm s}^2$. In
\Fig{Fwindq-visc-pall2_gam=1} we have chosen $GM=2$ and $c_{\rm
s}=1$, so $r_*=1$, which is consistent with the graph of $u$.

\subsection{The polytropic or adiabatic wind}

\begin{figure}[h!]\begin{center}\includegraphics[width=.99\textwidth]{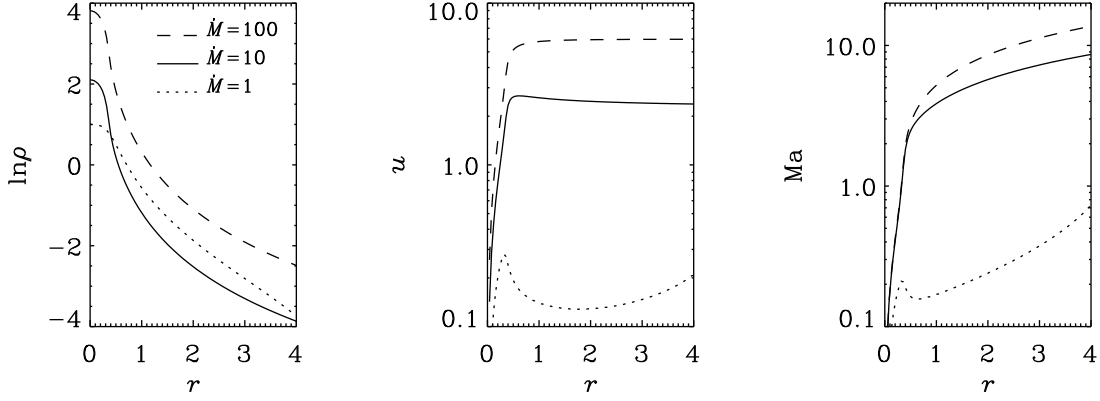}\end{center}\caption[]{
Polytropic Parker wind solutions for different values of $\dot{M}$. $GM=2$,
$c_{\rm s0}=1$, $r_0=0.4$.
}\label{Fwindq-visc-pall2}\end{figure}

In the following we make the assumption that the entropy is constant. In that
case it is particularly useful to solve for the {\it potential enthalpy},
$H\equiv h+\Phi$, which varies much less than either $h$ or $\Phi$.
Using $H$ as dependent variable is particularly useful if one solves
the equations all the way to the origin, $r=0$, where $\Phi$ tends to
become singular (or at least strongly negative if a smoothed potential
is used). In terms of $H$ the governing equations are
\EQ
{\partial u\over\partial t}+u{\partial u\over\partial r}=
-{\partial H\over\partial r}-u\dot{M}\xi,
\EN
\EQ
{\partial H\over\partial t}+u{\partial H\over\partial r}=
u{\partial\Phi\over\partial r}+c_{\rm s}^2\left[
-{1\over r^2}{\partial\over\partial r}(r^2u)+{\dot{M}\xi(r)\over\rho}
\right],
\EN
where $H=h+\Phi$ is the potential enthalpy, $h=p/\rho+e$ is the enthalpy,
and $c_{\rm s}^2=(\gamma-1)h=(\gamma-1)(H-\Phi)$ for a perfect gas,
where $c_{\rm s}$ is the adiabatic sound speed and $h=c_pT$ is
the enthalpy. These equations are also valid in the nonisothermal
case ($\gamma\neq1$). The isothermal case may be recovered by
putting $\gamma=1$ and replacing $h$ by $c_{\rm s}^2\ln\rho$. In
\Fig{Fwindq-visc-pall2} we show solutions for different values of
$\dot{M}$ and $\gamma=5/3$. Again we put $GM=2$ and $c_{\rm s0}=1$.

We note that, depending on the strength of the mass source, the
polytropic wind problem allows a variety of different velocity and Mach
number profiles, whereas for the isothermal wind problem there was only
one solution possible, independent of the strength of the mass source.
The velocity profile was always the same and also the density was the
same up to some scaling factor that changes with $\dot{M}$. This is
connected with the additional degree of freedom introduced through
the polytropic constant $K=p/rho^\gamma$. Since $c_{\rm s}$ is no longer
constant, the position of the sonic point is no longer fixed and different
solutions are possible.

\begin{figure}[h!]\begin{center}\includegraphics[width=.99\textwidth]{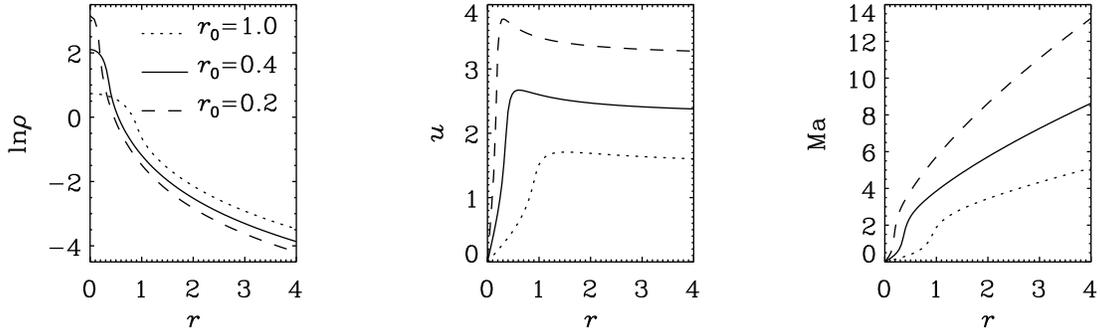}\end{center}\caption[]{
Density $\rho$, velocity $u$, and Mach number $\mbox{Ma}=u/c_{\rm s}$ for
the polytropic Parker wind solutions for different values of $r_0$. $GM=2$,
$c_{\rm s0}=1$, $\dot{M}=10$.
}\label{Fwindq-visc-pall3}\end{figure}

In \Fig{Fwindq-visc-pall3} we show solutions where $\dot{M}$ is kept
constant, but the depth of the potential well, $GM/r_0$, is changed
by varying the value of $r_0$. Note that the deeper the potential well,
the higher the wind speed. The density far away from the source is then
correspondingly smaller, so as to maintain the same mass flux.

As we have seen in \Sec{Spoly}, a polytropic equation of state
is unphysical. Therefore we now consider the case where the energy
equation is included. To be somewhat more general we consider first the
basic equations in conservative form with mass, momentum and energy
sources included, i.e.
\EQ
{\partial\rho\over\partial t}+
{\partial\over\partial x_j}\left(\rho u_j\right)
=\dot{M}\xi,
\EN
\EQ
{\partial\over\partial t}\left(\rho u_i\right)+
{\partial\over\partial x_j}\left(\rho u_i u_j+p\delta_{ij}-\tau_{ij}\right)
=\dot{I}_i\xi,
\EN
\EQ
{\partial\over\partial t}\left(\half\rho\uu^2+\rho e\right)+
{\partial\over\partial x_j}
\left[u_j\left(\half\rho\uu^2+\rho h\right)-u_i\tau_{ij}\right]
=\dot{E}\xi,
\EN
where $\dot{M}$, $\dot{\II}$, and $\dot{E}$ are the rates of mass,
momentum and energy injection into the system, $\tau_{ij}=2\nu\rho{\sf
S}_{ij}$ is the viscous stress tensor, and ${\sf S}_{ij}$ is the
(traceless) rate of strain tensor; see \Eq{rate-of-strain}. Rewriting
the energy equation in nonconservative form we have
\EQ
{\DD e\over\DD t}+{p\over\rho}\nab\cdot\uu\equiv T{\DD s\over\DD t}=2\nu\SSSS^2
+\left[\dot{E}-\uu\cdot\left(\dot{\II}-\uu\dot{M}\right)
-\left(\half\uu^2+e\right)\dot{M}\right]{\xi(r)\over\rho},
\EN
which can also be rewritten in terms of entropy, so the final system
of nonconservative equations with source terms is
\\

\SHADOWBOX{
\EQ
{\DD\ln\rho\over\DD t}+\nab\cdot\uu=\dot{M}{\xi(r)\over\rho},
\label{wind1}
\EN
\EQ
{\DD\uu\over\DD t}+c_{\rm s}^2(\nab\ln\rho+\nab s)
={1\over\rho}\nab\cdot(2\nu\rho\SSSS)
+\left(\dot{\II}-\uu\dot{M}\right){\xi(r)\over\rho},
\label{wind2}
\EN
\EQ
T{\DD s\over\DD t}=2\nu\SSSS^2
+\left[\dot{E}-\uu\cdot\dot{\II}+\left(\half\uu^2-e\right)\dot{M}\right]
{\xi(r)\over\rho},
\label{wind3}
\EN
}\\

\noindent
where $T$ can be replaced by $c_{\rm s}^2/(\gamma-1)$ (remember that
$c_p=1$), and $c_{\rm s}^2$ is given by \eq{adSoundSpeed}.

\begin{figure}[h!]\begin{center}\includegraphics[width=.99\textwidth]{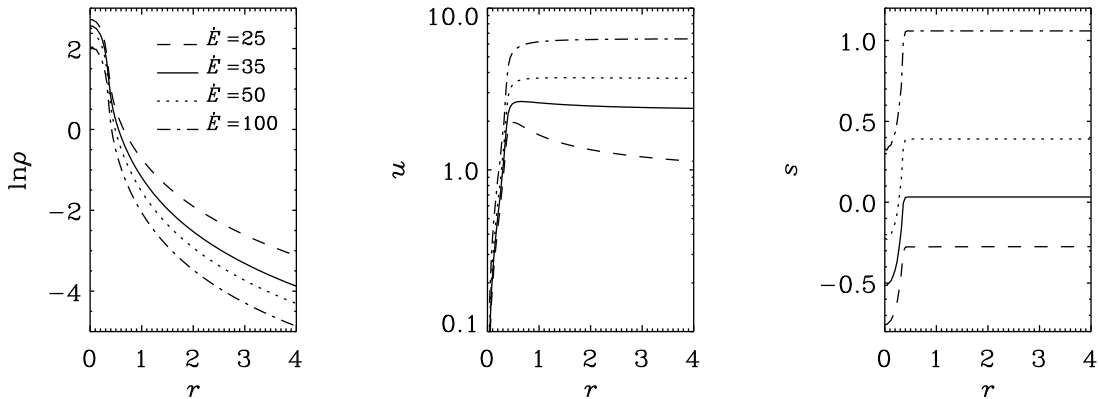}\end{center}\caption[]{
Wind solutions for different values of $\dot{E}$ and $\dot{M}=10$.
Note that the solution with $\dot{E}=35$ is quantitatively very
similar to the polytropic solution with the same value of $\dot{M}$.
$GM=2$, $c_{\rm s0}=1$, $r_0=0.4$.
}\label{Fwind-pall_edot}\end{figure}

In \Fig{Fwind-pall_edot} we present solutions of
Eqs.~(\ref{wind1})--(\ref{wind3}) for different values of $\dot{E}$ and
$\dot{M}$. The main effect of varying the value of $\dot{E}$ is to change
the value of the entropy in the wind. Outside the acceleration region,
however, the value of the entropy is fairly constant, so the polytropic
assumption appears to be reasonably good here.

While outflows of some very early-type stars are driven mostly by the
$\dot{\II}$ term (resulting from the radiation pressure in lines), the
winds of cool stars are driven mostly by the $\dot{E}$ term (resulting
from the hot coronae). Similar differences may also explain why some jets
are massive (stellar jets, for example), whilst others are not (jets from
active galactic nuclei, for example, or those anticipated in gamma-ray
bursters).


\subsection{Relevance to outflows and jets}

The pressure-driven outflows discussed in the previous section may take
the form of more collimated outflows once a magnetic field is involved.
This applies to the case of magnetized accretion discs. These discs are
generally magnetized both because of dynamo action within the disc and
because of external fields that were dragged into the disc from outside
due to the accretion flow.

At least in some types of jets the outflows may be driven by
hot coronae. Other possibilities for driving outflows involve the
magneto-centrifugal effect. It is well-known that outflows can be driven
from a magnetized disc if the angle between the field and the disc is
less than $60^\circ$ (Blandford \& Payne 1982). Recent work in this
field was directed to the question whether this angle is the result of
some self-regulating process (Ouyed, Pudritz, \& Stone 1997, Ouyed \&
Pudritz 1997a,b, 1999) and whether it can be obtained automatically from
a dynamo operating within the disc (Campbell 1999, 2000, Dobler \ea 1999,
M.\ v.\ Rekowski, R\"udiger, \& Elstner 2000). This latter question is
particularly interesting in view of the fact that jets in star-forming
regions are not really pointing in a similar direction (e.g.\ Hodapp \&
Ladd 1995), as one might expect from jet models that start off with a
prescribed large scale field.

\begin{figure}[h!]\begin{center}\includegraphics[width=.5\textwidth]{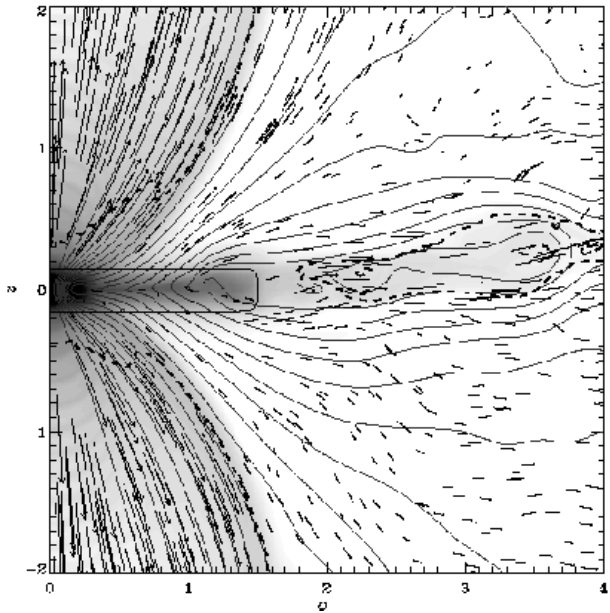}\end{center}\caption[]{
Poloidal velocity vectors and poloidal magnetic field lines superimposed
on a grey-scale image of the logarithmic density. Dark means high density.
The thick dashed line denotes the location where the poloidal flow speed
equals the fast magnetosonic speed. The thin solid line gives the
location of the disc surface. The slight asymmetry in the field is
a relic from the mixed-parity initial condition.
[Adapted from Brandenburg (2000).]
}\label{Fall}\end{figure}

In \Fig{Fall} we present a particular model of Dobler \ea (1999) and
Brandenburg (2000); see Brandenburg \ea (2000) for a full account
of this work. In these models the outflow is driven by mass sources
whose strength is proportional to the local density deficit relative to
that in the original equilibrium solution of the disc. Such a density
deficit was initially caused by slow gas motions that resulted from an
instability of the initial equilibrium solution, because a cool disc
embedded in a hot corona is nonrotating outside the disc, and it is the
resulting vertical shear profile that causes the instability (cf.\ Urpin
\& Brandenburg 1998). At later times, of course, the outflow makes the
corona corotating, but by that time the outflow is driven by a persistent
density deficit in the disc relative to the initial references solution.

In this model the magnetic field was generated by an $\alpha-\Omega$
dynamo operating within the disc. However $\alpha$ is negative in the
upper disc plane (see Brandenburg \ea 1995), and then the most preferred
field geometry is dipolar (Campbell 1999, v.~Rekowski, R\"udiger, \&
Elstner 2000). The field parity is sensitive to details in
the disc physics assumed in the particular model (aspect ratio, disc
thickness, the presence of outflows, and the conductivity in the disc
and the exterior). Nevertheless, both dipolar and quadrupolar fields are
equally well able to contribute to wind launching, at least in the outer
parts of the disc where the angle between the field and the disc plane
is less than $60^\circ$, the critical angle for magneto-centrifugal
wind launching (Blandford \& Payne 1982). We note, however, that the
more detailed analysis of Campbell (1999) suggests that the critical
angle can be significantly smaller.

In our models the outflow is only weakly collimated (if at all). This
is probably connected with the fact that here the fast magnetosonic
surface is rather close to the disc surface, making it difficult for the
field to become strong enough to channel the magnetic field. Instead,
the field lines themselves are still being controlled too strongly by
the outflow. However, outflows with rather large opening angles are
actually seen in some star-forming regions; see Greenhill \ea (1998).

\begin{figure}[h!]\begin{center}\includegraphics[width=.99\textwidth]{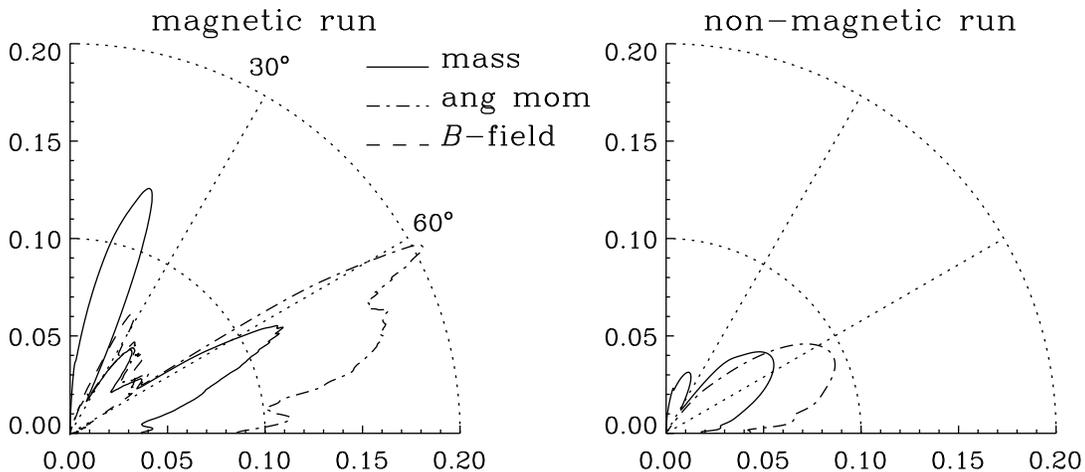}\end{center}\caption[]{
Comparison of the angular dependence of azimuthally integrated fluxes
for magnetic and nonmagnetic outflows. The solid line refers to mass
flux, the dashed line to angular momentum flux, and the dash-dotted line
(in the second panel) corresponds to the Poynting flux.  The units of
all quantities are thus $1/[t]$.
}\label{Fpcircle_both}\end{figure}

While most of the disc mass is ejected in a cone of half-opening angle
around $25^\circ$, most of the disc angular momentum is ejected at
rather low latitudes, almost in the direction of the disc plane away
from the central object. The timescales for these various processes are
comparable. In \Fig{Fpcircle_both} we show the azimuthally integrated
mass flux, angular momentum flux, and magnetic (Poynting) flux as a
function of polar angle, and compare with a nonmagnetic run. We find that
in the magnetic run the outflow is more strongly concentrated towards
the axis. Also, the amount of angular momentum loss (dash-dotted line)
is larger when the disc is magnetized. We emphasize in particular that
in the magnetic run significant amounts of magnetic field are eject
from the system. In the following section we discuss the significance
of such magnetic flux ejection for magnetizing the interstellar medium
into which the outflow is streaming.  This discussion is similar to
a corresponding discussion for the contamination of the intergalactic
medium via outflows from active galactic nuclei (Brandenburg 2000).

\subsection{Magnetic contamination from outflows}

It may at first appear somewhat unrealistic to expect significant
magnetization of the interstellar medium from outflows. However, the
following calculation shows that the effect may be quite significant.
Assume that every star did undergo a phase of strong accretion with
associated outflows, so $N=10^{11}$ for the whole galaxy. The duration
of intense outflow activity is $10^5$ years, say, but it could even be
$10^6$ years. The magnetic luminosity is $L_{\rm mag}=0.05\dot{M}_{\rm
w}c_{\rm s}^2$ (Brandenburg \ea 2000), where $c_{\rm s}\approx10\kms$ is
the average sound speed of the interstellar medium, and $\dot{M}_{\rm
w}=0.1\dot{M}_{\rm d}$ (see Pelletier \& Pudritz 1992), where
$\dot{M}_{\rm d}\approx10^{-7}M_\odot/\yr$ is a conservative estimate
for the disc accretion rate. Again this value may be larger. With the
above numbers the magnetic luminosity from all $N=10^{11}$ sources is
then $NL_{\rm mag}=7\times10^{39}\erg/\s$ and the total energy output
delivered from all stars at some early point in the life time is therefore
$E_{\rm mag}=\tau NL_{\rm mag}=2\times10^{52}\erg$. Diluting this over
a volume of a galaxy of $300\kpc^3$ (radius 10~kpc, height 1~kpc) gives
$2\times10^{-15}\erg/\cm^3$.  Multiplying this by $8\pi$ and taking the
square root gives $0.2\mu G$.  Expressed more concisely in a formula we
have for the rms magnetic field strength
\EQ
\bra{\BB^2}^{1/2}\approx
\left(8\pi\,{F_{\rm Poy}\over F_{\rm kin}}\,
{N\dot{M}_{\rm w}c_{\rm s}^2\over V}\,\delta t\right)^{1/2},
\EN
where the efficiency factor $F_{\rm Poy}/F_{\rm kin}$ (=0.05 in our
model) may be lower in systems where the disc dynamo is less strong.

The parameters for a corresponding estimate for outflows from young
galactic discs (active galactic nuclei) are as follows. Assuming
$N\sim10^4$ galaxies per cluster, each with $\dot{M}_{\rm
w}\approx0.1M_\odot/{\rm yr}=10^{25}\gs$, and $c_{\rm s}=1000\kms$ for
the sound speed in the intracluster gas, the rate of magnetic energy
injection for all galaxies together is $L_{\rm mag}=10^{44}\ergs$.
Distributing this over the volume of the cluster of $V\sim1\Mpc^3$, and
integrating over a duration of $\delta t=1\Gyr$, this corresponds to a
mean magnetic energy density of $\bra{\BB^2/8\pi}\approx10^{-13}\erg/{\rm
cm}^3$, so $\bra{\BB^2}^{1/2}\approx10^{-6}\G$, which is indeed of the
order of the field strength observed in galaxy clusters. We note that our
estimate has been rather optimistic in places ($\dot{M}_{\rm w}$ could be
lower, or the relevant $\delta t$ could be shorter, for example), but it
does show that outflows are bound to produce significant magnetization
of the intracluster gas and the interstellar medium (see also V\"olk \&
Atoyan 1999). In the latter case it will provide a good seed field for the
galactic dynamo. A dynamo is still necessary to shape the magnetic field
and to prevent if from decaying in the galactic turbulence. Similarly,
many galaxy clusters undergo merging and this too can enhance and
reorganize the magnetic field. The necessity for a recent merger event
would also be consistent with the fact that not all halos are observed to
have strong magnetic fields. Recent simulations by Roettiger, Stone \&
Burns (1999) suggest that after a merger the field strength may increase
by a factor of at least 20 (and this value increases with improving
observational resolution).

As an alternative consideration for causing the magnetization in clusters
of galaxies, {\it primordial} magnetic fields are sometimes discussed.
There are numerous mechanisms that could generate relatively strong fields
at an early time, for example during inflation (age $\sim10^{-36}\s$)
or during the electroweak phase transition (age $\sim10^{-10}\s$). Such
fields would now still be at a very small scale if one considers only
the cosmological expansion. However, depending on the degree of magnetic
helicity in this primordial field, the magnetic energy can be transferred
to larger scales that are now on the scale of galaxies. For a recent
discussion of these results see Brandenburg (2001a).

\section{Hydromagnetic turbulence and dynamos}

As mentioned in the beginning, accurate high order schemes are
essential in all applications to turbulent flows. Nevertheless, we
should mention that one often attempts solutions of the inviscid and
nonresistive equations using low-order finite differences combined with
monotonicity schemes that result in some kind of effective diffusion. The
piece-wise parabolic method (PPM) of Colella \& Woodward (1984) is an
example. However, unlike the Smagorinsky scheme (see Chan \& Sofia 1986,
1989, Steffen, Ludwig, \& Kr\"u{\ss} 1989, Fox, Theobald, \& Sofia 1991
for applications to convection simulations), PPM and similar methods
cannot be proven to converge to the original Navier-Stokes equation in
the limit of infinite resolution. Nevertheless, they are rather popular
in astrophysical gas simulations. These schemes are rather robust and
have also been applied to high resolution simulations of compressible
turbulence (Porter, Pouquet, \& Woodward 1992, 1994). While the results
from those simulations are generally quite plausible, the power spectrum
shows a $k^{-1}$ subrange at large wave numbers, which is still not fully
understood. This was sometimes regarded as an artifact of PPM, and should
therefore only occur at small scales. However, as the resolution was
increased further (up to $1024^3$), the $k^{-1}$ subrange just became
more extended.

A similar feature was found in cascade models of turbulence when the
ordinary $\nabla^2$ diffusion operator was replaced by a $-\nabla^4$
``hyperdiffusion'' operator (Lohse \& M\"uller-Groeling 1995). Whatever
the outcome of this puzzle is, it is clear that with schemes that cannot
be proven to converge to the actual Navier-Stokes equations in the limit
of infinite resolution, there would always remain some uncertainty and
debate. On the other hand, especially in the incompressible case the use of
hyperviscosity does generally allow the exploration of larger Reynolds numbers
and broader inertial ranges.

MHD simulations with the highest resolution to date have been performed by
Biskamp \& M\"uller (1999), who considered decaying turbulence with and
without magnetic helicity. They found that in the presence of magnetic
helicity the magnetic energy decay is significantly slower. In
particular, they found the magnetic energy decays like $t^{-1/2}$,
as opposed to $t^{-1}$ found earlier by Mac Low, Klessen, \& Burkert
(1998) for compressible turbulence.

Before we start discussing dynamo action in turbulence simulations
representative of more astrophysical settings, such as accretion discs
and the solar convection zone, let us first illustrate the mechanism of the
inverse cascade that is believed to be an important ingredient of large
scale magnetic field generation.

\subsection{Isotropic MHD turbulence}
\label{Sisotrop}

Most developments in the theory of turbulence have been carried out
under the assumptions of homogeneity and isotropy. This is certainly
true of the work on the inverse cascade (or turbulent cascades in
general), but it is also true of much of the work on the $\alpha$-effect
which -- like the inverse cascade -- describes the generation of large
scale fields. However, unlike the inverse cascade process, the energy
comes here directly from the velocity field at the scales of the
energy-carrying eddies and not from the velocity and magnetic field at
successively smaller scales, which are usually larger than the scale of the
energy-carrying eddies.

It is not easy to see whether any of these effects is actually responsible
for the large scale field generation in astrophysical bodies or even
the simulations. In simulations of accretion disc turbulence there is
certainly some evidence for the presence of an $\alpha$-effect, but it
is extremely noisy (Brandenburg \ea 1995, Brandenburg \& Donner 1997,
Ziegler \& R\"udiger 2000). Evidence for the inverse magnetic cascade
comes mostly from the magnetic energy spectra (Balsara \& Pouquet 1999,
Brandenburg 2001b), which show a marked peak at large scales, but this is
convincing only in cases where the flow is driven at a wavenumber that
is clearly larger than the smallest wavenumber in the box. In practice,
e.g.\ in convectively driven turbulence, the flow is driven at all scales
including the large scale making it difficult to see a marked peak at
the smallest wavenumber (see a corresponding discussion in Meneguzzi \&
Pouquet 1989).

\begin{figure}[h!]\begin{center}\includegraphics[width=.90\textwidth]{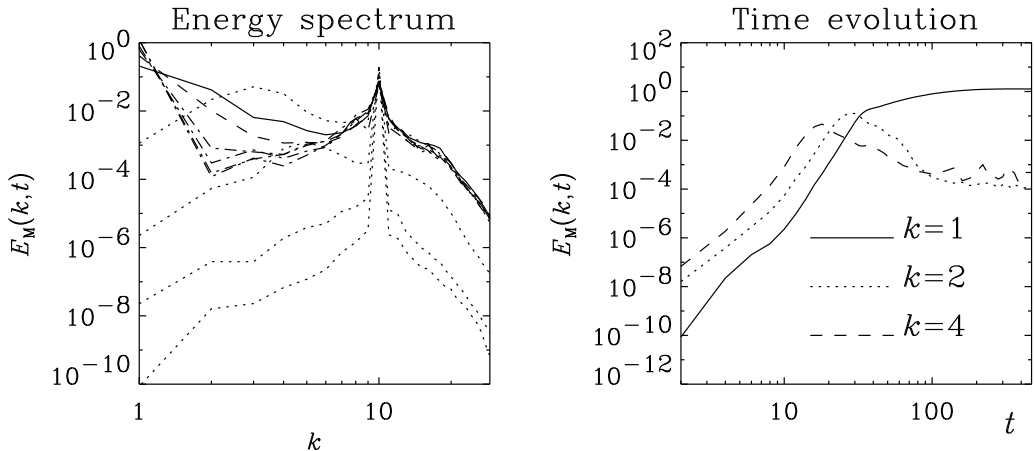}\end{center}\caption[]{
Spectral magnetic energy, $E_{\rm M}(k,t)$, as a function of wavenumber
$k$ for different times: dotted lines are for early times ($t=2,4,10,20$),
the solid and dashed lines are for $t=40$ and 60, respectively, and
the dotted-dashed lines are for later times ($t=80,100,200,400$). Here
magnetic energy is injected at wavenumber 10. Note the occurrence of a
sharp secondary peak of spectral magnetic energy at $k=10$. By the time
the energy at $k=1$ has reached equipartition the energies in $k=2$ and
$k=4$ become suppressed.
}\label{Finverse}\end{figure}

From the seminal papers of Frisch \ea (1975) and Pouquet, Frisch, L\'eorat
(1976) it is clear that amplification of large scale fields can also be
explained by an inverse cascade of magnetic helicity. In those papers the
authors also showed that the inverse cascade is a consequence of the fact
that the magnetic helicity, $\bra{\AAA\cdot\BB}$, is conserved by the
nonresistive equations. ($\AAA$ is the magnetic vector potential giving
the magnetic field as $\BB=\nab\times\AAA$.) The inverse magnetic cascade
effect too is rather difficult to isolate in simulations of astrophysical
turbulence. However, under somewhat more idealized conditions, for
example when magnetic energy is injected at high wave numbers, one
clearly sees how the magnetic energy increases at large scales; see
\Fig{Finverse}. Further details of this model have been published in
the proceedings of the helicity meeting in Boulder (Brandenburg 1999).

\begin{figure}[h!]\begin{center}\includegraphics[width=.90\textwidth]{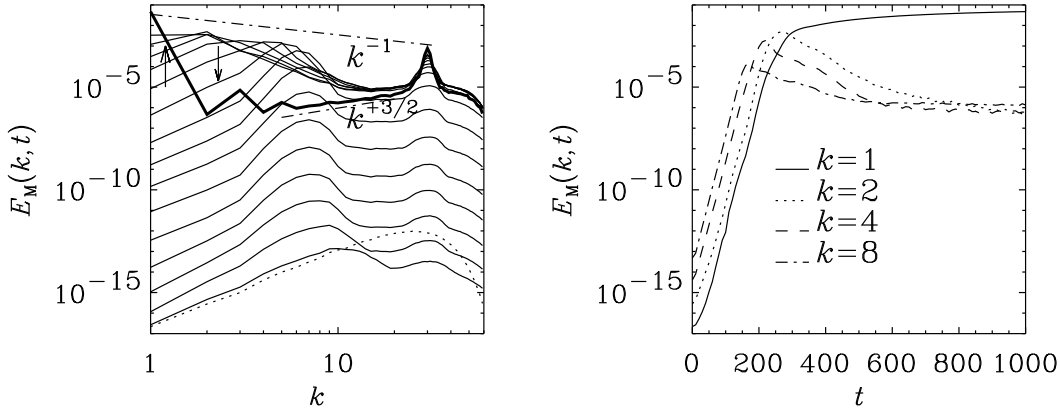}\end{center}\caption[]{
{\it Left}: magnetic energy spectra for a run with forcing at $k=30$. The
times range from 0 (dotted line) to 10, 30, ..., 290 (solid lines). The
thick solid line gives the final state at $t=1000$. Note that at early
times the spectra peaks at $k_{\max}\approx7$. The $k^{-1}$ and $k^{+3/2}$
slopes are given for orientation as dash-dotted lines. {\it Right}:
evolution of spectral magnetic energy for selected wavenumbers in a
simulation with hydrodynamical forcing at $k=30$.
}\label{FpMkt1+2}\end{figure}

In the model considered above the flow was forced magnetically. This
may be motivated by the recent realization that strong magnetic field
generation in accretion discs can be facilitated by {\it magnetic
instabilities}, such as the Balbus-Hawley instability. Other examples
of magnetic instabilities include the magnetic buoyancy instability,
which can lead to an $\alpha$-effect (e.g.\ Brandenburg \& Schmitt 1998,
Thelen 2000), and the reversed field pinch which also leads to a dynamo effect
(e.g.\ Ji \ea 1996). Before returning to the accretion disc dynamo in
\Sec{Sdiscs} we should emphasize that strong large scale field generation
is also possible with purely hydrodynamic forcing. Simulations in this
type were considered recently by Brandenburg (2001b). There are many
similarities compared with the case of magnetic forcing.  The evolution of
magnetic energy spectra in the presence of hydrodynamic forcing is shown
in \Fig{FpMkt1+2}. Like in the case of magnetic forcing (\Fig{Finverse})
there are marked peaks both at the forcing scale and at the largest scale
of the box. Furthermore, the evolution of spectral energy at the largest
scales shows similar behavior: the magnetic energy with wavenumber $k=8$
increases, reaches a maximum, and begins to decrease when the magnetic
energy at $k=4$ reaches a maximum. The same happens for the next larger
scales (wavenumbers $k=4$ and 2, until the scale of the box (with $k=1$)
is reached.

The {\it suppression} of magnetic energy at intermediate scales, $2\leq
k\leq8$, is quite essential for the development of a well-defined large
scale field. In a recent letter Brandenburg \& Subramanian (2000) showed
that this type of {\it self-cleaning} effect can also be simulated by
using ambipolar diffusion as nonlinearity and ignoring the Lorentz
force altogether. Without any nonlinearity, however, there would be
no interaction between different scales and the magnetic energy
would increase at all scales, especially at small scales, which would
soon swamp the large scale field structure with small scale fields.

The model presented in \Fig{FpMkt1+2} has large scale separation in the
sense that there is a large gap between the forcing wavenumber ($k=k_{\rm
f}=30$) and the wavenumber of the box ($k=k_1=1$). One sees that during
the growth phase there is a clear secondary maximum at $k=7$. This is
indeed expected for an $\alpha^2$ dynamo, whose maximum growth rate is
at $k_{\max}=\half\alpha/\eta_{\rm T}$, where $\eta_{\rm T}$ is the total
(turbulent plus microscopic) magnetic diffusion coefficient.

\begin{figure}[h!]\begin{center}\includegraphics[width=.90\textwidth]{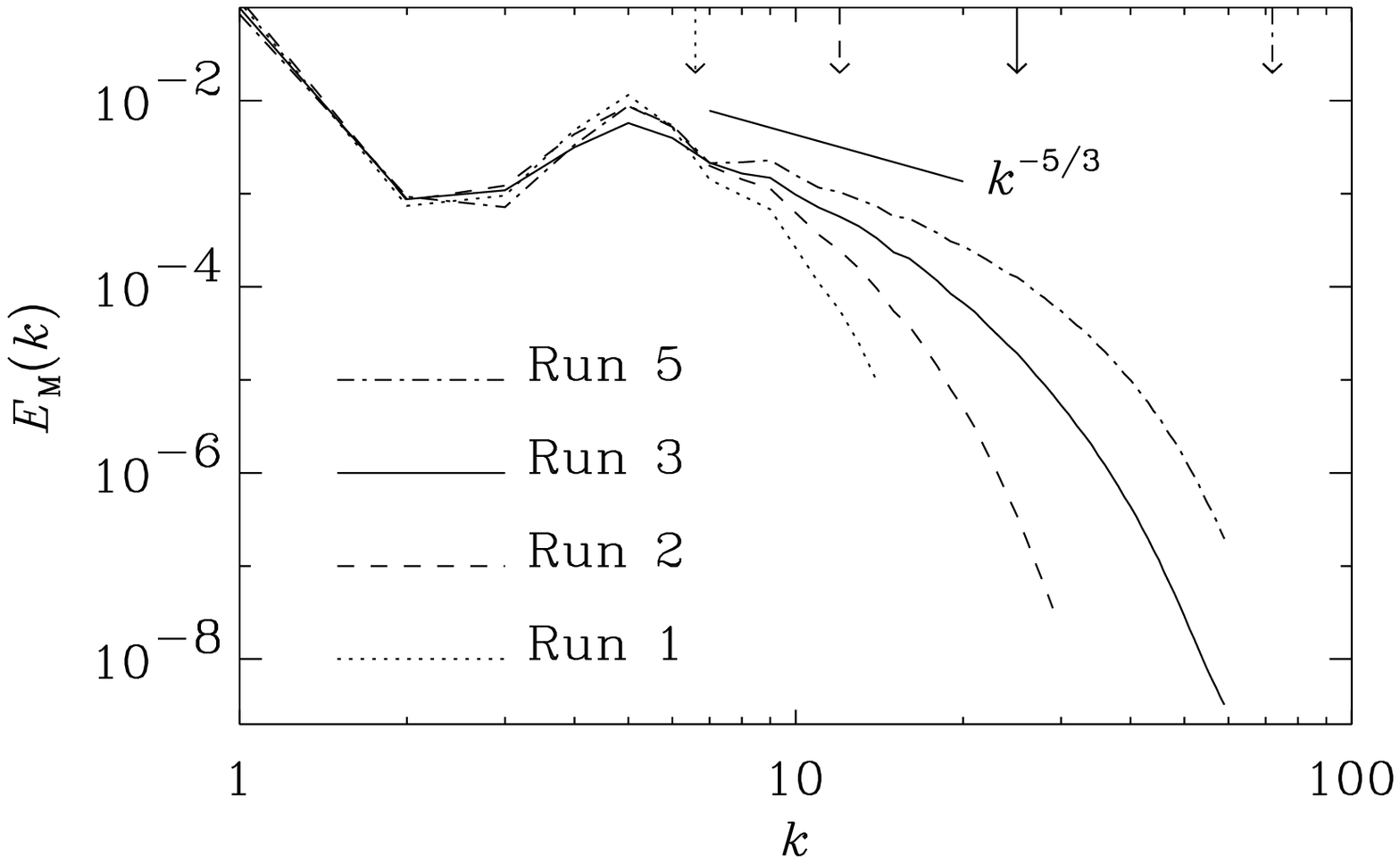}\end{center}\caption[]{
Comparison of time averaged magnetic energy spectra for Runs~1--3
($t=600-1000$) with a non-averaged spectrum for Run~5 (large magnetic
Prandtl number) taken at
$t=1600$. To compensate for different field strengths and to make the
spectra overlap at large scales, two of the three spectra have been
multiplied by a scaling factor. There are clear signs of the gradual
development of an inertial subrange for wavenumbers larger than the
forcing scale. The $k^{-5/3}$ slope is shown for orientation. The
dissipative magnetic cutoff wavenumbers, $\bra{\JJ^2/\eta^2}^{1/4}$,
are indicated by arrows at the top.
}\label{Fpspec_conv+hipr}\end{figure}

\begin{figure}[h!]\begin{center}\includegraphics[width=.90\textwidth]{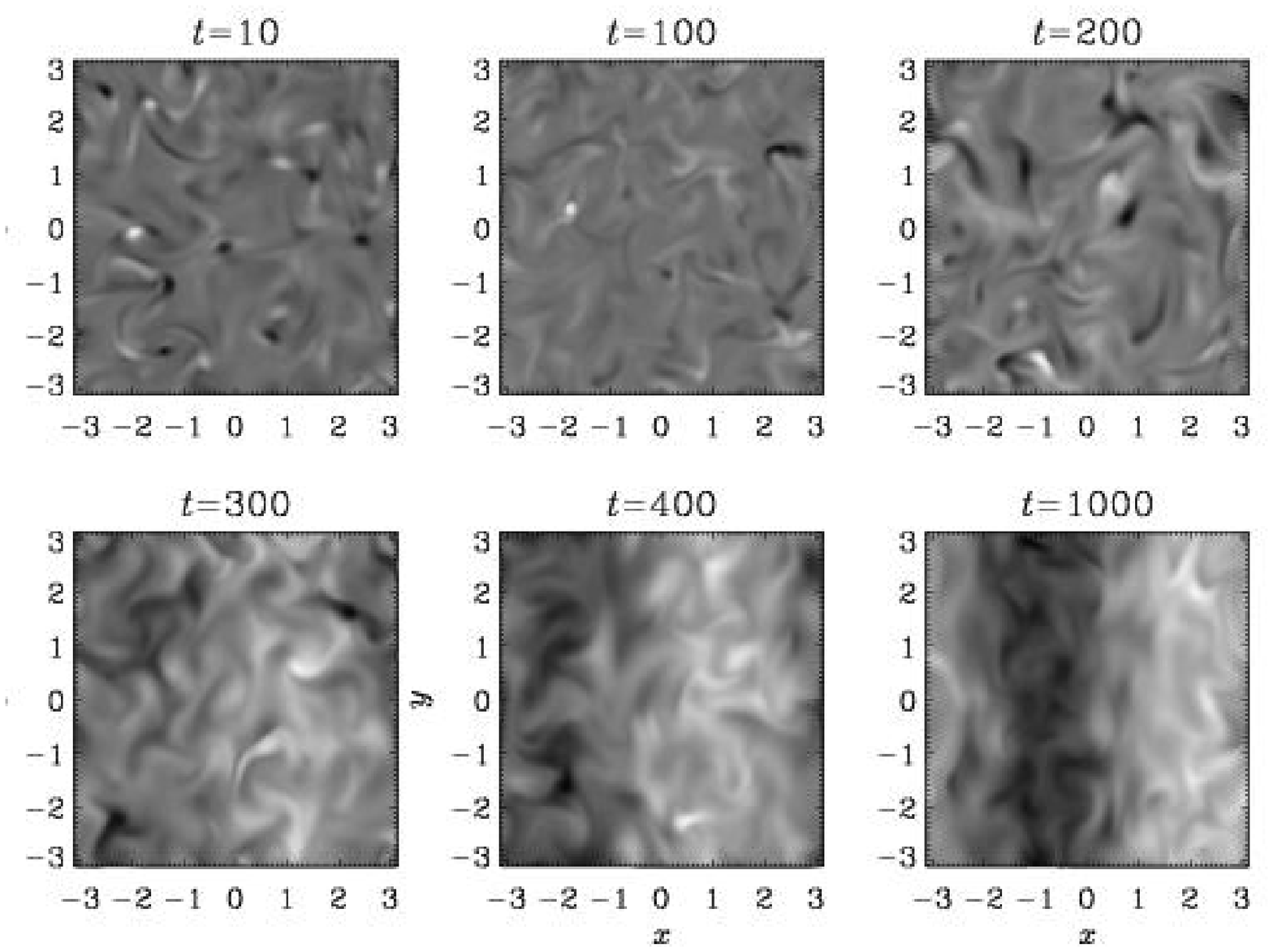}\end{center}\caption[]{
Gray-scale images of cross-sections of $B_x(x,y,0)$ for Run~3 of
Brandenburg 2001b at different times showing the gradual build-up of
the large scale magnetic field after $t=300$. Dark (light) corresponds
to negative (positive) values. Each image is scaled with respect to its
min and max values.
}\label{Fpimages_hor}\end{figure}

The disadvantage of a high forcing wavenumber is that for modest
resolution (here we used $120^3$ meshpoints) no inertial range can
develop. This is different if once forces at $k_{\rm f}=5$, keeping
otherwise the same resolution. In \Fig{Fpspec_conv+hipr} we show spectra
for different cases with $k_{\rm f}=5$ where we compare the results for
different values of the magnetic Reynolds and magnetic Prandtl number. In
\Fig{Fpimages_hor} we show cross-sections of one field component at
different times. In this model (Run~3 of Brandenburg 2001b) the forcing
is at $k_{\rm f}=5$, so there is now a clear tendency for the build-up
of an inertial range in $8\leq k\leq25$.

\subsection{The inverse cascade in decaying turbulence}

We now turn to the case of {\it decaying} turbulence, which is driven
only by an initial kick to the system. There are several circumstances
in astrophysics where this could be relevant: early universe, neutron
stars, and mergers of galaxy clusters. In all those cases one is
interested in the development of large scale fields. In the context of
the early universe the possibility of energy conversion from small to
large scale fields was pointed out by Brandenburg, Enqvist, \& Olesen
(1996) who found that fields generated at the horizon scale of
$3\cm$ after the electroweak phase transition would now have a scale on
the order of kiloparsecs, even though the cosmological expansion alone
would only lead to scales on the order of $1\AU$. These results were
only based on either two-dimensional simulations or three-dimensional
cascade model calculation (e.g.\ Biskamp 1994). Therefore we now turn
to fully three-dimensional simulations.

\begin{figure}[h!]\begin{center}\includegraphics[width=.90\textwidth]{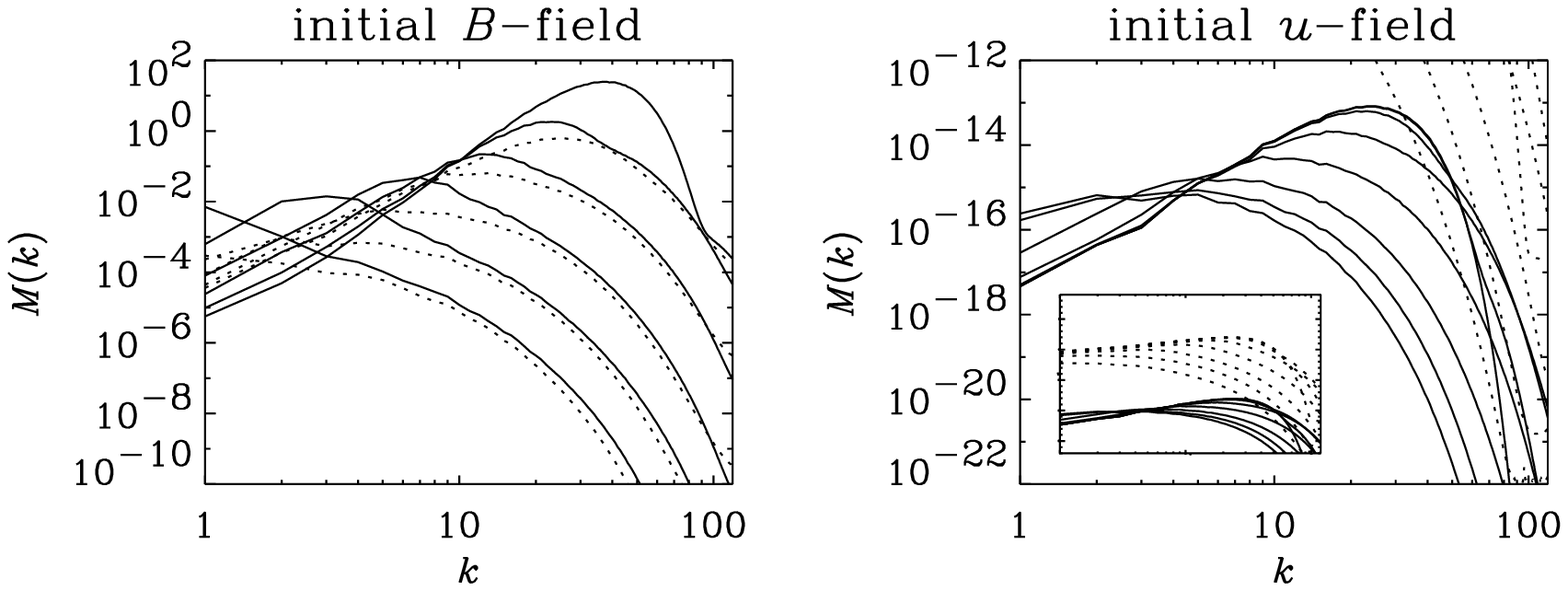}\end{center}\caption[]{
Power spectra of magnetic energy (solid lines) and kinetic energy (dotted
lines) in a decay run with helicity. The left hand panel is for a case
where the flow is only driven by an initial helical magnetic field.
In the right hand panel the field is weak and governed by strong decaying
fluid turbulence. The inset shows both velocity and magnetic spectra
in the same plot. The Prandtl number $\nu/\eta$ is equal to one in both
cases, but the mesh Reynolds number, which is kept constant at all times,
is different in the two cases: 20 in the left hand panel and 50 in the
right hand panel. The times are 0, 0.01, 0.1, etc, till $t=10^2$ in the
left hand panel and $t=10^3$ in the right hand panel.
}\label{Fpower_U1_Run1}\end{figure}

In the absence of any forcing and with no kinetic energy initially an
initial magnetic field can only decay. However, if initially most of the
magnetic energy is in the small scales, there is the possibility that
magnetic helicity and thereby also magnetic energy is transferred to large
scales. This is exactly what happens (\Fig{Fpower_U1_Run1}), provided
there is initially some net helicity. The inset of \Fig{Fpower_U1_Run1}
shows that in the absence of initial net helicity the field at large
scales remains unchanged, until diffusion kicks in and destroys the
remaining field at very late times.

If the magnetic field has the possibility to tap energy also from the
large scale velocity the situation is somewhat different again and there
is the possibility that a large scale magnetic field can also be driven
without net helicity. In that case the large scale field can increase due
to dynamo action from the incoherent $\alpha-\Omega$-effect (Vishniac \&
Brandenburg 1997). In astrophysical settings there is usually large scale
shear from which energy can be tapped. Before we discuss simulations
with imposed shear in more detail we first present a simple argument
that makes the link between the inverse cascade and helicity conservation.

\subsection{The connection with magnetic helicity conservation}

In the following we give a simple argument due to Frisch \ea (1975)
that helps to understand why the magnetic helicity conservation property
leads to the occurrence of an inverse cascade. We define in the following
magnetic energy and helicity spectra, $M(k)$ and $H(k)$, respectively.
Now, because of Schwartz inequality, we have
\EQ
|\BBBB(k)|^2=|(i\kk\times\AAAA)\cdot\BBBB|\geq|\kk||\AAAA\cdot\BBBB|
\EN
we have a lower bound on the spectral magnetic energy at each wavenumber
$k=\kk$. In terms of shell integrated magnetic energy and helicity spectra
this corresponds to
\EQ
M(k)\geq \half k|H(k)|,
\EN
where the 1/2-factor comes simply from the 1/2-factor in the definition
of the magnetic energy. Assuming that two wave numbers $q$ and $p$
interact such that they produce power at a new wave number $k$, then
\EQ
M(p)+M(q)=M(k),\quad H(p)+H(q)=H(k).
\EN
For simplicity we consider the case $p=q$, so
\EQ
2M(p)=M(k),\quad 2H(p)=H(k).
\EN
Assume also that initially the constraint was sharp (maximum helicity),
then
\EQ
M(p)=\half pH(p).
\EN
Now, from the constrain again we have
\EQ
\half kH(k)\leq M(k)=2M(p)=pH(p)=\half pH(k),
\EN
so
\EQ
k\leq p,
\EN
that is the wave number of the target result must be larger or equal to
the wave numbers of the initial field.

The argument given above is of course quite rough, because it ignores
for example the detailed angular dependence of the wave vectors. This
was taken into account properly already in the early paper by Pouquet,
Frisch, \& L\'eorat (1976), but this approach was based on closure
assumptions for the higher moments, which is in principle open to
criticism. Thus, numerical simulations, like those presented above,
are necessary for an independent confirmation that the inverse cascade
really works. In this connection one should mention that there are
some parallels with the inverse cascade of enstrophy in two-dimensional
hydrodynamic (nonmagnetic) turbulence. In that case the enstrophy (i.e.\
the mean squared vorticity) is conserved because of the absence of vortex
stretching in two dimensions. The inverse hydrodynamic cascade has some
significance in meteorology and perhaps in low aspect ratio convection
experiments, where one finds a peculiar energy and entropy spectrum that
is referred to as Bolgiano scaling; see Brandenburg (1992) and Suzuki \&
Toh (1995) for corresponding shell model calculations and Toh \& Iima
(2000) for direct simulations.

\subsection{Inverse cascade or $\alpha$-effect?}

In \Sec{Sisotrop} we made a distinction between inverse cascade and
$\alpha$-effect in the sense that, although both lead to large
scale field generation, in the inverse cascade there is a gradual
transfer of magnetic helicity and energy to ever larger scales,
whereas the $\alpha$-effect produces large scale magnetic fields
directly from small scale fields. Thus, the distinction is really
one between local and nonlocal inverse cascades.

\begin{figure}[h!]\begin{center}\includegraphics[width=.90\textwidth]{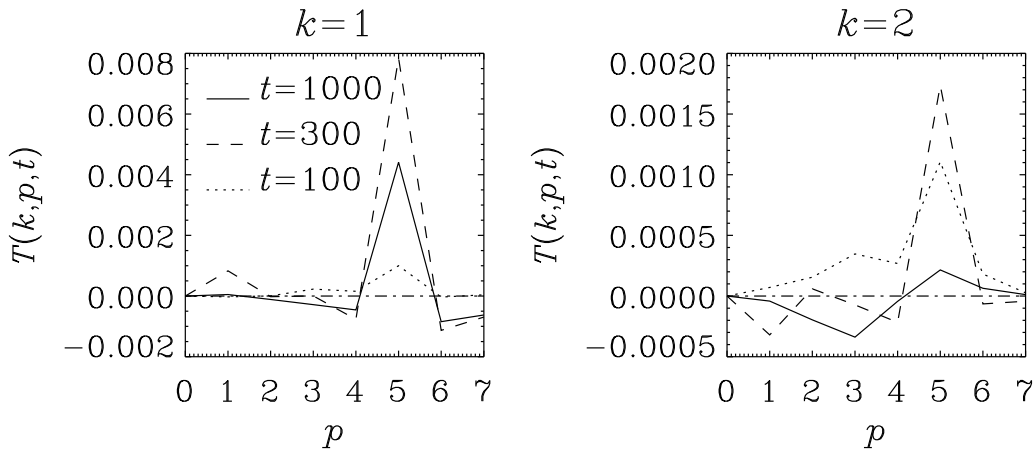}\end{center}\caption[]{
Spectral energy transfer function $T(k,p,t)$, normalized by $\bra{\BB^2}$
for three different times, for $k=1$ and 2. Run~3 of Brandenburg (2001b).
}\label{FpTM_spec_hor}\end{figure}

In \Fig{FpTM_spec_hor} we show the normalized spectral energy transfer
function $T(k,p,t)$ for $k=1$ and 2 as a function of $p$, and at
different times $t$. The index $k$ signifies the gain or losses of the
field at wavenumber $k$, and the index $p$ indicates the wavenumber
of the velocity from which the energy comes from. This function shows
that most of the energy of the large scale field at $k=1$ comes from
velocity and magnetic field fluctuations at the forcing scale, which is
here $k=k_{\rm f}=5$. At early times this is also true of the energy
of the magnetic field at $k=2$, but at late times, $t=1000$, the gain
from the forcing scale, $k=5$, has diminished, and instead there is now
a net loss of energy into the next larger scale, $k=3$, suggestive of
a direct cascade operating at $k=2$, and similarly at $k=3$.

Based on these results we may conclude that in the saturated state the
magnetic energy at $k=1$ is sustained by a {\it nonlocal} inverse cascade
from the forcing scale directly to the largest scale of the box. This is
characteristic of the $\alpha$-effect of mean-field electrodynamics, except
that here nonlinearity plays an essential role in isolating the large scale
from the small scale `magnetic trash', as Parker used to say.

A closer look at \Fig{FpMkt1+2}, where $k=k_{\rm f}=30$, suggests
that once the scale separation is large enough the energy is at first
transferred not to the scale of the box, but instead to a somewhat smaller
scale (here at wavenumber $k=7$).  Following the corresponding discussion
in Brandenburg (2001b), this wavenumber is close to the wavenumber,
$k_{\max}=\half|\alpha|/\eta_{\rm T}$, where the $\alpha^2$ dynamo
grows fastest.

In the following section we address the issue of magnetic helicity
conservation which has important consequences for the timescale after
which the large scale field begins to develop. This has also a bearing
on the widely discussed controversy of the so-called `catastrophic
$\alpha$-quenching' of Vainshtein \& Cattaneo (1992).

\subsection{Approximate helicity conservation}

The magnetic helicity, $H=\bra{\AAA\cdot\BB}$, is conserved by
the nonresistive MHD equations. For a closed or periodic box
$\bra{\AAA\cdot\BB}$ satisfies the equation
\EQ
{\dd\over\dd t}\bra{\AAA\cdot\BB}=-2\eta\mu_0\bra{\JJ\cdot\BB},
\label{magheldef}
\EN
where $\bra{\JJ\cdot\BB}$ is the current helicity, and angular
brackets denote volume averages. Note that
for a periodic box $\bra{\AAA\cdot\BB}$ is gauge invariant,
i.e.\ $\bra{\AAA\cdot\BB}$ does not change after a gauge
transformation, $\AAA\rightarrow\AAA+\nab\varphi$. This is a direct
consequence of the solenoidality of the magnetic field, because
$\bra{\nab\varphi\cdot\BB}=-\bra{\varphi\nab\cdot\BB}=0$ owing to
$\nab\cdot\BB=0$.

\begin{figure}[h!]\begin{center}\includegraphics[width=.90\textwidth]{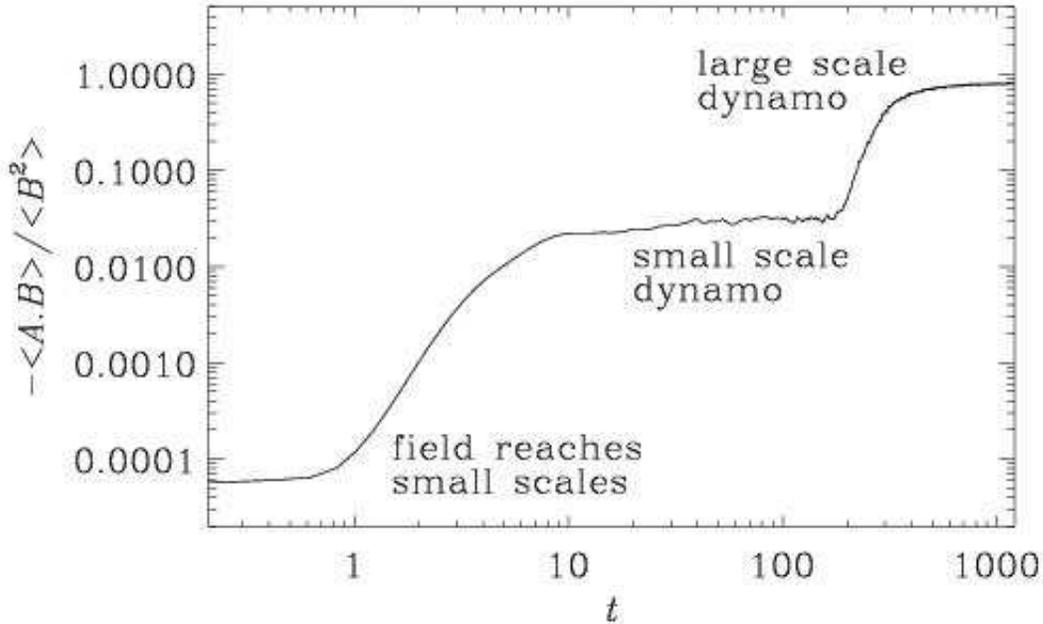}\end{center}\caption[]{
Evolution of the (negative) magnetic helicity length scale in a
double-logarithmic plot. Note the presence of three distinct phases:
very approximate helicity conservation near zero, followed by a phase
of larger magnetic helicity scale (three orders of magnitude), and
finally a phase where the magnetic helicity scale has reached the
scale of the box.
}\label{Fpabm}\end{figure}

In order to judge whether $\bra{\AAA\cdot\BB}$ is small or large we
calculate the length scale
\EQ
\ell_{\rm H}=|\bra{\AAA\cdot\BB}|/\bra{\BB^2}.
\EN
In \Fig{Fpabm} we see that the evolution of $\ell_{\rm H}$ proceeds in
three distinct phases: (i) a very short period ($t<1$) where $\ell_{\rm
H}$ is very small and comparable to the numerical noise level, so
magnetic helicity almost perfectly conserved,
(ii) an intermediate interval ($2<t<200$) where $\ell_{\rm H}$ is much
larger, but still only roughly equal to the mesh size of the calculation,
and then (iii) a regime where $\ell_{\rm H}$ is of order unity. The latter
is only possible because of the presence of helicity in the system, which
leads to a large scale magnetic field configuration that is nearly force-free.

\subsection{Resistively limited growth of the large scale field}

The approximate conservation of magnetic helicity has an important
consequence for the generation of large scale fields: in order to build
up a large scale field with magnetic helicity one has to {\it change}
the value of $\bra{\AAA\cdot\BB}$ from its initial value of zero to
its certain final value. This final value of $\bra{\AAA\cdot\BB}$ is
such that the length scale $\ell_{\rm H}$ is close to the maximum value
possible for a certain geometry. It also implies that when interpreting
the results in terms of mean-field electrodynamics ($\alpha$-effect and
turbulent diffusion), the $\alpha$-effect must be quenched early on, well
before the field has reached its final value. Nevertheless, this strong
quenching, which was first anticipated by Vainshtein \& Cattaneo (1992)
and later confirmed numerically by Cattaneo \& Hughes (1996), does not
prevent the field from reaching its final super-equipartition field strength.

Before we come to the details we mention already now that the helicity
constraint is probably too severe to be acceptable for astrophysical
conditions, so one must look for possible escape routes. The most
plausible way of relaxing the helicity constraint is in allowing for
open boundary conditions (Blackman \& Field 2000, Kleeorin \ea 2000),
but the situation can still be regarded as inconclusive. Given that much
of the work on large scale dynamos so far assumes periodic boundaries,
we shall now consider this particular case in more detail. In the
periodic case the final field geometry can be, for example, of the form
$\meanBB=B_0(\cos k_1z,\sin k_1z,0)$, where $k_1=1$ is the smallest
possible wavenumber in the box, $B_0$ is the field amplitude, and
$\bra{\meanBB^2}=B_0^2$. Alternatively, the field may vary in the $x$
or $y$ direction, and there may be an arbitrary phase shift; examples of
these possibilities have been reported in Brandenburg (2001b). Anyway,
for $\meanBB=B_0(\cos k_1z,\sin k_1z,0)$ the corresponding vector
potential is $\meanAA=-(B_0/k_1)(\cos k_1z,\sin k_1z,0)+\nab\phi$,
where $\phi$ is an arbitrary gauge which does not affect the value of
$\bra{\meanAA\cdot\meanBB}$. In this example we have
\EQ
\bra{\meanAA\cdot\meanBB}=-\bra{\meanBB^2}/k_1,
\label{meanAB}
\EN
where we have included the $k_1$ factor, even though in the present
case $k_1$=1. (The minus sign in \Eq{meanAB} would turn into a plus if
the forcing had negative helicity.) The mean current density is given
by $\meanJJ=-B_0k_1(\cos k_1z,\sin k_1z,0)$, so the current helicity of
the mean field is given by
\EQ
\mu_0\bra{\meanJJ\cdot\meanBB}=-k_1\bra{\meanBB^2}.
\label{meanJB}
\EN
Before we can use \Eqs{meanAB}{meanJB} in \Eq{magheldef} we need to
relate the magnetic and current helicities of the mean field to those
of the actual field. We can generally split up the two helicities into
contributions from large and small scales, i.e.\
\EQ
\bra{\AAA\cdot\BB}=\bra{\meanAA\cdot\meanBB}+\bra{\aaa\cdot\bb},
\label{ABsplit}
\EN
\EQ
\bra{\JJ\cdot\BB}=\bra{\meanJJ\cdot\meanBB}+\bra{\jj\cdot\bb}.
\label{JBsplit}
\EN
As the large scale magnetic field begins to saturate, the magnetic
helicity has to become constant and so \Eq{magheldef} dictates that
$\bra{\JJ\cdot\BB}$ must go to zero in the steady state. Consequently,
the contribution from $\bra{\jj\cdot\bb}$ must be as large as that
of $\bra{\meanJJ\cdot\meanBB}$, and of opposite sign, so that the two
cancel. This, together with \Eq{meanJB}, allows us immediately to write
down an expression for the equilibrium strength of the mean field;
\EQ
\bra{\meanBB^2}=\mu_0|\bra{\jj\cdot\bb}|/k_1,
\label{Bestimate1}
\EN
which is now valid for both signs of the helicity of the forcing. The
{\it residual} helicity (Pouquet, Frisch, L\'eorat 1976),
\EQ
H_{\rm res}=\bra{\oo\cdot\uu}-\bra{\jj\cdot\bb}/\rho_0,
\EN
is small in the nonlinear saturated state and nearly vanishing.
[We mention that this is
also the case in the models of Brandenburg \& Subramanian
(2000).] Furthermore, for forced turbulence with a well defined
forcing wavenumber the kinetic helicity may be estimated as
$\bra{\oo\cdot\uu}\approx k_{\rm f}\bra{\uu^2}$.  Together with
\Eq{Bestimate1} we have
\EQ
\bra{\meanBB^2}={k_f\over k_{\rm 1}}\,B_{\rm eq}^2,
\label{Bestimate2}
\EN
where $B_{\rm eq}^2=\mu_0\bra{\rho\uu^2}$ and $B_{\rm eq}$ is the
equipartition field strength. Thus, the mean field can exceed (!) the
equipartition field by the factor $(k_f/k_{\rm 1})^{1/2}$. This estimate
agrees well with the results of the simulations; see Brandenburg (2001b).

Using \Eqs{ABsplit}{JBsplit} together with \Eqs{meanAB}{meanJB}
we can rewrite \Eq{magheldef} in the form
\EQ
{\dd\over\dd t}\bra{\meanBB^2}=-2\eta k_1^2\bra{\meanBB^2}
+2\eta k_1\mu_0|\bra{\jj\cdot\bb}|,
\label{meanBBevol}
\EN
where we have taken into account the contribution of the small
scale current helicity which is of similar magnitude as the large
scale current helicity. For the magnetic helicity, on the other hand,
the small scale contribution is negligible, because
\EQ
|\bra{\aaa\cdot\bb}|
\approx\mu_0|\bra{\jj\cdot\bb}|/k_{\rm f}^2
\approx\mu_0|\bra{\meanJJ\cdot\meanBB}|/k_{\rm f}^2
\approx|\bra{\meanAA\cdot\meanBB}|\left({k_1\over k_{\rm f}}\right)^2
\ll|\bra{\meanAA\cdot\meanBB}|.
\EN
After the saturation at small and intermediate scales the small scale
current helicity is approximately constant and can be estimated as
\EQ
|\bra{\jj\cdot\bb}|\approx\rho_0|\bra{\oo\cdot\uu}|
\approx k_{\rm f}\bra{\rho\uu^2}=k_{\rm f}B_{\rm eq}^2/\mu_0.
\EN
The solution of \eq{meanBBevol} is given by
\EQ
\bra{\meanBB^2}
=\epsilon_0 B_{\rm eq}^2\left[1-e^{-2\eta k_1^2(t-t_{\rm sat})}\right],
\label{approxB}
\EN
where $\epsilon_0$ is a coefficient which, in the present model
with a well-defined forcing wavenumber, can be approximated by
$\epsilon_0\approx k_{\rm f}/k_1$.

\begin{figure}[h!]\begin{center}\includegraphics[width=.90\textwidth]{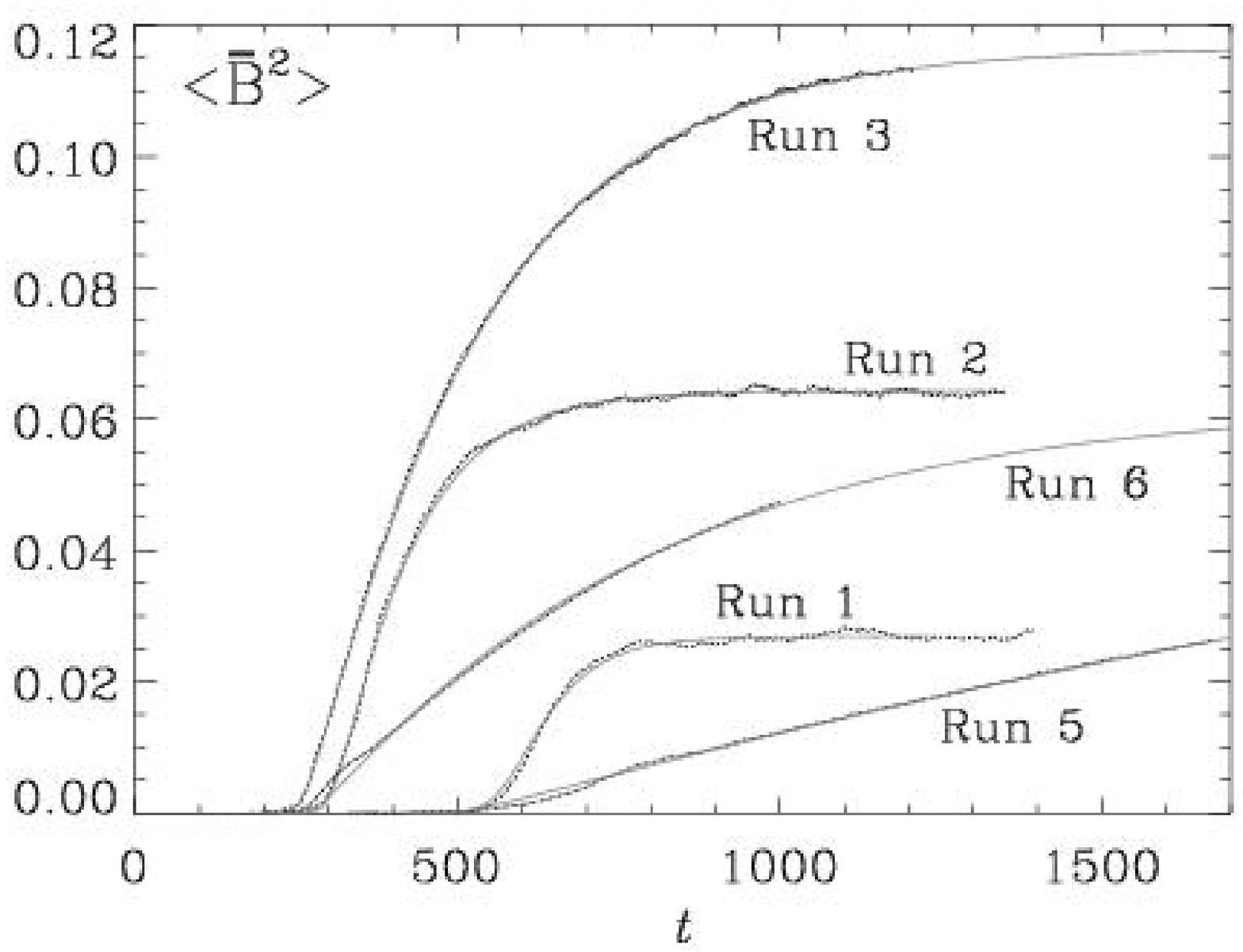}\end{center}\caption[]{
Evolution of $\bra{\meanBB^2}$ for Runs~1--3, 5, and 6, compared
with the solution \eq{approx_quench} of the dynamo equations using
\Eq{approx_quench}.
}\label{Fpjbm_decay_nfit}\end{figure}

This is indeed also the limiting behavior found for $\alpha^2$-dynamos
with simultaneous $\alpha$ and $\eta$ quenching of the form
\EQ
\alpha={\alpha_0\over1+\alpha_B\meanBB^2\!/B_{\rm eq}^2},\quad
\eta_{\rm t}={\eta_{\rm t0}\over1+\eta_B\meanBB^2\!/B_{\rm eq}^2},
\label{quench_both}
\EN
where $\alpha_B=\eta_B$ is assumed. Assuming that the magnetic energy
density of the mean field, $\meanBB^2$, is approximately uniform
(which is well satisfied in the simulations) we can obtain the solution
$\meanBB=\meanBB(t)$ of \Eq{dyneq} in the form
\EQ
\meanBB^2/(1-\meanBB^2\!/B_{\rm fin}^2)^{1+\lambda/\eta k_1^2}
=B_{\rm ini}^2\,e^{2\lambda t},
\label{approx_quench}
\EN
where
\EQ
\lambda=|\alpha_0|k_1-\eta_{\rm T0}k_1^2
\EN
is the kinematic growth rate of the dynamo, $B_{\rm ini}$ is the initial
field strength, and $B_{\rm fin}$ is the final field strength of the
large scale field, which is related to $\alpha_B$ and $\eta_B$ via
\EQ
\alpha_B=\eta_B={\lambda\over\eta k_1^2}
\left({B_{\rm eq}\over B_{\rm fin}}\right)^2.
\label{quench_formula}
\EN
The full derivation is given in \App{Saquench}. The significance
of this result is that it provides an excellent fit to the numerical
simulations; see \Fig{Fpjbm_decay_nfit} where we present the evolution of
$\bra{\meanBB^2}$ for the different runs of Brandenburg (2001b). Equation
\eq{approx_quench} can therefore be used to extrapolate to astrophysical
conditions. The time it takes to convert the small scale field generated
by the small scale dynamo to a large scale field, $\tau_{\rm eq}$,
increases linear with the magnetic Reynolds number, $R_{\rm m}$. Apart
from some coefficients of order unity the ratio of $\tau_{\rm eq}$ to the
turnover time is therefore just $R_{\rm m}$. For the sun this ratio would
be $10^8-10^{10}$. However, before interpreting this result further one
really has to know whether or not the presence of open boundary conditions
could alleviate the issue of very long timescales for the mean magnetic
field. Furthermore, it is not clear whether the long timescales discussed
above have any bearing on the cycle period in the case of oscillatory
solutions. The reason this is not so clear is because for a cyclic
dynamo the magnetic helicity in each hemisphere stays always of the same
sign and is only slightly modulated. It is likely that this modulation
pattern is advected precisely with the meridional circulation, in which
case the helicity could be nearly perfectly conserved in a lagrangian
frame. This could provide an interesting clue for why the solar dynamo
is migrating. The relation between meridional circulation and dynamo
wave propagation has been advocated by Durney (1995) and Choudhuri,
Sch\"ussler, \& Dikpati (1995), but helicity conservation would strongly
lock the two aspects.

It is clear that virtually all astrophysical bodies are open, allowing for
constant loss of magnetic helicity. In the case of the sun significant
amounts of magnetic helicity are indeed observed at the solar surface
(Berger \& Ruzmaikin 2000). Significant losses of magnetic helicity are
particularly obvious in the case of accretion discs which are almost
always accompanied by strong outflows that can sometimes be collimated
into jets. Thus, dynamo action from accretion disc turbulence would be
a good candidate for clarifying the significance of open boundaries on
the nature of the dynamo. Another reason why accretion disc turbulence
is a fruitful topic for understanding dynamo action is because the shear
is extremely strong. In the following we discuss some recent progress
that has been made in this field.

\subsection{Joule dissipation from mean and fluctuating fields}

\begin{figure}[h!]\begin{center}\includegraphics[width=.90\textwidth]{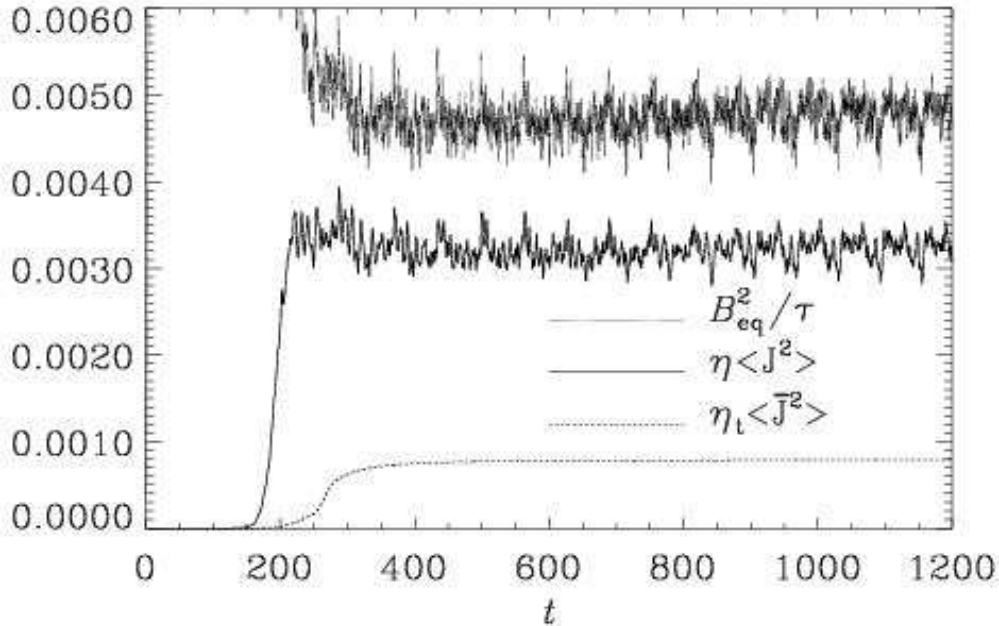}\end{center}\caption[]{
Joule dissipation for Run~3 (solid line), compared with the Joule
dissipation estimated for a corresponding mean-field model (dashed
line). An estimate for the rate of total energy dissipation, $B_{\rm 
eq}^2/\tau$, is also given.
}\label{Fpjoule}\end{figure}

In an MHD flow the mean magnetic Joule dissipation per unit volume
is given by
\EQ
Q_{\rm Joule}=\eta\mu_0\bra{\JJ^2}.
\EN
Whilst in may astrophysical flows $\eta$ may be very small, $|\JJ|$
can be large so that $Q_{\rm Joule}$ remains finite even in the limit
$\eta\rightarrow0$. One example where this is very important is accretion
discs, where Joule dissipation (together with viscous dissipation) are
important in heating the disc. These viscous and resistive processes are
indeed the only significant sources of energy supply in discs, and yet
the luminosities of discs that result from the conversion of magnetic 
and kinetic energies into heat and radiation can be enormous. Much of the
work on discs involves mean-field modelling, so it would be interesting
to see how the Joule dissipation, $Q_{\rm Joule}^{\rm(mf)}$, predicted
from a mean-field model,
\EQ
Q_{\rm Joule}^{\rm(mf)}=\eta_{\rm t}\mu_0\bra{\meanJJ^2},
\EN
relates to the actual Joule dissipation. In \Fig{Fpjoule} we show the
evolution of actual and mean-field Joule dissipation and compare with an
estimate for the rate of total energy dissipation, $B_{\rm eq}^2/\tau$,
where $\tau$ is the turnover time. Here we have taken into account 
that $\eta_{\rm t}$ is `catastrophically' quenched using the formulae
of Brandenburg (2001) with the parameters for Run~3.

There is no reason a priori that the magnetic energy dissipations from
the mean-field model should agree with the actual one. It turns out
that the mean-field dissipation is a fourth of the actual one, so it is
definitely significant. It would therefore be interesting so see
how those two dissipations compare with each other in other models.

\subsection{Possible pitfalls in connection with hyperresistivity}

In many astrophysical applications hyperresistivity and hyperviscosity
are sometimes used in order to concentrate the effects of magnetic
diffusion and viscosity to the smallest possible scale. The purpose of
this section is to highlight possible spurious artifacts associated with
this procedure. As we have seen above, large scale dynamos can depend
on the microscopic magnetic diffusivity and must therefore be affected
when it is replaced by hyperresistivity. The resulting modifications
that are to be expected are easily understood: on the right hand side of
\Eq{magheldef} the term $\bra{\JJ\cdot\BB}$ needs to be replaced by
$\bra{(\nabla^4\AAA)\cdot\BB}$. This leads to a change of the relative
importance of small and large scale contributions, which therefore
changes \Eq{Bestimate2} to
\EQ
\bra{\meanBB^2}=\left({k_f\over k_{\rm 1}}\right)^3\,B_{\rm eq}^2.
\EN
Thus, the final field strength will be even larger than before: instead
of a factor of 5 superequipartition (for $k_{\rm f}=5$) one now expects
a factor of 125. Recent simulations by Brandenburg \& Sarson (2001)
have indeed confirmed this tendency. The main conclusion is that
hyperresistivity can therefore be used to address certain issues
regarding large magnetic Reynolds numbers that are otherwise still
inaccessible. On the other hand, the results are in some ways
distorted and need therefore be interpreted carefully.

\subsection{Remarks on accretion disc turbulence}
\label{Sdiscs}

We have already mentioned the possibility of dynamo action in accretion
discs. Accretion discs have been postulated some 30 years ago in order to
explain the incredibly high luminosities of quasars. Only in the past few
years has direct imaging of accretion discs become possible, mostly due
to the Hubble Space Telescope. Accretion discs form in virtually all
collapse processes, such as galaxy and star formation. In the latter
case the central mass is of the order of one solar mass, while in the
former it is around $10^8$ solar masses and is concentrated in such a
small volume that that
it must be a black hole. If the surrounding matter was nonrotating, it
would fall radially towards the center. But this is unrealistic and even
the slightest rotation relative to the central object would become
important eventually as matter falls closer to the center.

\begin{figure}[h!]\begin{center}\includegraphics[width=.90\textwidth]{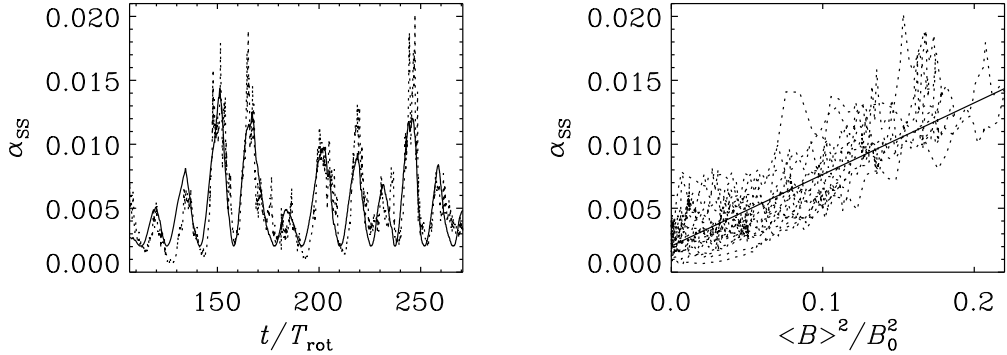}\end{center}\caption[]{
Dependence of $\alpha_{\rm SS}$ on time and the mean magnetic field
strength for a local accretion disc model
(Run~B of Brandenburg \ea 1996a). Here $B_0=\bra{\mu_0\rho
c_{\rm s}^2}^{1/2}$ is the thermal equipartition field strength and
$T_{\rm rot}=2\pi/\Omega$ the local rotation period.  In the left hand
panel the dotted line represents the actual data and the solid line
gives the fit obtained by correlating $\alpha_{\rm SS}$ with the mean
magnetic field (right hand panel).
}\label{Fvisc}\end{figure}

\begin{figure}[h!]\begin{center}\includegraphics[width=.90\textwidth]{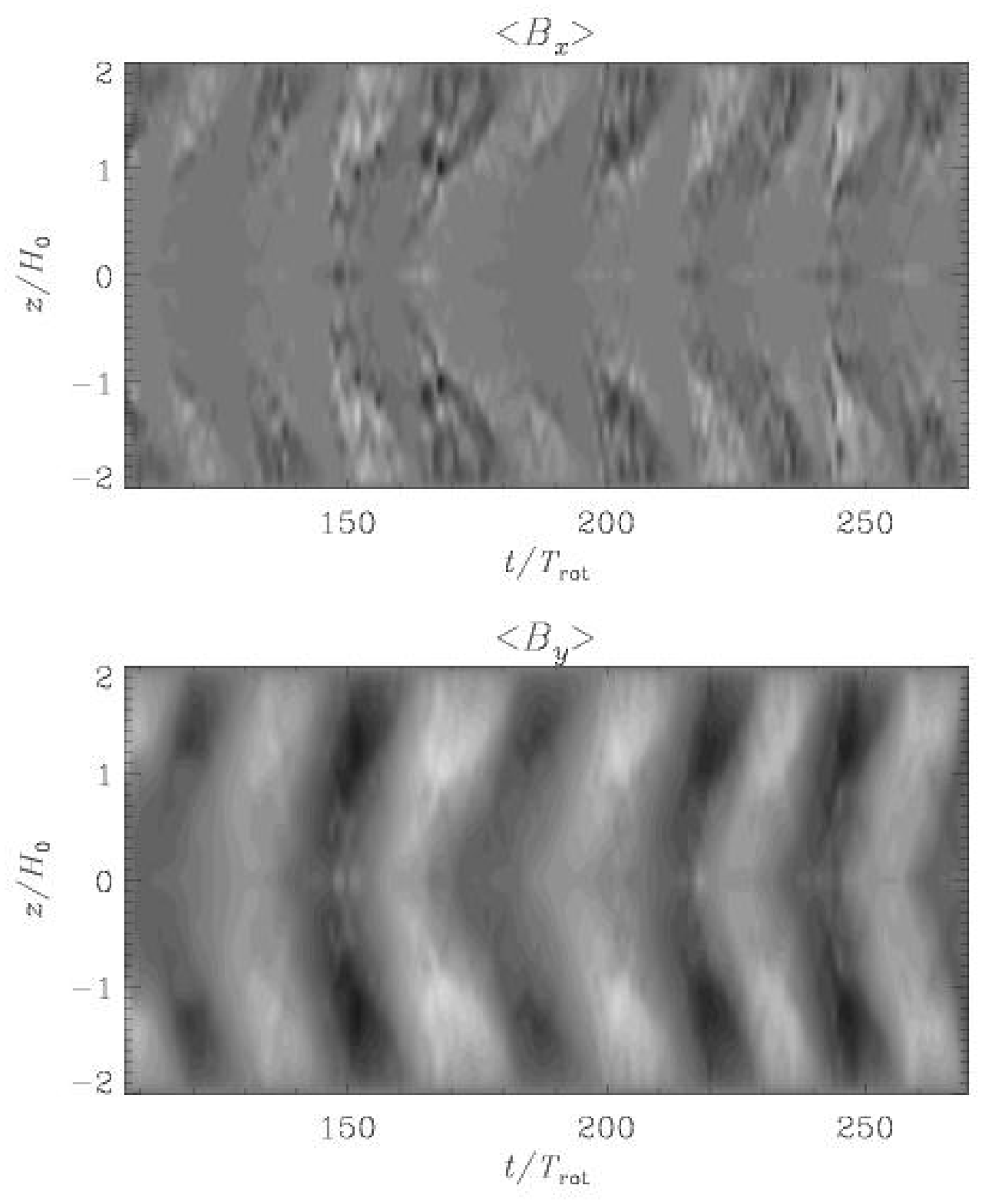}\end{center}\caption[]{
Butterfly (space-time) diagram of the poloidal and toroidal
magnetic field components averaged over the two horizontal ($x$ and $y$)
directions for the local accretion disc model of (Brandenburg \ea 1996a,
Run~B). Note that the poloidal field is much more noisy than
the toroidal field, and that there is a clear outward migration
of magnetic field.
}\label{Fpbutter}\end{figure}

If there was no effective diffusive process in discs, the angular momentum
of the matter would stay with the gas parcels, and since the gravity force
is balanced by the corresponding centrifugal force, the gas would never
accrete. However, the angular velocity of the gas follows a $r^{-3/2}$
Kepler law, so the gas is differentially rotating and one may expect shear
instabilities to occur that would drive turbulence and hence turbulent
dissipation. Unfortunately, however, the story is not so simple. Discs
are both linearly stable (Stewart 1975) and probably also nonlinearly
stable (Hawley, Gammie, \& Balbus 1996). Nevertheless, in the presence
of a magnetic field there is a powerful {\it linear} instability (Balbus
\& Hawley 1991), and subsequent work has shown that this instability is
indeed capable of driving the instability and hence turbulence.

One of the key outcomes of such simulations is the rate of turbulent
dissipation, which determines the rate of angular momentum transport and
correspondingly the rate at which orbital kinetic and potential energy
is liberated in the form of heat. This is normally expressed in terms
of a turbulent viscosity (e.g.\ Frank, King, \& Raine 1992), but it may
equally well be expressed in terms of the horizontal components of the
Reynolds and Maxwell stress tensors. The stress may then be normalized
by $\bra{\rho c_{\rm s}^2}$ to give a nondimensional measure (called
$\alpha_{\rm SS}$) of the ability of the turbulence to transport angular
momentum outward (if $\alpha_{\rm SS}>0$). This $\alpha_{\rm SS}$ is
indeed always positive, see \Fig{Fvisc}, but it fluctuates significantly
about a certain mean value. These fluctuations are in fact correlated
with the energy in the mean magnetic field, $\bra{\BB}^2$, as is shown
in the right hand panel of \Fig{Fvisc}. This mean magnetic field shows
regular reversals combined with a migration away from the midplane,
as can be seen in \Fig{Fpbutter}.

The evolution of the mean magnetic field found in the simulations is
reminiscent of the behavior known from mean-field $\alpha-\Omega$
dynamos. Further details regarding this correspondence (relation between
the value of $\alpha$ and cycle period, field parity for different
boundary conditions, etc.) can be found in recent reviews of the subject
(e.g., Brandenburg 1998, 2000).

\subsection{Connection with the solar dynamo problem}

The disc simulations have shown that a global large scale field can be
obtained even in cartesian geometry. The detailed behavior of this large
scale field depends of course on the boundary conditions adopted
(Brandenburg 1998), and will therefore be different in different
geometries. Nevertheless, the very fact that large scale dynamo action is
possible already in simple cartesian geometry is interesting.

\begin{figure}[h!]\begin{center}\includegraphics[width=.90\textwidth]{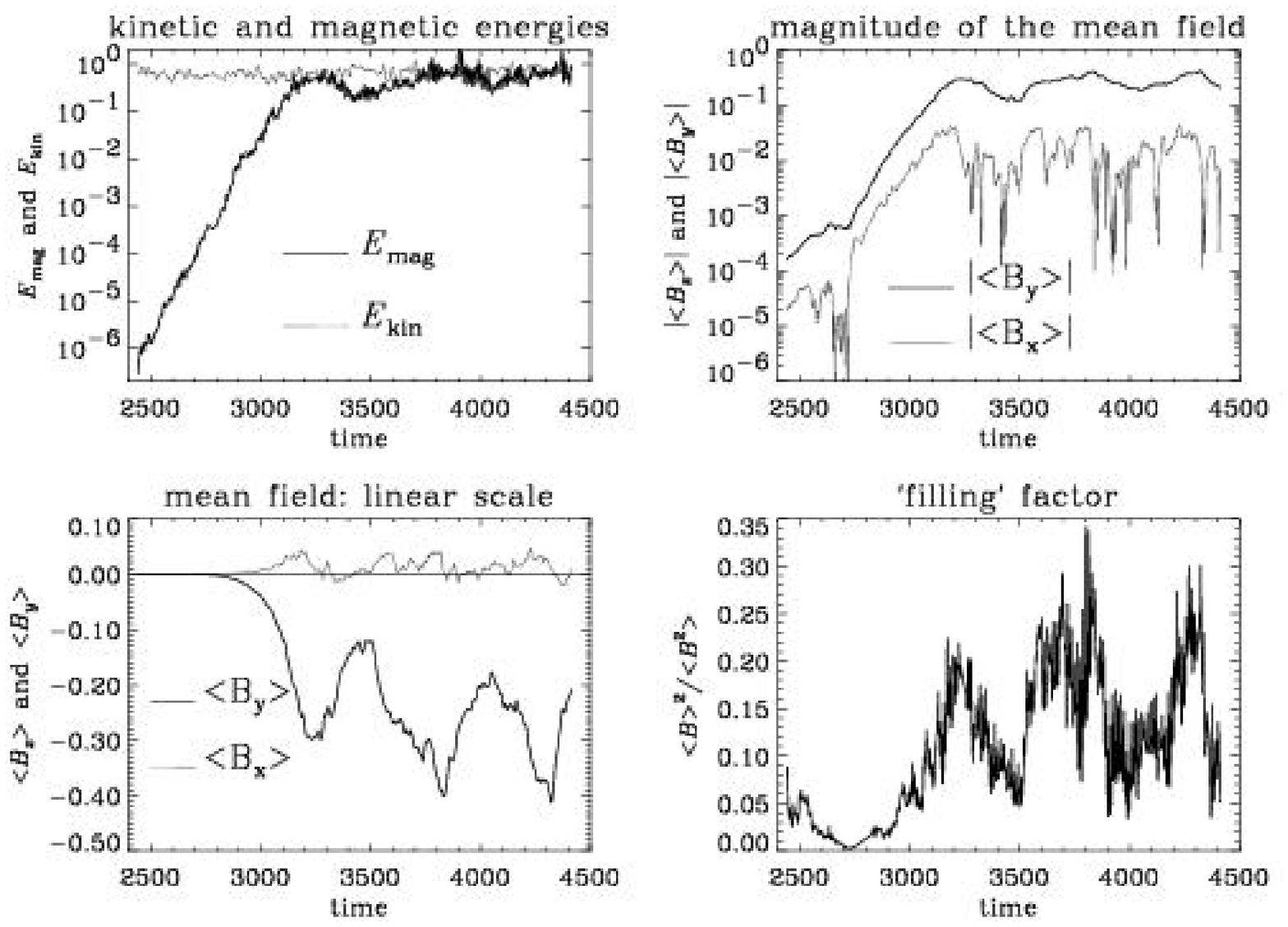}\end{center}\caption[]{
Evolution of several quantities for a convective dynamo model with shear:
kinetic and magnetic energies (dotted and solid lines; first
panel), mean latitudinal and toroidal fields (dotted and solid lines;
second panel), mean magnetic field in a linear scale (third panel), and
the filling factor (fourth panel). Energies and mean magnetic fields are
given in units of the equipartition value, $B_{\rm eq}=4\pi\bra{\uu^2}$.
}\label{Fp3}\end{figure}

In \Fig{Fp3} we show the evolution of magnetic and kinetic energies
as well as the magnitudes of the large scale field for a simulation
of a convectively driven dynamo in the presence of large scale shear
(Brandenburg, Nordlund, \& Stein 2001). It turns out that the ratio of
the magnetic energies in large scale fields relative to the total field,
$\bra{\BB}^2/\bra{\BB^2}$, which is a measure of the filling factor of
the magnetic field, is around 15\% when the field has reached saturation,
i.e.\ when the field growth has stopped. This is similar to the case of
isotropic nonmirror-symmetric turbulence considered in \Sec{Sisotrop}. On
the average, however, the magnetic field is then directed into the
negative $y$-direction (corresponding to the negative azimuthal direction
in spherical geometry), but there is a weak and more noisy cross-stream
field component directed in the positive $x$-direction (pointing north).

When the field has reached saturation, the mean field direction is
approximately constant. Although the magnitude of this mean field
fluctuates somewhat, the sign is always the same. Thus, this simulation
shows no cycles, which are so characteristic of the solar dynamo.
However, since those features, including the field geometry depend
strongly on boundary condition and on the location of the boundary
conditions, this disagreement is to be expected, and one would really need
to resort to global simulations in spherical geometry.

\subsection{Dynamics of the overshoot layer}

Late-type stars with outer convection zones have an interface between
the convection zone proper and the radiative interior. This leads to
some additional dynamics that is important to include, especially in
connection with the dynamo problem. This interface is the layer where
magnetic flux can accumulate, i.e.\ {\it not} necessarily the layer
where the dynamo operates; see the discussion in Brandenburg (1994). The
accumulation is a consequence of turbulent pumping down the turbulence
intensity gradient and the effect was seen clearly in video animations
reported by Brandenburg \& Tuominen (1991) and was analyzed in detail
by Nordlund \ea (1992). Tobias \ea (1998) have studied the effect in
isolation starting with an initial magnetic field distribution as opposed
to a dynamo-generated field.

\begin{figure}[h!]\begin{center}\includegraphics[width=.50\textwidth]{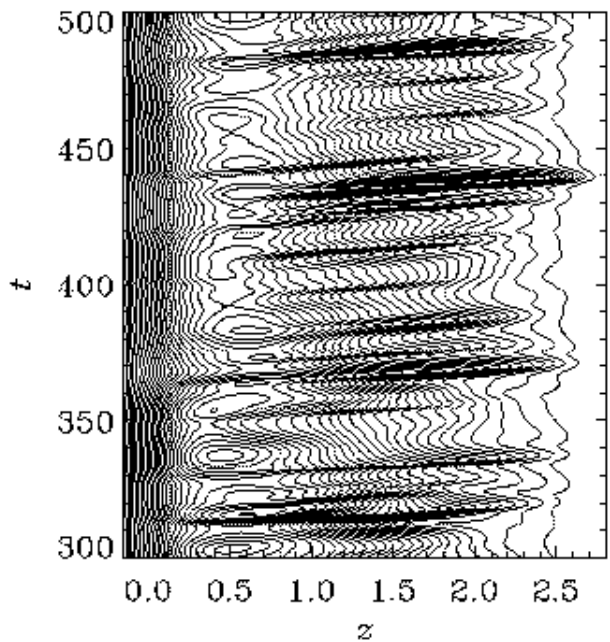}\end{center}\caption[]{
Space-time diagram of the vertical rms velocities for a nonmagnetic
convection zone model of Brandenburg \ea (2001). Note the propagation
of isolated plumes in more or less regular time intervals. Note also that
the wavy pattern extends well into the convection zone proper ($0.5\leq
z\leq1$), and that the plumes appear to propagate at an approximately
constant speed towards the bottom. This speed is around 0.1, which is
comparable to the rms velocity in the runs.
}\label{Fovershoot_MM3c2}\end{figure}

\begin{figure}[h!]\begin{center}\includegraphics[width=.90\textwidth]{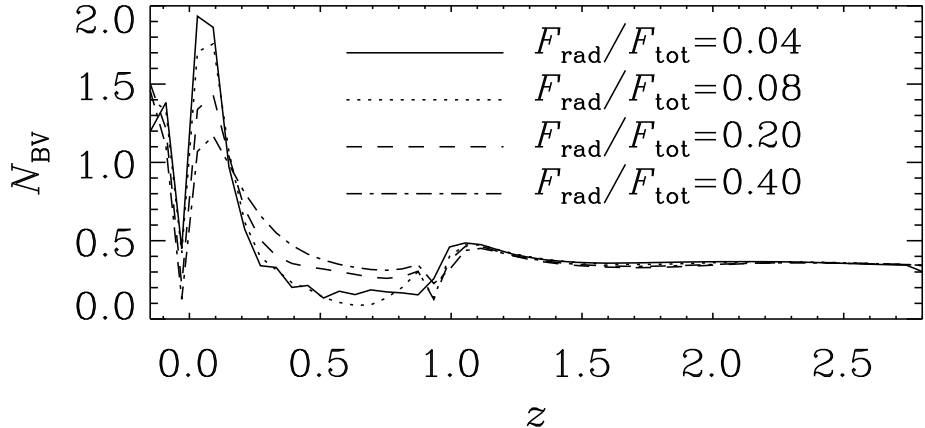}\end{center}\caption[]{
Modulus of the Brunt-V\"ais\"al\"a frequency for a run with polytropic
index $m=-0.9$ and a nondimensional input flux ${\cal F}_{\rm conv}=0.01$.
The various curves are for different values
of the radiative flux, but fixed nominal convective flux.
Small values of ${\cal F}_{\rm rad}/{\cal F}_{\rm tot}$
are typical for the upper layers of the solar convection zone.
}\label{Fbrunt-vaisala}\end{figure}

The flow dynamics changes drastically as one enters this overshoot
layer. The stabilizing buoyancy effect provides a restoring force on a
downward moving element, which can give rise to gravity waves
that could be driven by individual plumes. This leads to a marked
wavy pattern that can extend deep into the lower overshoot layer, as
seen in \Fig{Fovershoot_MM3c2} where we have plotted the vertical rms
velocity as a function of depth and time. These waves extend a major
fraction into the stably stratified layer beneath the convection zone,
but are damped eventually. The typical period of such events is seen
to be around 20 (in units of $\sqrt{d/g}$), where $d$ is the depth of the
unstable (convective) layer. This is comparable with the
mean Brunt-V\"ais\"al\"a frequency,
\EQ
N_{\rm BV}^2=-\grav\cdot\nab(s/c_p).
\EN
which is around 0.3 in the overshoot layer; see \Fig{Fbrunt-vaisala}.

Technically the presence of a lower overshoot layer provides a formidable
challenge, because there the dynamics is governed by the very slow thermal
timescale. This can lead to problems if the properties of the upper
surface layers change which could affect the entropy in deeper parts of
the convection zone. This in turn will also affect the stratification
in the rest of the radiative interior. Since this can only happen on
a thermal timescale (there is no turbulence in these layers to speed
things up) it takes very long before one arrives at a new statistically
stable state. This problem is of course also encountered when starting
with initial conditions that are derived under too unrealistic
assumptions. There are essentially two different approaches to this:
one either considers a toy model where dynamical and thermal timescales
are artificially brought closer together or one adopts an implicit code
which allows the use of somewhat longer timesteps. The former approach
implies that one adopts input fluxes that exceed those in real stars,
but the good news is that turbulent velocities and temperature fluctuations
vary with changing flux in exactly the way that is expected from mixing
length theory; see Brandenburg, Nordlund, \& Stein (2001) for details.

\section{Conclusions}

Many phenomena in astrophysics show direct manifestations of turbulence.
As in the case of accretion discs, without turbulence there would be no
enhanced dissipation, no heating of the disc, and hence no emission.
Magnetic fields are of major importance as is evidenced again by the
example of accretion discs, where the turbulence is a direct consequence of
the presence of magnetic fields (Balbus-Hawley instability). Although
magnetic fields in discs could have primordial origin and could just have
been compressed during their formation, it is also clear that discs are
actually a favorable candidate for producing strong large scale magnetic
fields, as shown by the local simulations discussed above.

Other bodies where strong dynamo action is possible are stars.
Simulations of convection have shown that strong small scale magnetic
fields are possible (Meneguzzi \& Pouquet 1989, Nordlund \ea 1992,
Brandenburg \ea 1996b, Cattaneo 1999), but if there is strong shear an
intense large scale field can also emerge.

The use of nonconservative high order schemes has proved useful in all
those applications. They are easy to implement and to modify, but they are
also reasonably accurate. In this chapter we have illustrated the behavior
of such schemes using various test problems. Using potential
enthalpy and entropy as
the main thermodynamic variables has a number of advantages, especially
in connection with strongly stratified flows near a central object with
a deep potential well, which is relevant to studying outflow phenomena.
Contrary to common belief, high order schemes are not more vulnerable
to Gibbs phenomena near discontinuities. Instead, in simple advection
tests high order methods are able to produce smoother solutions with
less viscosity, which is important for accurate modeling of turbulence.

In the last part of this chapter we have briefly mentioned some
astrophysical applications of simulations using high order schemes where
hydromagnetic turbulence played an important role. In the next few years
we may expect a dramatic increase in the quality and predictive power
of such simulations, as larger computers become available. Already now
a number of very promising simulations are emerging. There is important
work addressing the stability of astrophysical jets in three dimensions
(Ouyed, Clarke, \& Pudritz 2000). Also worthwhile mentioning are recent
high resolution simulations by Hawley (2000) of three-dimensional
accretion tori in global geometry.  What remains to be done in this
field is a proper connection between disc physics and the launching
mechanism of jets. This would require incorporating proper thermodynamics
allowing for radiative cooling and magnetic heating in particular. Global
simulations would also be highly desirable to address the global stellar
dynamo problem. It would be interesting to see how the dynamo works in
fully convective stars, for example. This problem is in some ways simpler
than the solar dynamo problem, because one does not need to worry about
the lower overshoot layer where the relevant timescales are much longer
than in the convection zone proper.

\vspace{4mm}\noindent{\bf Acknowledgements:}~
I am indebted to {\AA}ke Nordlund for teaching me many of the methods
and techniques that are reflected to some extend in the present work.
I thank Wolfgang Dobler and Petri K\"apyl\"a for their careful reading
of the manuscript and for pointing out a number of mistakes in the
manuscript.

\section*{References}

\begin{list}{}{\leftmargin 3em \itemindent -3em\listparindent \itemindent
\itemsep 0pt \parsep 1pt}\item[]

Arber, T. D., Longbottom, A. W., Gerrard, C. L., \& Milne, A. M.\yjcp{2001}{171}{151}
{181}{A staggered grid, Lagrangian-Eulerian remap code for 3D MHD simulations}

Balbus, S. A. \& Hawley, J. F.\yapj{1991}{376}{214}
{222}{A powerful local shear instability in weakly magnetized disks.
I. Linear analysis}

Balsara, D., \& Pouquet, A.\ypp{1999}{6}{89}
{99}{The formation of large-scale structures in supersonic
magnetohydrodynamic flows}

Benson, D. J.\yjour{1992}{Comp. Meth. Appl. Mech. Eng.}{99}{235}
{394}{Computational methods in Lagrangian and Eulerian hydrocodes}

Berger, M. A., \& Ruzmaikin, A.\yjgr{2000}{105}{10481}
{10490}{Rate of helicity production by solar rotation}

Biskamp, D.\ypr{1994}{E50}{2702}
{2711}{Cascade models for magnetohydrodynamic turbulence}

Biskamp, D., \& M\"uller, W.-C.\yprl{1999}{83}{2195}
{2198}{Decay laws for three-dimensional magnetohydrodynamic turbulence}

Bigazzi, A.\ybook{1999}{Models of small-scale and large-scale dynamo action}
{PhD thesis, Universit\`a Degli Studi Dell'Aquila}

Blackman, E. G., \& Field, G. F.\yapj{2000}{534}{984}
{988}{Constraints on the magnitude of $\alpha$ in dynamo theory}

Blandford, R. D. \& Payne, D. G.\ymn{1982}{199}{883}
{903}{Hydromagnetic flows from accretion discs and the production of radio jets}

Brandenburg, A.\yprl{1992}{69}{605}
{608}{Energy spectra in a model for convective turbulence}

Brandenburg, A.\yproc{1994}{117}
{159}{Solar Dynamos: Computational Background}
{Lectures on Solar and Planetary Dynamos}
{M. R. E. Proctor \&  A. D. Gilbert}
{Cambridge University Press}

Brandenburg, A.\yproc{1998}{61}
{86}{Disc Turbulence and Viscosity}
{Theory of Black Hole Accretion Discs}
{M. A. Abramowicz, G. Bj\"ornsson \& J. E. Pringle}
{Cambridge University Press}

Brandenburg, A.\yproc{1999}{65}
{73}{Helicity in large-scale dynamo simulations}
{Magnetic Helicity in Space and Laboratory Plasmas}
{M. R. Brown, R. C. Canfield, A. A. Pevtsov}
{Geophys. Monograph {\bf 111}, American Geophysical Union, Florida}

Brandenburg, A.\yjour{2000}{Phil.\ Trans.\ Roy.\ Soc.\ Lond.\ A}{358}{759}
{776}{Dynamo-generated turbulence and outflows from accretion discs}

Brandenburg, A.\yjour{2001a}{Science}{292}{2440}
{2441}{Magnetic mysteries}

Brandenburg, A.\yapj{2001b}{550}{824}
{840}{The inverse cascade and nonlinear alpha-effect in simulations
of isotropic helical hydromagnetic turbulence}

Brandenburg, A., \& Donner, K. J.\ymn{1997}{288}{L29}
{L33}{The dependence of the dynamo alpha on vorticity}

Brandenburg, A., \& Schmitt, D.\yana{1998}{338}{L55}
{L58}{Simulations of an alpha-effect due to magnetic buoyancy}

Brandenburg, A., \& Subramanian, K.\yana{2000}{361}{L33}
{L36}{Large scale dynamos with ambipolar diffusion nonlinearity}

Brandenburg, A., Tuominen, I.\yproc{1991}{223}
{233}{The solar dynamo}
{The Sun and cool stars: activity, magnetism, dynamos, IAU Coll. 130}
{I. Tuominen, D. Moss \& G. R\"udiger}
{Lecture Notes in Physics {\bf 380}, Springer-Verlag}

Brandenburg, A., Enqvist, K., Olesen, P.\ypr{1996}{D 54}{1291}
{1300}{Large-scale magnetic fields from hydromagnetic turbulence
in the very early universe}

Brandenburg, A., Nordlund, \AA., \& Stein, R. F.\yproc{2000}{85}
{105}{Astrophysical convection and dynamos}
{Geophysical and Astrophysical Convection}
{P. A. Fox \& R. M. Kerr}{Gordon and Breach Science Publishers}

Brandenburg, A., Nordlund, \AA., \& Stein, R. F.\tana{2001}
{Simulation of a convective dynamo with imposed shear}

Brandenburg, A., Nordlund, \AA., Stein, R. F., \&
Torkelsson, U.\yapj{1995}{446}{741}
{754}{Dynamo generated turbulence and large scale magnetic fields
in a Keplerian shear flow}

Brandenburg, A., Nordlund, \AA., Stein, R. F., \&
Torkelsson, U.\yapjl{1996a}{458}{L45}
{L48}{The disk accretion rate for dynamo generated turbulence}

Brandenburg, A., Jennings, R. L., Nordlund, \AA.,
Rieutord, M., Stein, R. F., \& Tuominen, I.\yjfm{1996b}{306}{325}
{352}{Magnetic structures in a dynamo simulation}

Brandenburg, A., Dobler, W., Shukurov, A., \&
von Rekowski, B.\sana{2000}
{Pressure-driven outflow from a dynamo active disc}
{\sf astro-ph/0003174}

Brandenburg, A., Chan, K. L., Nordlund, \AA., Stein, R. F.\tana{2001}
{The effect of the radiative background flux in convection simulations}

Brandenburg A., \& Sarson, G. R.\ybook{2001}
{The effect of hyperdiffusivity on turbulent dynamos with helicity}
{\url{http://www.nordita.dk/~brandenb/tmp/graeme/paper.ps.gz}}

Campbell, C. G.\ymn{1999}{310}{1175}
{1184}{Launching of accretion disc winds along dynamo-generated magnetic fields}

Campbell, C. G.\ymn{2000}{317}{501}
{527}{An accretion disc model with a magnetic wind and turbulent viscosity}

Canuto, C., Hussaini, M. Y., Quarteroni, A., \& Zang, T. A.\ybook{1988}
{Spectral Methods in Fluid Dynamics}{Springer, Berlin}

Cattaneo, F.,\yapj{1999}{515}{L39}
{L42}{On the origin of magnetic fields in the quiet photosphere}

Cattaneo, F., \& Hughes, D. W.\ypr{1996}{E 54}{R4532}
{R4535}{Nonlinear saturation of the turbulent alpha effect}

Chan, K. L., \& Sofia, S.\yapj{1986}{307}{222}
{241}{Turbulent compressible convection in a deep atmosphere. II.
Tests on the validity and limitation of the numerical approach}
 
Chan, K. L., \& Sofia, S.\yapj{1989}{336}{1022}
{1040}{Turbulent compressible convection in a deep atmosphere. III.
Results of three-dimensional computations}

Choudhuri, A. R., Sch\"ussler, M., \& Dikpati, M.\yana{1995}{303}{L29}
{L32}{The solar dynamo with meridional circulation}

Collatz, L.\ybook{1966}{The numerical treatment of differential equations}
{Springer-Verlag, New York}, p.~164.

Colella, P., \& Woodward, P. R.\yjcp{1984}{54}{174}
{201}{The piecewise parabolic method (PPM) for gas-dynamical simulations}

Dobler, W., Brandenburg, A., \& Shukurov, A.\yproc{1999}{347}
{352}{Pressure-driven outflow and magneto-centrifugal wind from a
dynamo active disc}
{Plasma Turbulence and Energetic Particles in Astrophysics}
{M.\ Ostrowski \& R.\ Schlickeiser}
{Publ.\ Astron.\ Obs.\ Jagiellonian Univ., Cracow}

Dodd, R. K., Eilbeck, J. C., Gibbon, J. D., \& Morris, H. C.\ybook{1982}
{Solitons and nonlinear wave equations}{Academic Press: London}

Durney, B. R.\ysph{1995}{166}{231}
{260}{On a Babcock-Leighton dynamo model with a deep-seated generating
layer for the toroidal magnetic field. II}

Fox, P. A., Theobald, M. L., \& Sofia, S.\yapj{1991}{383}{860}
{881}{Compressible magnetic convection: formulation and
two-dimensional models}

Frank, J., King, A. R., \& Raine, D. J.\ybook{1992}{Accretion power
in astrophysics}{Cambridge: Cambridge Univ. Press}

Frisch, U., Pouquet, A., L\'eorat, J., Mazure, A.\yjfm{1975}{68}{769}
{778}{Possibility of an inverse cascade of magnetic helicity
in hydrodynamic turbulence}

Galsgaard, K. \& Nordlund, \AA.\yjgr{1996}{101}{13445}
{13460}{Heating and activity of the solar corona: I.
boundary shearing of an initially homogeneous magnetic-field}

Galsgaard, K. \& Nordlund, \AA.\yjgr{1997a}{102}{219}
{230}{Heating and activity of the solar corona: II.
Kink instability in a flux tube}

Galsgaard, K. \& Nordlund, \AA.\yjgr{1997b}{102}{231} 
{248}{Heating and activity of the solar corona: III.
Dynamics of a low beta plasma with three-dimensional null points}

Glatzmaier, G. A., \& Roberts, P. H.\ynat{1995}{377}{203}
{209}{A three-dimensional self-consistent computer simulation
of a geomagnetic field reversal}

Glatzmaier, G. A., \& Roberts, P. H.\ysci{1996}{274}{1887}
{1891}{Rotation and magnetism of Earth's inner core}

Grauer, R., Marliani, C., \& Germaschewski, K.\yprl{1998}{80}{4177}
{4180}{Adaptive mesh refinement for singular solutions of the
incompressible Euler equations}

Greenhill, L.~J., Gwinn, C.~R., Schwartz, C., Moran, J.~M., \&
Diamond, P.~J.\ynat{1998}{396}{650}
{653}{Coexisting conical bipolar and equatorial outflows from a high-mass
protostar}

Hawley, J. F.\yapj{2000}{528}{462}
{479}{Global magnetohydrodynamical simulations of accretion tori}

Hawley, J. F., Gammie, C. F., \& Balbus, S. A.\yapj{1995}{440}{742}
{763}{Local three-dimensional magnetohydrodynamic simulations
of accretion discs}

Hawley, J. F., Gammie, C. F., \& Balbus, S. A.\yapj{1996}{464}{690}
{703}{Local three dimensional simulations of an accretion disk
hydromagnetic dynamo}

Hodapp, K.-W. \& Ladd, E. F.\yapj{1995}{453}{715}
{720}{Bipolar jets from extremely young stars observed
in molecular hydrogen emission}

Ji, H., Prager, S. C., Almagri, A. F., Sarff, J. S., \&
Toyama, H.\ypp{1996}{3}{1935}
{1942}{Measurement of the dynamo effect in a plasma}

Kerr, R. M., \& Brandenburg, A.\yprl{1999}{83}{1155}
{1158}{Evidence for a singularity in ideal magnetohydrodynamics:
implications for fast reconnection}

Kleeorin, N. I, Moss, D., Rogachevskii, I., \& Sokoloff, D.\yana{2000}{361}{L5}
{L8}{Helicity balance and steady-state strength of the dynamo generated
galactic magnetic field}

Korpi, M. J., Brandenburg, A., Shukurov, A., Tuominen, I.,
\& Nordlund, \AA.\yapjl{1999}{514}{L99}
{L102}{A supernova regulated interstellar medium: simulations of the
turbulent multiphase medium}

Lele, S. K.\yjcp{1992}{103}{16}
{42}{Compact finite difference schemes with spectral-like resolution}

LeVeque, R. J., Mihalas, D., Dorfi, E. A., \& M\"uller, E.\ybook{1998}
{Computational methods for astrophysical fluid flow}
{Springer, Berlin}

Lohse, D., \& M\"uller-Groeling, A.\yprl{1995}{74}{1747}
{1750}{Bottleneck effects in turbulence: Scaling phenomena in r versus p space}

Mac Low, M.-M., Klessen, R. S., \& Burkert, A.\yprl{1998}{80}{2754}
{2757}{Kinetic energy decay rates of supersonic and super-Alfv\'enic
turbulence in star-forming clouds}

Meneguzzi, M., \& Pouquet, A.\yjfm{1989}{205}{297}
{312}{Turbulent dynamos driven by convection}

Nordlund, \AA.\yana{1982}{107}{1}
{10}{Numerical Simulations of the Solar Granulation I. Basic Equations
and Methods}

Nordlund, \AA.\ysph{1985}{100}{209}
{235}{Solar convection}

Nordlund, \AA., \& Galsgaard, K.\ybook{1995}
{A 3D MHD code for Parallel Computers}
{\url{http://www.astro.ku.dk/~aake/NumericalAstro/papers/kg/mhd.ps.gz}}

Nordlund, \AA., \& Stein, R. F.\yjour{1990}{Comput. Phys. Commun.}{59}{119}
{125}{3-D Simulations of Solar and Stellar Convection and Magnetoconvection}

Nordlund, \AA., Galsgaard, K., \& Stein, R. F.\yproc{1994}{471}
{498}{Magnetoconvection and Magnetoturbulence}
{Solar surface magnetic fields}{R. J. Rutten \& C. J. Schrijver}
{NATO ASI Series, Vol. {\bf 433}}

Nordlund, \AA., Brandenburg, A., Jennings, R. L., Rieutord, M.,
Ruokolainen, J., Stein, R. F., \& Tuominen, I.\yapj{1992}{392}{647}
{652}{Dynamo action in stratified convection with overshoot}

Ouyed, R., Pudritz, R.\,E. \& Stone, J.\,M.\ynat{1997}{385}{409}
{414}{Episodic jets from black holes and protostars}

Ouyed, R., \& Pudritz, R.\,E.\yapj{1997a}{482}{712}
{732}{Numerical simulations of astrophysical jets from keplerian discs.
I. Stationary models}

Ouyed, R., \& Pudritz, R.\,E.\yapj{1997b}{484}{794}
{809}{Numerical simulations of astrophysical jets from keplerian discs.
II. Episodic outflows}

Ouyed, R., \& Pudritz, R.\,E.\ymn{1999}{309}{233}
{244}{Numerical simulations of astrophysical jets from keplerian discs.
III. The effects of mass loading}

Ouyed, R., Clarke, D. A. \& Pudritz, R.\,E.\papj{2000}
{3-Dimensional simulations of astrophysical jets from keplerian accretion 
Disks I: stability issues}

Padoan, P., Nordlund, \AA., \& Jones, B. J. T.\yapj{1997}{288}{145}
{152}{The universality of the stellar mass function}

Padoan, P., \& Nordlund, \AA.\yapj{1999}{526}{279}
{294}{A super-Alfvénic model of dark clouds}

Passot, T., \& Pouquet, A.\yjfm{1987}{181}{441}
{466}{Numerical simulation of compressible homogeneous flows
in the turbulent regime}

Pelletier, G., \& Pudritz, R. E.\yapj{1992}{394}{117}
{138}{Hydromagnetic disk winds in young stellar objects and active 
galactic nuclei}

Peterkin, R. E., Frese, M. H., \& Sovinec, C. R.\yjcp{1998}{140}{148}
{171}{Transport of magnetic flux in an arbitrary coordinate ALE code}

Porter, D. H., Pouquet A., \& Woodward, P. R.\yprl{1992}{68}{3156}
{3159}{Three-dimensional supersonic homogeneous turbulence: a numerical study}

Porter, D. H., Pouquet A., \& Woodward, P. R.\ypf{1994}{6}{2133}
{2142}{Kolmogorov-like spectra in decaying 2-dimensional supersonic flows}

Pouquet, A., Frisch, U., L\'eorat, J.\yjfm{1976}{77}{321}
{354}{Strong MHD helical turbulence and the nonlinear dynamo effect}

Rast, M. P., Nordlund, \AA.,  Stein, R.F., \& Toomre, J.\yapjl{1993}{408}{L53}
{L56}{Ionization effects in three--dimensional granulation simulations}

Rast, M. P., \& Toomre, J.\yapj{1993a}{419}{224}
{239}{Compressible convection with ionization.
I. Stability, flow asymmetries, and energy transport}

Rast, M. P., \& Toomre, J.\yapj{1993b}{419}{240}
{254}{Compressible convection with ionization.
II. Thermal boundary-layer instability}

Rekowski, M. v., R\"udiger, G., \& Elstner, D.\yana{2000}{353}{813}
{822}{Structure and magnetic configurations of accretion disk-dynamo models}

Roettiger, K., Stone, J. M., \& Burns, J. O.\yapj{1999}{518}{594}
{602}{Magnetic field evolution in merging clusters of galaxies}

R\"ognvaldsson, \"O. E., Nordlund, \AA., \& Sommer-Larsen, J.\ybook{2001}
{Cooling flows and disk galaxy formation}{preprint}

Rogallo, R. S.\ybook{1981}{Numerical experiments in homogeneous turbulence}
{NASA Tech. Memo. 81315}

S\'{a}nchez-Salcedo, F. J., \& Brandenburg, A.\yapjl{1999}{522}{L35}
{L38}{Deceleration by dynamical friction in a gaseous medium}

S\'{a}nchez-Salcedo, F. J., \& Brandenburg, A.\ymn{2001}{322}{67}
{78}{Dynamical friction of bodies orbiting in a gaseous sphere}

Sod, G. A.\yjcp{1978}{27}{1}
{31}{A survey of several finite difference methods for systems of
nonlinear hyperbolic conservation laws}

Stanescu, D. \& Habashi, W. G.\yjcp{1998}{143}{674}
{681}{$2N$-storage low dissipation and dispersion Runge-Kutta schemes
for computational acoustics}

Steffen, M., Ludwig, H.-G., \& Kr\"u\ss, A.\yana{1989}{123}{371}
{382}{A numerical study of solar granular convection in cells of different
horizontal dimensions}

Stein, R.F., \& Nordlund, \AA.\yapjl{1989}{342}{L95}
{L98}{Topology of convection beneath the solar surface}

Stein, R.F., \& Nordlund, \AA.\yapj{1998}{499}{914}
{933}{Simulations of solar granulation. I. General properties}

Stone, J. M., Norman, M.\yapjS{1992a}{80}{753}
{790}{ZEUS-2D: A radiation magnetohydrodynamics code for astrophysical flows
in two space dimensions: I. The hydrodynamic algorithms and tests}

Stone, J. M., Norman, M.\yapjS{1992b}{80}{791}
{818}{ZEUS-2D: A radiation magnetohydrodynamics code for astrophysical flows
in two space dimensions: II. The magnetohydrodynamic algorithms and tests}

Stone, J. M., Hawley, J. F., Gammie, C. F.,
\& Balbus, S. A.\yapj{1996}{463}{656}
{673}{Three dimensional magnetohydrodynamical simulations of vertically
stratified accretion disks}

Stewart, J. M.\yana{1975}{42}{95}
{101}{The hydrodynamics of accretions discs I: Stability}

Suzuki, E. \& Toh, S.\ypr{1995}{E 51}{5628}
{5635}{Entropy cascade and temporal intermittency in a shell model for
convective turbulence}

Thelen, J.-C.\ymn{2000}{315}{155}
{164}{A mean electromotive force induced by magnetic buoyancy instabilities}

Tobias, S. M., Brummell, N. H., Clune, T. L.,
\& Toomre, J.\yapjl{1998}{502}{L177}
{L177}{Pumping of magnetic fields by turbulent penetrative convection}

Toh, S., \& Iima, M.\ypr{2000}{E 61}{2626}
{2639}{Dynamical aspect of entropy transfer in free convection turbulence}

Urpin, V., \& Brandenburg, A.\ymn{1998}{294}{399}
{406}{Magnetic and vertical shear instabilities in accretion discs}

Vainshtein, S. I., \& Cattaneo, F.\yapj{1992}{393}{165}
{171}{Nonlinear restrictions on dynamo action}

Vishniac, E. T., \& Brandenburg, A.\yapj{1997}{475}{263}
{274}{An incoherent $\alpha-\Omega$ dynamo in accretion disks}

V\"olk, H. J. \& Atoyan, A. M.\yjour{1999}{Astroparticle Phys.}{11}{73}
{82}{Clusters of galaxies: magnetic fields and nonthermal emission}

Williamson, J. H.\yjcp{1980}{35}{48}
{}{Low-storage Runge-Kutta schemes}

Ziegler, U., \& R\"udiger, G.\yana{2000}{356}{1141}
{1148}{Angular momentum transport and dynamo-effect in stratified,
weakly magnetic disks}

\end{list}

\section*{Appendix}
\appendix

\section{Centered, onesided and semi-onesided derivatives}
\label{Sonesided}

In \Sec{Scentered} we gave the centered finite difference formulae
for schemes of different order. Here we first describe the method
for determining the finite difference formulae for second order, but
the generalization to higher order is straightforward. We also give
the corresponding expressions for one-sided and semi-onesided finite
difference formulae.

We want to write the derivative $f'(x)$ at the point $x_i$ as
\EQ
f'_i=a_{-1}f_{i-1}+a_0f_0+a_1f_1,
\EN
where $f'_i=f'(x_i)$, $f_{i-1}=f(x_i-\delta x)$, and $f_{i+1}=f(x_i+\delta
x)$. To determine the coefficients $a_{-1}$, $a_0$ and $a_1$ we expand
$f(x)$ up to second order
\EQ
f(x)=c_0+c_1 x+c_2 x^2.
\EN
The first derivative is then
\EQ
f'(x)=c_1+2c_2 x.
\label{fprime}
\EN
In particular, the value at $x=0$ is just $f'(0)=c_1$.
Likewise, we have $f''(0)=2c_2$
To determine all coefficients we make use of our knowledge
at the neighboring points around $x_i$, i.e.\ we use the function
values $f(x_i-\delta x)\equiv f_{i-1}$, $f(x_i)\equiv f_i$,
and $f(x_i+\delta x)\equiv f_{i+1}$, so we have
\EQA
f_{i-1}&=&c_0+c_1(-\delta x)+c_2(-\delta x)^2,\\
f_{i  }&=&c_0,\\
f_{i+1}&=&c_0+c_1(+\delta x)+c_2(+\delta x)^2.
\ENA
This can be written in matrix form
\EQ
\pmatrix{f_{i-1}\cr f_i\cr f_{i+1}}
=\pmatrix{
(-1)^0 & (-1)^1 & (-1)^2\cr
0^0 & 0^1 & 0^2\cr
1^0 & 1^1 & 1^2\cr }
\pmatrix{c_0\cr c_1\delta x\cr c_2\delta x^2}
\EN
(where $(-1)^0=0^0=1^0=1$), or
\EQ
\ff=\MM\cc,
\EN
and so we obtain the coefficients as
\EQ
\cc=\MM^{-1}\ff.
\EN
To calculate $f'$ we need the value of $c_1$, see \Eq{fprime}, and so
the coefficients $a_n$ needed to express the derivative are
$a_{-1}=(\MM^{-1})_{10}$, $a_0=(\MM^{-1})_{11}$, and $a_1=(\MM^{-1})_{12}$,
i.e.\ all points of the inverted matrix in the second row. The resulting
formula for $f'_i$ is well-known,
\EQ
f'_i=(-f_{i-1}+f_{i+1})/(2\delta x).
\EN
The corresponding result for the second derivative is
\EQ
f''_i=(f_{i-1}-2f_i+f_{i+1})/(\delta x^2).
\EN
On the boundaries we have to calculate for derivative using only
points inside the domain, which is explained in the next subsection
for second order accuracy, but again the generalization to higher
order is straightforward and only the results will be listed.

\subsection{One-sided 2nd order derivatives}
Again, we want to write the derivative $f'(x)$ as
\EQ
f'_i=a_0f_0+a_1f_1+a_2f_2,
\EN
but now
\EQ
\pmatrix{f_i\cr f_{i+1}\cr f_{i+2}}
=\pmatrix{
0^0 & 0^1 & 0^2\cr
1^0 & 1^1 & 1^2\cr
2^0 & 2^1 & 2^2 }
\pmatrix{c_0\cr c_1\delta x\cr c_2\delta x^2}
\EN
Thus, one arrives at
\EQ
f'_i=(-3f_i+4f_{i+1}-f_{i+2})/(2\delta x).
\EN
Correspondingly, for the second derivative we have
\EQ
f''_i=(2f_i-5f_{i+1}+4f_{i+2}-f_{i+3})/\delta x^2.
\EN

\subsection{4th order derivatives}
First derivatives
\EQ
f'_i=(f_{i-2}-8f_{i-1}+8f_{i+1}-f_{i+2})/(12\delta x)
\EN
\EQ
f'_i=(-3f_{i-1}-10f_i+18f_{i+1}-6f_{i+2}+f_{i+3})/(12\delta x)
\EN
\EQ
f'_i=(-25f_i+48f_{i+1}-36f_{i+2}+16f_{i+3}
-3f_{i+4})/(12\delta x)
\EN
Second derivatives
\EQ
f''_i=(-f_{i-2}+16f_{i-1}-30f_i
+16f_{i+1}-f_{i+2})/(12\delta x^2)
\EN
\EQ
f''_i=(11f_{i-1}-20f_i+6f_{i+1}+4f_{i+2}
-f_{i+3})/(12\delta x^2)
\EN
\EQ
f''_i=(35f_i-104f_{i+1}+114f_{i+2}-56f_{i+3}
+11f_{i+4})/(12\delta x^2)
\EN

\subsection{6th order derivatives}
First derivatives
\EQ
f'_i=(-f_{i-3}+9f_{i-2}-45f_{i-1}
+45f_{i+1}-9f_{i+2}+f_{i+3})/(60\delta x)
\EN
\EQ
f'_i=(2f_{i-2}-24f_{i-1}-35f_i+80f_{i+1}-30f_{i+2}+8f_{i+3}
-f_{i+4})/(60\delta x)
\EN
\EQ
f'_i=(-10f_{i-1}-77f_i+150f_{i+1}-100f_{i+2}+50f_{i+3}
-15f_{i+4}+2f_{i+5})/(60\delta x)
\EN
\EQ
f'_i=(-147f_i+360f_{i+1}-450f_{i+2}+400f_{i+3}-225f_{i+4}+72f_{i+5}-10f_{i+6}
)/(60\delta x)
\EN
Second derivatives
\EQ
f''_i=(2f_{i-3}-27f_{i-2}+270f_{i-1}-490f_i
+270f_{i+1}-27f_{i+2}+2f_{i+3})/(180\delta x^2)
\EN
\EQ
f''_i=(-13f_{i-2}+228f_{i-1}-420f_i+200f_{i+1}+15f_{i+2}
-12f_{i+3}+2f_{i+4})/(180\delta x^2)
\EN
\EQ
f''_i=(137f_{i-1}-147f_i-255f_{i+1}+470f_{i+2}-285f_{i+3}
+93f_{i+4}-13f_{i+5})/(180\delta x^2)
\EN
\EQ
f''_i=\left(812f_i-3132f_{i+1}+5265f_{i+2}
-5080f_{i+3}+2970f_{i+4}-972f_{i+5}+137f_{i+6}\right)/(180\delta x^2)
\EN

\section{The $2N$-RK3 scheme}
\label{S2NRK3}

If $N$ is the number of variables to be updated from one timestep to
the next, the $2N$-schemes require only $2\times N$ variables to be
stored in memory at any time. This is better than for the standard
Runge-Kutta schemes. The general iteration formula is
\EQ
w_i=\alpha_i w_{i-1}+hF(t_{i-1},u_{i-1}),\quad
u_i=u_{i-1}+\beta_i w_i.
\label{iterform}
\EN
For a third order scheme we have $i=1,...,3$.
In order to advance the variable $u$ from $u^{(n)}$ at time $t^{(n)}$
to $u^{(n+1)}$ at time $t^{(n+1)}=t^{(n)}+h$ we set in \Eq{iterform}
\EQ
u_0=u^{(n)}\quad\mbox{and}\quad u_3=u^{(n+1)},
\EN
with $u_1$ and $u_2$ being intermediate steps. In order to be able to
calculate the first step, $i=1$, for which no $w_{i-1}\equiv w_0$ exists,
we have to require
$\alpha_1=0$. Thus, we are left with 5 unknowns, $\alpha_2$, $\alpha_3$,
$\beta_1$, $\beta_2$, and $\beta_3$. We write down \Eq{iterform} in
explicit form for $i=1,...,3$,
\EQ
w_1=hF(t_0,u_0),\quad u_1=u_0+\beta_1 w_1.
\label{iterform1}
\EN
\EQ
w_2=\alpha_2 w_1+hF(t_1,u_1),\quad u_2=u_1+\beta_2 w_2.
\label{iterform2}
\EN
\EQ
w_3=\alpha_3 w_2+hF(t_2,u_2),\quad u_3=u_2+\beta_3 w_3.
\label{iterform3}
\EN
Written in explicit form, we have for $i=1$
\EQ
u_1=u_0+\beta_1 hF(t_0,u_0).
\label{u1eqn}
\EN
The $i=2$ step yields
\EQ
w_2=\alpha_2 hF(t_0,u_0)+hF(t_1,u_1),
\EN
\EQ
u_2=u_0+h\left[(\beta_1+\beta_2\alpha_2)F(t_0,u_0)+\beta_2F(t_1,u_1)\right].
\label{u2eqn}
\EN
and the $i=3$ step gives
\EQ
w_3=h\left[\alpha_2\alpha_3F(t_0,u_0)+\alpha_3F(t_1,u_1)+F(t_2,u_2)\right]
\EN
\EQA
u_3=u_0+\left[\beta_1+\alpha_2(\beta_2+\alpha_3\beta_3)\right]hF(t_0,u_0)
+(\beta_2+\beta_3\alpha_3)hF(t_1,u_1)+\beta_3hF(t_2,u_2).
\label{u3eqn}
\ENA
The corresponding times can be calculated by putting $F=1$.
This yields
\EQ
t_1=t_0+\beta_1 h,
\label{increment1}
\EN
\EQ
t_2=t_0+h\left[\beta_1+\beta_2(1+\alpha_2)\right],
\label{increment2}
\EN
\EQ
t_3=t_0+h\left[\beta_1+\beta_3+(1+\alpha_2)(\beta_2+\alpha_3\beta_3)\right].
\label{increment3}
\EN
The last expression can also be written in the form
\EQ
t_3=t_0+h\left\{\beta_1+\beta_2(1+\alpha_2)
+\beta_3\left[1+(1+\alpha_2)\alpha_3)\right]\right\}
\EN

Next we need to determine the conditions that the scheme is indeed of
third order. This can be done by considering the differential equation
\EQ
\dd u/\dd t=u,\quad u(0)=u_0,
\label{Adudt}
\EN
for $u=u(t)$, where $u_0$ is the initial value of $u$. The exact solution
of \Eq{Adudt} is $u_0 e^t$. Its Taylor expansion for $t=t_0+h$ is
\EQ
u(t_0+h)=u_0[1+h+\half h^2+{\textstyle{1\over6}}h^3+...].
\label{expansion}
\EN
The solution based on \Eq{u3eqn} is
\EQA
u_3=u_0+\left[\beta_1+\alpha_2(\beta_2+\alpha_3\beta_3)\right]h u_0
+(\beta_2+\alpha_3\beta_3)h u_1+\beta_3h u_2
\label{u3eqn2}
\ENA
In order to compare with \Eq{expansion}
we need the explicit expressions for $u_1$ and $u_2$, which are
\EQ
u_1=(1+h\beta_1)u_0
\label{Eqn_u_1}
\EN
\EQ
u_2=\left\{1+h\left[\beta_1+(1+\alpha_2)\beta_2\right]
+h^2\beta_1\beta_2\right\}u_0.
\label{Eqn_u_2}
\EN
and so we can write
\EQA
u_3=u_0+\gamma_1h+\gamma_2h^2+\gamma_3h^3
\ENA
with
\\

\SHADOWBOX{
\EQ
\gamma_1=\beta_1+\beta_3+(1+\alpha_2)(\beta_2+\alpha_3\beta_3)
\EN
\EQ
\gamma_2=\beta_1\beta_2+\beta_3[\beta_2(1+\alpha_2)+\beta_1(1+\alpha_3)]
\EN
\EQ
\gamma_3=\beta_1\beta_2\beta_3
\EN
In order for the scheme to be third order we have to require
$\gamma_1=1$, $\gamma_2=1/2$, and $\gamma_3=1/6$; see \Eq{expansion}.
Thus, we have now {\it three} equations for {\it five} unknowns.
We now have to come up with two more equations to solve for the
five unknowns.
}\\

If we assume that the intermediate timesteps are evaluated in
equidistant time intervals, we have to require that the time increments
in \Eqs{increment1}{increment2} are $1/3$ and $2/3$, respectively.
This yields
\EQ
\beta_1=\beta_2(1+\alpha_2)=1/3,
\EN
with two particular solutions\footnote{I thank Petri K\"apyl\"a
from Oulu for pointing out the second of these solutions.}
\EQA
\left\{
\begin{array}{l}
\alpha_2=-2/3,\quad
\alpha_3=-1,\quad
\beta_1=1/3,\quad
\beta_2=1,\quad
\beta_3=1/2,\\
\alpha_2=-1/3,\quad
\alpha_3=-1,\quad
\beta_1=1/3,\quad
\beta_2=1/2,\quad
\beta_3=1.
\end{array}
\right.
\ENA
These are in fact the simplest $2N$-RK3 schemes that also lead to comparatively
small residual errors.

Alternatively, one can move the times closer to the end time of the time step
and evaluate the right hand side at times $t_1-t_0={1\over2}h$ and
$t_2-t_0=h$. This gives the particular solution
\EQ
\alpha_2=-1/4,\quad
\alpha_3=-4/3,\quad
\beta_1=1/2,\quad
\beta_2=2/3,\quad
\beta_3=1/2.
\EN
Again, there could be other solutions.

Another possibility is to require that the inhomogeneous equation
\EQ
\dd u/\dd t=t^n,\quad u(0)=0,
\label{Bdudt}
\EN
is solved exactly up to some $n$. The exact solutions for $t=h$ are
$u=\half h^2$ for $n=1$ and $u=\onethird h^3$ for $n=2$.

The case $n=0$ was already considered in \Eqss{increment1}{increment3}.
For $n=1$ we have $F(t_1,u_1)=\onethird h$ and $F(t_2,u_2)=\twothird h$,
so \Eq{u3eqn} gives
\EQ
u_3=u_0+(\beta_2+\beta_3\alpha_3)\onethird h^2+\beta_3\twothird h^2,
\EN
or
\EQ
u_3=u_0+\onethird[\beta_2+\beta_3(\alpha_3+2)]h^2.
\EN
Comparing with the exact solution this yields the additional equation
\EQ
\beta_2+\beta_3(\alpha_3+2)={\textstyle{3\over2}}.
\EN
For $n=2$ we have $F(t_1,u_1)={1\over9}h^2$ and $F(t_2,u_2)={4\over9}h^2$,
so \Eq{u3eqn} gives
\EQ
u_3=u_0+(\beta_2+\beta_3\alpha_3){\textstyle{1\over9}}h^3
+\beta_3{\textstyle{4\over9}}h^3,
\EN
or
\EQ
u_3=u_0+{\textstyle{1\over9}}[\beta_2+\beta_3(\alpha_3+4)]h^3.
\EN
Again, comparing with the exact solution one obtains
\EQ
\beta_2+\beta_3(\alpha_3+4)=3.
\EN
This gives the solution
\EQ
\alpha_2=-17/32,\quad
\alpha_3=-32/27,\quad
\beta_1=1/4,\quad
\beta_2=8/9,\quad
\beta_3=3/4,
\EN
which implies that the right hand sides are evaluated at the times
$t_1-t_0={1\over4}h$ and $t_2-t_0={2\over3}h$.
In \Tabs{Ttab_2N-RK3}{Ttab_2N-RK3-err} this scheme is referred to
as ``inhomogeneous''.

Yet another idea (W.\ Dobler, private communication) is to obtain the
additional two equations by requiring that the quadratic differential equation
$\dd u/\dd t=u^2$ with $u_0=1$ is solved exactly. The solution is
$u=(1-t)^{-1}$, of which we only need the expansion up to $h^2$, so we
have $u_3\approx1+h+h^2$. Again, we use \Eq{u3eqn}, but now with $F=u^2$,
\EQ
u_3=1+\left[\beta_1+\alpha_2(\beta_2+\alpha_3\beta_3)\right]h
+(\beta_2+\alpha_3\beta_3)h u_1^2+\beta_3h u_2^2.
\label{Eqn_u3_new}
\EN
We need $u_1^2$ and $u_2^2$ only up to the term linear in $h$.
Using \Eqs{Eqn_u_1}{Eqn_u_2} we have
\EQ
u_1^2=1+2\beta_1 h+{\cal O}(h^2),\quad
u_2^2=1+2\left[\beta_1+(1+\alpha_2)\beta_2\right]h+{\cal O}(h^2).
\EN
Inserting this in \Eq{Eqn_u3_new} yields
\EQ
u_3=1+\delta_1 h+\delta_2 h^2+{\cal O}(h^3),
\EN
with
\EQ
\delta_1=\beta_1+\alpha_2(\beta_2+\alpha_3\beta_3)
+(\beta_2+\alpha_3\beta_3)+\beta_3
\EN
and
\EQ
\delta_2=2\beta_1(\beta_2+\alpha_3\beta_3)
+2\beta_3\left[\beta_1+(1+\alpha_2)\beta_2\right].
\EN
Thus, the two additional equations are
\EQ
\beta_1+(1+\alpha_2)(\beta_2+\alpha_3\beta_3)+\beta_3=1,
\EN
\EQ
\beta_1\left[\beta_2+(1+\alpha_3)\beta_3\right]
+(1+\alpha_2)\beta_2=1/2.
\EN
The numerical solution is
\EQ
\alpha_2=-0.36726297,\quad
\alpha_3=-1.0278248,\quad
\beta_1=0.30842796,\quad
\beta_2=0.54037434,\quad
\beta_3=1.
\EN
which implies that the right hand sides are evaluated at the times
$t_1-t_0=0.308h$ and $t_2-t_0=0.650h$.
In \Tabs{Ttab_2N-RK3}{Ttab_2N-RK3-err} this scheme is referred to
as ``quadratic''.


\section{Derivation of the jacobian for transformation on a sphere}
\label{Sjacobian}

Here we give the explicit derivation of \Eqs{transform_sphere1}{transform_sphere2}.
We first use the transformation in the form
\EQ
x=\tx^2/\tr,\quad y=\tx\ty/\tr,
\quad\mbox{if $\tx\ge\ty$},
\EN
\EQ
x=\tx\ty/\tr,\quad y=\ty^2/\tr,
\quad\mbox{if $\tx\le\ty$}.
\EN
To obtain the jacobian we differentiate with respect to $x$, so
\EQ
1=2\xtx-\rtx,\quad
0=\xtx+\ytx-\rtx,\quad
\mbox{if $\tx\ge\ty$},
\EN
\EQ
1=\xtx+\ytx-\rtx,\quad
0=2\ytx-\rtx,\quad
\mbox{if $\tx\le\ty$}.
\EN
We now differentiate with respect to $y$,
\EQ
0=2\xty-\rty,\quad
1=\xty+\yty-\rty,\quad
\mbox{if $\tx\ge\ty$},
\EN
\EQ
0=\xty+\yty-\rty,\quad
1=2\yty-\rty,\quad
\mbox{if $\tx\le\ty$}.
\EN
The derivatives of $\tr$ can be written as
\EQ
\rtx=\xt\xtx+\yt\ytx,
\EN
\EQ
\rty=\xt\xty+\yt\yty.
\EN
In all cases we have
\EQ
{x\over y}={\tx\over\ty},
\EN
so
\EQ
+1=\xtx-\ytx,
\EN
\EQ
-1=\xty-\yty,
\EN
and so
\EQ
1=2\xtx-\xt\xtx-\yt\left(\xtx-1\right),\quad
\mbox{if $\tx\ge\ty$},
\EN
\EQ
0=\xtx+\ytx-\xt\xtx-\yt\left(\xtx-1\right),\quad
\mbox{if $\tx\ge\ty$},
\EN
so
\EQ
1=\xtx+\yt,\quad
\mbox{if $\tx\ge\ty$},
\EN
and so
\EQ
\xtx=1-\yt,\quad\ytx=-\yt,\quad
\mbox{if $\tx\ge\ty$},
\EN
and correspondingly
\EQ
1=\xtx+\ytx-\xt\xtx-\yt\left(\xtx-1\right),\quad
\mbox{if $\tx\le\ty$},
\EN
\EQ
0=2\xtx-\xt\xtx-\yt\left(\xtx-1\right),\quad
\mbox{if $\tx\le\ty$},
\EN
so
\EQ
1=\ytx+\yt,\quad
\mbox{if $\tx\le\ty$},
\EN
and so
\EQ
\xtx=2-\yt,\quad\ytx=1-\yt,\quad
\mbox{if $\tx\le\ty$}.
\EN
Hence note that there is a discontinuity
of the jacobian along the diagonals.
Now for the $y$-derivatives we have
\EQ
0=2\xty-\xt\xty-\yt\left(\xty+1\right),\quad
\mbox{if $\tx\ge\ty$},
\EN
\EQ
1=\xty+\yty-\xt\xty-\yt\left(\xty+1\right),\quad
\mbox{if $\tx\ge\ty$},
\EN
so
\EQ
0=\xty-\yt,\quad
\mbox{if $\tx\ge\ty$},
\EN
and so
\EQ
\xty=+\yt,\quad\yty=1+\yt,\quad
\mbox{if $\tx\ge\ty$},
\EN
and correspondingly
\EQ
0=\xty+\yty-\xt\xty-\yt\left(\xty+1\right),\quad
\mbox{if $\tx\le\ty$},
\EN
\EQ
-1=2\xty-\xt\xty-\yt\left(\xty+1\right),\quad
\mbox{if $\tx\le\ty$},
\EN
so
\EQ
0=\xty+1-\yt,
\quad\mbox{if $\tx\le\ty$},
\EN
and so
\EQ
\xty=-1+\yt,\quad\yty=+\yt,
\quad\mbox{if $\tx\le\ty$}.
\EN
So, in summary
\EQ
\pmatrix{\mxtx&\mxty\cr\mytx&\myty}=
\pmatrix{+1-\myt&+1-\mxt\cr-1+\mxt&+1+\myt}
\quad\mbox{if $\tx\ge\ty$},
\EN
\EQ
\pmatrix{\mxtx&\mxty\cr\mytx&\myty}=
\pmatrix{+1+\mxt&-1+\myt\cr+1-\myt&+1-\mxt}
\quad\mbox{if $\tx\le\ty$}.
\EN

\section{Derivation of the incremental jacobian for second derivatives}
\label{Sjacobi2nd}

Here we present the explicit derivation of \Eq{2nd_jacobi3}.
To calculate the second derivative of a function $f$ that is represented
on a coordinate mesh $\tilde{\xx}$, is given by
\EQ
{\partial^2 f\over\partial x_i\partial x_j}=
{\partial\over\partial x_i}\left({\partial f\over\partial x_j}\right)=
{\partial\over\partial x_i}\left({\partial f\over\partial\tilde x_k}
{\partial\tilde x_k\over\partial x_j}\right)
\EN
so
\EQ
{\partial^2 f\over\partial x_i\partial x_j}=
{\partial^2 f\over\partial\tilde x_l\partial\tilde x_k}
{\partial\tilde x_l\over\partial x_i}
{\partial\tilde x_k\over\partial x_j}+
{\partial f\over\partial\tilde x_k}
{\partial^2\tilde x_k\over\partial x_i\partial x_j},
\EN
or
\EQ
{\partial^2 f\over\partial x_i\partial x_j}=
{\partial^2 f\over\partial\tilde x_p\partial\tilde x_q}{\sf J}_{pi}{\sf J}_{qj}
+{\partial f\over\partial\tilde x_k}{\sf K}_{kij},
\EN
which is just \Eq{2nd_jacobi1}, using \Eq{2nd_jacobi2} for the definition
of ${\sf K}_{kij}$ of the second order jacobian.

To obtain the second order jacobian by successive tensor multiplication
we differentiate twice the evolution equation for $\xx$,
\EQ
x_k^{(n+1)}=x_k^{(n)}+u_k^{(n)}\,\delta t,
\EN
so
\EQ
{\partial^2 x^{(n+1)}_k\over\partial x^{(n+1)}_i\partial x^{(n+1)}_j}=
{\partial^2 x^{(n)}_k\over\partial x^{(n+1)}_i\partial x^{(n+1)}_j}+
{\partial^2 u^{(n)}_k\over\partial x^{(n+1)}_i\partial x^{(n+1)}_j}\delta t.
\EN
The expression on the left hand side is just the derivative of a
Kronecker delta, see \Eq{1st_jacobi2}, so it is zero. Thus we have
\EQ
0={\partial^2 x^{(n)}_k\over\partial x^{(n+1)}_i\partial x^{(n+1)}_j}+
{\partial\over\partial x^{(n+1)}_i}\left(
{\partial u^{(n)}_k\over\partial x^{(n)}_q}
{\partial x^{(n)}_q\over\partial x^{(n+1)}_j}\right)\delta t,
\EN
or
\EQ
0={\partial^2 x^{(n)}_q\over\partial x^{(n+1)}_i\partial x^{(n+1)}_j}
\left(\delta_{kq}+{\partial u^{(n)}_k\over\partial x^{(n)}_q}\delta t\right)
+{\partial^2 u^{(n)}_k\over\partial x^{(n)}_p\partial x^{(n)}_q}
{\sf J}_{pi}^{(n)}{\sf J}_{qj}^{(n)}\delta t,
\EN
which can be written as
\EQ
0={\sf M}_{kq}{\sf K}^{(n)}_{qij}+u_{k,pq}{\sf J}_{pi}^{(n)}{\sf J}_{qj}^{(n)}\delta t,
\EN
so 
\EQ
{\sf K}^{(n)}_{kij}=-\delta t\left(\MMMM^{-1}\right)_{kl}\,u_{lpq}
{\sf J}_{pi}^{(n)}{\sf J}_{qj}^{(n)},
\EN
which is just \Eq{2nd_jacobi5}.

We now need to derive the equation that relates the incremental second
order jacobians to the second order jacobian of the previous timestep. To
this end we begin with the second order jacobian at time $2\delta t$, so
\EQ
{\sf K}^{(0\rightarrow2)}_{kij}\equiv
{\partial^2x_k^{(0)}\over\partial x_i^{(2)}\partial x_j^{(2)}}=
{\partial\over\partial x_i^{(2)}}\left(
{\partial x_k^{(0)}\over\partial x_q^{(1)}}
{\partial x_q^{(1)}\over\partial x_j^{(2)}}\right)
\EN
or
\EQ
{\sf K}^{(0\rightarrow2)}_{kij}\equiv
{\partial^2x_k^{(0)}\over\partial x_p^{(1)}\partial x_q^{(1)}}
{\partial x_p^{(1)}\over\partial x_i^{(2)}}
{\partial x_q^{(1)}\over\partial x_j^{(2)}}+
{\partial^2x_q^{(1)}\over\partial x_i^{(2)}\partial x_j^{(2)}}
{\partial x_k^{(0)}\over\partial x_q^{(1)}}
\EN
or
\EQ
{\sf K}_{kij}^{\rm(0\rightarrow 2)}=
{\sf K}_{kpq}^{\rm(0\rightarrow 1)}{\sf J}_{pi}^{(1)}{\sf J}_{qj}^{(1)}+
{\sf J}_{kl}^{\rm(0\rightarrow 1)}{\sf K}_{lij}^{(1)}.
\EN
For the next step we have
\EQ
{\sf K}^{(0\rightarrow3)}_{kij}\equiv
{\partial^2x_k^{(0)}\over\partial x_i^{(3)}\partial x_j^{(3)}}=
{\partial\over\partial x_i^{(3)}}\left(
{\partial x_k^{(0)}\over\partial x_q^{(1)}}
{\partial x_q^{(1)}\over\partial x_p^{(2)}}
{\partial x_p^{(2)}\over\partial x_j^{(3)}}\right),
\EN
so
\EQ
{\sf K}_{kij}^{\rm(0\rightarrow 3)}=
{\sf K}_{kpq}^{\rm(0\rightarrow 1)}
{\sf J}_{pr}^{(1)}{\sf J}_{qs}^{(1)}
{\sf J}_{ri}^{(2)}{\sf J}_{sj}^{(2)}+
{\sf J}_{kl}^{\rm(0\rightarrow 1)}{\sf K}_{lrs}^{(1)}
{\sf J}_{ri}^{(2)}{\sf J}_{sj}^{(2)}+
{\sf J}_{kr}^{(0)}{\sf J}_{rs}^{(1)}{\sf K}_{sij}^{(2)}.
\EN
This can be written as
\EQ
{\sf K}_{kij}^{\rm(0\rightarrow 3)}=
{\sf K}_{kpq}^{\rm(0\rightarrow 2)}{\sf J}_{pi}^{(2)}{\sf J}_{qj}^{(2)}+
{\sf J}_{kl}^{\rm(0\rightarrow 2)}{\sf K}_{lij}^{(2)}
\EN
which, for the general step from $0$ to $n$, becomes \Eq{2nd_jacobi3}.

\section{Solution for $\alpha$ and $\eta_{\rm t}$ quenched $\alpha^2$-dynamo}
\label{Saquench}

Here we present the explicit derivation of \Eq{quench_formula}. According
to mean-field theory for non-mirror symmetric isotropic homogeneous
turbulence with no mean flow the mean magnetic field is governed by
the equation
\EQ
{\partial\over\partial t}\meanBB=\nab\times\left(\alpha\meanBB
-\eta_{\rm T}\eta_0\meanJJ\right),
\label{dyneq}
\EN
where bars denote the mean fields and $\eta_{\rm T}=\eta+\eta_{\rm t}$
is the total (microscopic plus turbulent) magnetic diffusion. Both
$\alpha$-effect and turbulent diffusion are assumed to be quenched in
the same way, so
\EQ
\alpha={\alpha_0\over1+\alpha_{\rm B}\,\meanBB^2/B_{\rm eq}^2},\quad
\eta_{\rm t}={\eta_{\rm t0}\over1+\eta_{\rm B}\,\meanBB^2/B_{\rm eq}^2}.
\EN
In the following we assume $\alpha_{\rm B}=\eta_{\rm B}$ and denote
\EQ
\alpha_{\rm B}/B_{\rm eq}^2=1/B_0^2=\eta_{\rm B}/B_{\rm eq}^2
\EN
We emphasize that only the {\it turbulent} magnetic diffusivity is
quenched, not of course the total one. It is only because of the
presence of microscopic diffusion that saturation is possible.

In the simulations $\meanBB^2$ is to a good approximation spatially
uniform. Defining the magnetic energy as $M(t)=\half\bra{\meanBB^2}$
we have
\EQ
\meanBB^2=2M,
\EN
which is only a function of time.

Consider the particular example where the large scale field
varies only in the $z$ direction \Eq{dyneq} becomes
\EQ
\dot{\meanBB}_x=-\alpha\meanBB_y'+\eta_{\rm T}\meanBB_x'',
\EN
\EQ
\dot{\meanBB}_y=+\alpha\meanBB_x'+\eta_{\rm T}\meanBB_y'',
\EN
where dots and primes denote differentiation with respect to $t$ and $z$,
respectively. Since $\alpha<0$, the solution can be written in the form
\EQ
\meanBB_x(y,t)=b_x(t)\cos(z+\phi),
\EN
\EQ
\meanBB_y(y,t)=b_y(t)\sin(z+\phi),
\EN
where $b_x(t)$ and $b_z(t)$ are positive functions of time that satisfy
\EQ
\dot{b}_x=|\alpha| b_y-\eta_{\rm T} b_x,
\label{mf1}
\EN
\EQ
\dot{b}_y=|\alpha| b_x-\eta_{\rm T} b_y.
\label{mf2}
\EN
We now choose the special initial condition, $b_x=b_y\equiv b$, so we have
only one equation for the variable $b$. Note also that in the quenching factor
$\meanBB^2=b_x^2+b_y^2=2b^2$. Thus, we have
\EQ
{\dd b\over\dd t}
={\alpha_0k_1-\eta_{\rm t0}k_1^2\over1+2b^2/B_0^2}\,b
-\eta k_1^2b.
\EN
Multiplying with $b$ yields
\EQ
\half{\dd b^2\over\dd t}
={\alpha_0k_1-\eta_{\rm t0}k_1^2\over1+2b^2/B_0^2}b^2
-\eta k_1^2b^2.
\EN
Using the definition $M=\half b^2$ we have
\EQ
\half{1\over M}{\dd M\over\dd t}
={\alpha_0k_1-\eta_{\rm t0}k_1^2\over1+2M/M_0}-\eta k_1^2,
\EN
where $M_0=\half B_0^2$. Thus, we have
\EQ
\left(1+2{M\over M_0}\right)
\,\half{1\over M}{\dd M\over\dd t}=\alpha_0k_1-\eta_{\rm T0}k_1^2
-2\eta k_1^2{M\over M_0},
\EN
where we have defined $\eta_{\rm T0}=\eta_{\rm t0}+\eta$. We define the
abbreviations $\lambda=\alpha_0k_1-\eta_{\rm T0}k_1^2$ for the kinematic growth
rate of the dynamo and $\lambda_{\rm t0}=\eta_{\rm t0}k_1^2$ for the turbulent
decay rate if there were no dynamo action, and arrive thus at the integral
\EQ
\int_{M_{\rm ini}}^M{1+2M'/M_0\over
1-(2\eta k_1^2/\lambda)\,M'/M_0}
\,{\dd M'\over M'}=2\lambda t.
\EN
We now also define the abbreviation
\EQ
{\cal M}=2M'/M_0
\EN
and have
\EQ
\int_{M_{\rm ini}}^M{1+{\cal M}\over
1-(\eta k_1^2/\lambda)\,{\cal M}}
\,{\dd{\cal M}\over{\cal M}}=2\lambda t.
\EN
which can be split into two integrals,
\EQ
\int_{M_{\rm ini}}^M{\dd{\cal M}\over
[1-(\eta k_1^2/\lambda)\,{\cal M}]{\cal M}}
+\int_{M_{\rm ini}}^M{\dd{\cal M}\over
1-(\eta k_1^2/\lambda)\,{\cal M}}
=2\lambda t.
\label{int1}
\EN
To solve these integrals we note that
\EQ
\int{\dd x\over1-x/x_0}=-x_0\ln(x_0-x),
\EN
\EQ
\int{\dd x\over(1-x/x_0)x}=\int{\dd x\over x_0-x}+\int{\dd x\over x}
=\ln x-\ln(1-x/x_0)
=\ln\left({x\over x_0-x}\right),
\EN
so \Eq{int1} becomes
\EQ
\ln\left({{\cal M}\over\lambda/\eta k_1^2-{\cal M}}\right)
+{\lambda\over\eta k_1^2}
\ln\left({1\over\lambda/\eta k_1^2-{\cal M}}\right)
=2\lambda(t-t_0),
\EN
where $t_0$ is an integration constant.
Exponentiation yields
\EQ
{{\cal M}\over(\lambda/\eta k_1^2-{\cal M})^{1+\lambda/\eta k_1^2}}
=e^{2\lambda(t-t_0)},
\EN
which is, in terms of the original variables,
$\meanBB^2/B_0^2=2b^2/B_0^2=2M/M_0={\cal M}$, we have
\EQ
{\meanBB^2\over B_0^2}
\left/\left({\lambda\over\eta k_1^2}
-{\meanBB^2\over B_0^2}\right)^{1+\lambda/\eta k_1^2}\right.
=e^{2\lambda(t-t_0)}.
\label{implres1}
\EN
The final field strength,
\EQ
B_{\rm fin}=\lim_{t\rightarrow\infty}|\meanBB(t)|,
\EN
is given by requiring the denominator to vanish, which yields
\EQ
{B_{\rm fin}^2\over B_0^2}={\lambda\over\eta k_1^2}.
\EN
Rewriting \Eq{implres1} in terms of $B_{\rm fin}$ we have
\EQ
{\meanBB^2\over B_{\rm fin}^2}\left/
\left(1-{\meanBB^2\over B_{\rm fin}^2}\right)^{1+\lambda/\eta k_1^2}\right.
=\left({\lambda\over\eta k_1^2}\right)^{1+\lambda/\eta k_1^2}
e^{2\lambda(t-t_0)}.
\label{implres2}
\EN
We can express $t_0$ in terms of the initial field strength, $B_{\rm
ini}=|\meanBB(0)|$, and if the initial field strength is much weaker
than the final field strength, i.e.\ $B_{\rm ini}\ll B_{\rm fin}$,
then we can rewrite \Eq{implres2}, in the form
\\

\SHADOWBOX{
\EQ
\meanBB^2/(1-\meanBB^2/B_{\rm fin}^2)^{1+\lambda/\eta k_1^2}
=B_{\rm ini}^2\,e^{2\lambda t}.
\label{implres3}
\EN
}\\

\noindent
Thus, for early times we have the familiar relation
\EQ
|\meanBB|\approx B_{\rm ini}\,e^{\lambda t}\quad(|\meanBB|\ll B_{\rm fin}),
\EN
whereas for late times near the final field strength we have
\EQ
(1-\meanBB^2/B_{\rm fin}^2)^{-(1+\lambda/\eta k_1^2)}\approx
(B_{\rm ini}/B_{\rm fin})^2\,e^{2\lambda t}
\quad(|\meanBB|\approx B_{\rm fin}),
\EN
or
\EQ
\meanBB^2\approx B_{\rm fin}^2\,
\left[1-e^{-2\lambda(t-t_{\rm sat})}\right]^{\eta k_1^2/(\lambda+\eta k_1^2)}
\quad(|\meanBB|\approx B_{\rm fin}),
\EN
where $t_{\rm sat}=\lambda^{-1}\ln(B_{\rm fin}/B_{\rm ini})$ is the time
it takes to reach saturation. If the Reynolds number is large
we have $\lambda\gg\eta k_1^2$, so
\EQ
\meanBB^2\approx B_{\rm fin}^2\,
\left[1-e^{-2\eta k_1^2(t-t_{\rm sat})}\right]
\quad(|\meanBB|\approx B_{\rm fin}),
\EN
which is identical to the result obtained from helicity conservation.

Note that the solution \eq{implres3} is governed by four parameters:
$B_{\rm ini}$, $B_{\rm fin}$, $\lambda$, and $\eta k_1^2$. The latter
is known from the input data to the simulation, $B_{\rm ini}$ and
$\lambda$ can be determined from the linear growth phase of the dynamo
(characterized by properties of the small scale dynamo!) and so $B_{\rm
fin}$ is the only parameter that is determined by the nonlinearity of the
dynamo and can easily be determined from the simulations. Once $B_{\rm
fin}$ is measured from numerical experiments we know immediately the
quenching parameters
\EQ
\alpha_{\rm B}=\eta_{\rm B}={\lambda\over\eta k_1^2}\,
\left({B_{\rm eq}\over B_{\rm fin}}\right)^2,
\EN
and since $B_{\rm fin}^2/B_{\rm eq}^2\approx k_{\rm f}/k_1$ we have
\EQ
\alpha_{\rm B}=\eta_{\rm B}\approx{\lambda\over\eta k_1 k_{\rm f}},
\EN
which shows that $\alpha_{\rm B}$ and $\eta_{\rm B}$ are proportional
to the magnetic Reynolds number.

\vfill\bigskip\noindent{\it
$ $Id: paper.tex,v 1.38 2001/09/27 07:24:17 brandenb Exp $ $}
\end{document}

%% file: axel.tex
\newcommand{\EQ}{\begin{equation}}
\newcommand{\EN}{\end{equation}}
\newcommand{\EQA}{\begin{eqnarray}}
\newcommand{\ENA}{\end{eqnarray}}
\newcommand{\eq}[1]{(\ref{#1})}
\newcommand{\Eq}[1]{eq.~(\ref{#1})}
\newcommand{\Eqs}[2]{eqs.~(\ref{#1}) and~(\ref{#2})}

\newcommand{\Eqss}[2]{eqs.~(\ref{#1})--(\ref{#2})}
\newcommand{\eqss}[2]{(\ref{#1})--(\ref{#2})}
\newcommand{\Sec}[1]{\S\,\ref{#1}}
\newcommand{\App}[1]{Appendix~\ref{#1}}
\newcommand{\Fig}[1]{figure~\ref{#1}}

\newcommand{\Tab}[1]{table~\ref{#1}}

\newcommand{\Figs}[2]{figures~\ref{#1} and \ref{#2}}
\newcommand{\Tabs}[2]{tables~\ref{#1} and \ref{#2}}
\newcommand{\bra}[1]{\langle #1\rangle}

\newcommand{\meanAA}{\overline{\mbox{\boldmath $A$}}}
\newcommand{\meanBB}{\overline{\mbox{\boldmath $B$}}}
\newcommand{\meanJJ}{\overline{\mbox{\boldmath $J$}}}

\newcommand{\rr}{\hat{\mbox{\boldmath $r$}} {}}

\newcommand{\pp}{\hat{\mbox{\boldmath $\phi$}} {}}

\newcommand{\AAAA}{\hat{\mbox{\boldmath $A$}} {}}
\newcommand{\BBBB}{\hat{\mbox{\boldmath $B$}} {}}

\newcommand{\rrr}{\mbox{\boldmath $r$} {}}
\newcommand{\xx}{\mbox{\boldmath $x$} {}}

\newcommand{\uu}{\mbox{\boldmath $u$} {}}

\newcommand{\UU}{\mbox{\boldmath $U$} {}}

\newcommand{\bb}{\mbox{\boldmath $b$} {}}

\newcommand{\BB}{\mbox{\boldmath $B$} {}}

\newcommand{\II}{\mbox{\boldmath $I$} {}}
\newcommand{\AAA}{\mbox{\boldmath $A$} {}}
\newcommand{\aaa}{\mbox{\boldmath $a$} {}}
\newcommand{\jj}{\mbox{\boldmath $j$} {}}
\newcommand{\JJ}{\mbox{\boldmath $J$} {}}
\newcommand{\cc}{\mbox{\boldmath $c$} {}}

\newcommand{\ff}{\mbox{\boldmath $f$} {}}
\newcommand{\EE}{\mbox{\boldmath $E$} {}}
\newcommand{\FF}{\mbox{\boldmath $F$} {}}

\newcommand{\CC}{\mbox{\boldmath $C$} {}}

\newcommand{\MM}{\mbox{\boldmath $M$} {}}

\newcommand{\kk}{\mbox{\boldmath $k$} {}}

\newcommand{\grav}{\mbox{\boldmath $g$} {}}
\newcommand{\nab}{\mbox{\boldmath $\nabla$} {}}

\newcommand{\oo}{\mbox{\boldmath $\omega$} {}}
\newcommand{\ttau}{\mbox{\boldmath $\tau$} {}}

\newcommand{\KKKK}{\mbox{\boldmath ${\sf K}$} {}}
\newcommand{\JJJJ}{\mbox{\boldmath ${\sf J}$} {}}

\newcommand{\SSSS}{\mbox{\boldmath ${\sf S}$} {}}
\newcommand{\MMMM}{\mbox{\boldmath ${\sf M}$} {}}

\newcommand{\diag}{{\rm diag}  \, {}}

\newcommand{\DD}{{\rm D} {}}

\newcommand{\dd}{{\rm d} {}}

\newcommand{\yana}[5]{~ #1~ #5. {\em Astron. Astrophys. }{\bf #2}, #3--#4.}

\newcommand{\ysci}[5]{~ #1~ #5. {\em Science }{\bf #2}, #3--#4.}

\newcommand{\ysph}[5]{~ #1~ #5. {\em Solar Phys. }{\bf #2}, #3--#4.}

\newcommand{\ymn}[5]{~ #1~ #5. {\em Monthly Notices Roy. Astron. Soc. }
{\bf #2}, #3--#4.}

\newcommand{\ynat}[5]{~ #1~ #5. {\em Nature }{\bf #2}, #3--#4.}
\newcommand{\yjfm}[5]{~ #1~ #5. {\em J. Fluid Mech. }{\bf #2}, #3--#4.}

\newcommand{\ypr}[5]{~ #1~ #5. {\em Phys. Rev. }{\bf #2}, #3--#4.}
\newcommand{\yprl}[5]{~ #1~ #5. {\em Phys. Rev. Lett. }{\bf #2}, #3--#4.}

\newcommand{\yjcp}[5]{~ #1~ #5. {\em J. Comp. Phys. }{\bf #2}, #3--#4.}
\newcommand{\yjgr}[5]{~ #1~ #5. {\em J. Geophys. Res. }{\bf #2}, #3--#4.}

\newcommand{\yapj}[5]{~ #1~ #5. {\em Astrophys. J. }{\bf #2}, #3--#4.}
\newcommand{\yapjS}[5]{~ #1~ #5. {\em Astrophys. J. }{\bf #2}, #3--#4}
\newcommand{\ypp}[5]{~ #1~ #5. {\em Phys. Plasmas }{\bf #2}, #3--#4.}

\newcommand{\ypf}[5]{~ #1~ #5. {\em Phys. Fluids }{\bf #2}, #3--#4.}

\newcommand{\yapjl}[5]{~ #1~ #5. {\em Astrophys. J. Letters }{\bf #2}, #3--#4.}

\newcommand{\yjour}[6]{~ #1~ #6. {\em #2} {\bf #3}, #4--#5.}
\newcommand{\yproc}[7]{~ #1~ #4. In {\em #5} (ed. #6), pp. #2--#3. #7.}

\newcommand{\ybook}[3]{~ #1~ {\em #2}. #3.}

\newcommand{\sana}[2]{ ~#1~ ``#2,'' {\em Astron. Astrophys.} (submitted).}
\newcommand{\tana}[2]{ ~#1~ ``#2,'' {\em Astron. Astrophys.} (to be submitted).}

\newcommand{\papj}[3]{:~#1, ``#2,'' {\em Astrophys. J.} (in press, scheduled for the #3 issue)}

\newcommand{\ea}{{\em et al.\ }}

\def\half{{\textstyle{1\over2}}}

\def\onethird{{\textstyle{1\over3}}}
\def\twothird{{\textstyle{2\over3}}}

\newcommand{\G}{\,{\rm G}}

\newcommand{\s}{\,{\rm s}}

\newcommand{\cm}{\,{\rm cm}}

\newcommand{\kms}{\,{\rm km/s}}

\newcommand{\gs}{\,{\rm g/s}}

\newcommand{\kpc}{\,{\rm kpc}}
\newcommand{\Mpc}{\,{\rm Mpc}}
\newcommand{\yr}{\,{\rm yr}}

\newcommand{\Gyr}{\,{\rm Gyr}}
\newcommand{\erg}{\,{\rm erg}}
\newcommand{\ergs}{\,{\rm erg/s}}

\newcommand{\AU}{\,{\rm AU}}

\newcommand{\SHADOWBOX}[1]{\shadowbox{\parbox{15cm}{#1}}}